\def\pdate{01.12.04}
\documentclass[aps,prb,twocolumn,groupedaddress,showpacs,%
               nobalancelastpage,floatfix]{revtex4}
\usepackage{amsmath,amssymb,amsfonts}
\usepackage{bm,graphicx}
\begin{document}
\title{Auxiliary-Fermion Approach to Critical Fluctuations \\
       in the 2D Quantum AF Heisenberg Model}
\author{Jan Brinckmann}
\author{Peter W{\"o}lfle}
\affiliation{Institut f{\"u}r Theorie der Kondensierten Materie,
             Universit{\"a}t Karlsruhe, 
             D-76128 Karlsruhe, Germany}
\date{\pdate}
\begin{abstract}
  The nearest-neighbor quantum-antiferromagnetic (AF) Heisenberg model
  for spin $1/2$ on a two-dimensional square lattice is studied in the
  auxiliary-fermion representation. Expressing spin operators by
  canonical fermionic particles requires a constraint on the fermion
  charge $Q_i=1$ on each lattice site $i$\,, which is imposed
  approximately through the thermal average. The resulting interacting
  fermion system is first treated in mean-field theory (MFT), which
  yields an AF ordered ground state and spin waves in quantitative
  agreement with conventional spin-wave theory. At finite temperature
  a self-consistent approximation beyond mean field is required in
  order to fulfill the Mermin--Wagner theorem. We first discuss a
  fully self-consistent approximation, where fermions are renormalized
  due to fluctuations of their spin density, in close analogy to FLEX.
  While static properties like the correlation length,
  $\xi(T)\propto\exp(a\,J/T)$\,, come out correctly, the dynamical
  response lacks the magnon-like peaks which would reflect the
  appearance of short-range order at low $T$\,. This drawback, which
  is caused by overdamping, is overcome in a `minimal self-consistent
  approximation' (MSCA), which we derive from the equations of motion.
  The MSCA features dynamical scaling at small energy and temperature
  and is qualitatively correct both in the regime of order-parameter
  relaxation at long wavelengths $\lambda>\xi$ and in the
  short-range-order regime at $\lambda<\xi$\,. We also discuss the
  impact of vertex corrections and the problem of pseudo-gap formation
  in the single-particle density of states due to long-range
  fluctuations. Finally we show that the (short-range) magnetic order
  in MFT and MSCA helps to fulfill the constraint on the local fermion
  occupancy.
\end{abstract}
\maketitle
\newlength{\mywidth}
\newcommand{\mygraph}[2]{%
  \settowidth{\mywidth}{\includegraphics[#1]{#2}}%
  \addtolength{\mywidth}{2pt}%
  \parbox{\mywidth}{\hspace*{1pt}\includegraphics[#1]{#2}%
                    \hspace*{1pt}}%
  }

\section{Introduction}
\label{sec-intro}
Antiferromagnetic correlations play an important role in the physics
of high-temperature superconductors. Magnetism is most pronounced in
the undoped parent compounds, where the almost uncoupled
two-dimensional CuO-planes form magnetic Mott insulators with one
spin-1/2 electron per Cu-site. The simplest model for a CuO-plane is
the quantum Heisenberg antiferromagnet (QHAF) in two dimensions (2D)
on a square lattice,
\begin{equation} \label{eqn-ham}
  H= \frac{1}{2}\sum_{i,j}J_{ij}\,\mathbf{S}_i\mathbf{S}_j
\end{equation}
The super-exchange coupling $J_{ij}= J>0$ is non-zero only if the
Cu-sites $i,j$ are nearest neighbors. In the undoped compounds studied
experimentally, it assumes values around $J\simeq 130\,$meV\,.

This model has frequently been considered in the past\cite{note-manous}.
Its ground state at zero temperature $T=0$ appears to be a N\'{e}el
state with an order parameter $|\langle\mathbf{S}\rangle|$ moderately
reduced from its classical value $1/2$ by quantum fluctuations. At
finite temperature $T>0$ the theorem of Mermin and Wagner prohibits
any long-range order. For sufficiently low $T$ the magnetic
correlation length $\xi$ grows exponentially in the two-dimensional
system and diverges at $T=0$\,,
\begin{equation}  \label{eqn-xiresult}
  \xi(T)\propto\exp(a\,J/T)
\end{equation}
$\xi$ is measured in units of the lattice spacing, which is set to
$1$\,. In the following we also use $\hbar\equiv 1$\,, $k_B\equiv
1$\,. The exponential factor in $\xi(T)$ has been obtained from
quantum-monte-carlo (QMC) calculation\cite{bea98,kimtro98,makdin91} of
the QHAF, high-temperature series expansion\cite{elst95},
Schwinger-boson mean-field theory\cite{aaa88} (SBMFT), and modified
spin-wave theory\cite{tak87}. The exponent comes
out\cite{kimtro98,bea98,aaa88} as $a\simeq 1.15$\,. An exponential
behavior like Eq.(\ref{eqn-xiresult}) also results from the
(classical) non-linear sigma model\cite{nelpel77,amitremark}
(NL$\sigma$M) as well as the quantum NL$\sigma$M (QNL$\sigma$M) in the
so-called renormalized classical regime\cite{chn88,chn89,hn93}.

The theoretical predictions from the QHAF and the (Q)NL$\sigma$M are
consistent with experiments on undoped cuprates above their respective
N\'{e}el temperature $T_N$\,, confirming the almost two-dimensional
nature of fluctuations in these systems. In particular, a correlation
length behaving like Eq.(\ref{eqn-xiresult}) has been seen in
quasi-elastic (energy integrated) neutron scattering on
Sr$_2$Cu$_3$O$_4$Cl$_2$ (Ref.\ \onlinecite{kim01a}),
Sr$_2$CuO$_2$Cl$_2$ (Ref.\ \onlinecite{gre94}), and La$_2$CuO$_4$
(Ref.\ \onlinecite{bir99,keim92,end88})\,. The
observed\cite{kim01a,bir99,gre94} exponent $a= 1.15$ agrees with
theory. A three-dimensional (algebraic) behavior might be
observable\cite{kim01a,gre94} at temperatures very close to $T_N$\,.

The dynamics of spin fluctuations are measured through the dynamical
structure factor $S(\mathbf{q},\omega)$\,. In general two regimes are
expected, depending on the wave vector $\mathbf{k}= \mathbf{q} -
\mathbf{Q}$ relative to the N\'{e}el ordering vector $\mathbf{Q}=
(\pi,\pi)$\,: For $k\xi\ll 1$\,, i.e., for distances much longer than
the correlation length, the spectrum $S$ shows a single peak at
$\omega=0$ with some width $\sim\omega_0$ that depends weakly on
$k\xi$\,. The energy scale $\omega_0$ represents the time scale for
relaxation of the order parameter in the disordered phase $T>0$\,.
Critical slowing down leads to a vanishing of $\omega_0$ like
$\omega_0\sim\xi^{-z}$ as $T\to 0$\,. The dynamical exponent $z$ comes
out as $z= 1$\,. For $k\xi\gg 1$ the peak position shifts to finite
$\omega$ and becomes $k$-dependent, reminiscent of damped spin waves.
These originate from the short-range magnetic order, visible only
at distances small compared to the correlation length. Moreover, the
dynamical scaling hypothesis\cite{chn89} states that for $\xi\gg 1$
and all $k\ll\pi$ spin fluctuations are governed by the single energy
scale $\omega_0$\,; the structure factor obeys the scaling form
\begin{equation}  \label{eqn-hypdyn}
  S(\mathbf{q}, \omega)= 
    \frac{1}{\omega_0}S^{st}(\mathbf{q})\Phi(k\xi, \omega/\omega_0)
\end{equation}
with the static structure factor (equal-time correlation function)
\begin{equation}  \label{eqn-hypstat}
  S^{st}(\mathbf{q})= \langle S^x_{\mathbf{q}} S^x_{-\mathbf{q}} \rangle
    = S^{st}(\mathbf{Q})\varphi(k\xi)\,.
\end{equation}
$\varphi$ and $\Phi$ denote (a priori unknown) scaling functions. The
dynamical structure factor has been calculated\cite{makjar92} using
the QMC plus maximum-entropy method for the QHAF, molecular dynamics
(MD) on the classical Heisenberg model\cite{wysbis90}, the
SBMFT\cite{aaa88a,kop90}, and MD on a classical lattice
model\cite{thc89} equivalent to the QNL$\sigma$M in the renormalized
classical regime\cite{chn89}. The form (\ref{eqn-hypdyn}) of the
structure factor, the presence of propagating modes in the `shape
function' $\Phi$ for $k\xi> 1$\,, and the energy scale
$\omega_0\sim\xi^{-1}$ have been well confirmed. The latter has also
been calculated from coupled-mode theory\cite{grempel88}.
Experimentally, inelastic neutron scattering (INS) on several cuprate
parent compounds\cite{yam89,hay90,kim01} at temperatures $T>T_N$ shows
a quasi-elastic peak at zero energy, the observed\cite{kim01}
dynamical exponent is $z=1$\,. The latter is also found in magnetic
resonance\cite{carr97}. The INS data is consistent to spin-wave like
excitations at higher energies\cite{yam89} in the paramagnetic phase.

Results for the dynamics of spin fluctuations come primarily from
purely numerical approaches like QMC and MD\,. For an analytical
treatment of the Heisenberg model, the non-canonical commutation
relations of spin operators $S^x, S^y, S^z$ pose a severe difficulty.
This can be circumvented by re-writing $S^\mu$ in canonical boson or
fermion creation and annihilation operators. These act on an enlarged
Hilbert space that contains unphysical states, which have to be
removed by imposing a constraint. Using Schwinger bosons $b_n$\,, $n=
1,2$\,, the spin-1/2 operator and constraint read\cite{aaa88}
\begin{displaymath}
  \mathbf{S}= \frac{1}{2}\sum_{n,n'= 1,2}
    b^\dagger_{n}\bm{\sigma}^{n n'}b_{n'}
    \;\;,\;\;\;
  \sum_{n= 1,2}b^\dagger_{n}b_{n}= 1
\end{displaymath}
Alternatively, one can use Abrikosov auxiliary
fermions\cite{abr65,affmar88} $f_\uparrow, f_\downarrow$ that carry
spin-1/2\,,
\begin{equation} \label{eqn-spinop}
  \mathbf{S}_i= \frac{1}{2}\sum_{\alpha,\alpha'=\pm 1}
            f^\dagger_{i\alpha}\bm{\sigma}^{\alpha\alpha'}f_{i\alpha'}
    \;\;,\;\;\;
  Q_i= \sum_\alpha f^\dagger_{i\alpha}f_{i\alpha}= 1
\end{equation}
Here a lattice-site index $i$ has been added; $\alpha= +1, -1$
corresponds to $\uparrow, \downarrow$\,.

The fermion representation (\ref{eqn-spinop}) of the Heisenberg model
(\ref{eqn-ham}) is actually the 1/2-filled (undoped) limit of the
standard slave-boson formulation of the $t$-$J$-model. The latter has
been extensively studied using, e.g., mean-field
theory\cite{note-rvb}. These works focus on the superconducting and
normal-state properties of the $t$-$J$-model at finite hole filling
(doping), and its application to the doped cuprate
superconductors\cite{note-rvb}. At low hole concentration and in
particular at 1/2-filling the slave-boson mean-field theory gives
antiferromagnetic (AF) long-range order\cite{inui88,inaba96}. The
strong magnetic correlations at finite temperature in the 2D system,
however, have not been considered in this type of approach.

The purpose of this paper is to study the antiferromagnetic spin-1/2
Heisenberg model (QHAF) on a strictly two dimensional square lattice
at finite temperature, using the auxiliary-fermion representation
(\ref{eqn-spinop})\,. In contrast to the boson
approach\cite{aaa88,aaa88a}, mean-field theory with fermions leads to
a finite N\'eel temperature $T_N= 0.5\,J$\,. While mean-field results
are good in the AF ordered ground state (see Sect.~\ref{sec-mf-mf}
below), at $T>0$ a self-consistent T-matrix approximation beyond
mean-field is required in order to suppress the phase transition. Two
approximation schemes will be proposed and compared to the established
results reviewed above. A suitable approximation should be able to
fulfill the Mermin--Wagner theorem and reproduce the exponentially
growing magnetic correlation length, also it should yield a
qualitatively correct description of the dynamics including critical
slowing down and short-range order. Once the important elements of
such an approximation are identified, the theory can possibly be
extended to less well understood cases, e.g., the $t$-$J$-model at
finite hole filling by introducing `slave' bosons for the doped holes.

The present study is related to certain self-consistent diagrammatic
schemes for the 2D Hubbard model, like the FLEX
approximation\cite{bicscawhi89}: When the magnetic correlation length
is growing (e.g., due to a reduction of hole filling) the density of
states of electrons is expected to develop a pseudo precursor
gap\cite{kamsch90b} around the Fermi level, reflecting the proximity
to an AF ordered state which is characterized by zero spectral weight
at the Fermi energy. It has been demonstrated\cite{vilktrem97} that an
approximation with `reduced self-consistency', i.e., where part of the
electron lines in the self energy are replaced by bare ones, favor the
formation of such a spectral-weight suppression. A similar observation
is made here: our approximation with reduced self-consistency
(presented in Sect.~\ref{sec-red}) is able to produce a propagating
spin-wave-like mode, and the spectrum of auxiliary fermions shows a
pseudo gap. In the fully self-consistent approximation (built in close
analogy to FLEX), on the other hand, `spin-wave' mode and pseudo gap
are absent. The related problem of strong collective Cooper-pair
fluctuations in the normal state of 2D superconductors has been
studied in the past\cite{kadmar61,patton71,levin97}. The connection of
these approaches to the present work is considered in
Sect.~\ref{sec-red} and Appendix \ref{sec-app-emotion}\,.

The order of the paper is as follows: The auxiliary-fermion
formulation is introduced in detail in the next Section
\ref{sec-slave}\,, and the mean-field theory is discussed in
Sect.~\ref{sec-mf}\,. At zero temperature the mean-field theory
resembles linear spin-wave theory\cite{man91} (SWT). In
Sect.~\ref{sec-flex} the fully self-consistent diagrammatic
approximation is presented. Static and dynamical properties are
calculated by analytical and numerical solution of the self-consistent
integral equations. In Sect.~\ref{sec-red} the above-mentioned
approximation with reduced self-consistency is explored in detail. It
leads to results that qualitatively agree very well with what is known
about the QHAF and the QNL$\sigma$M at low temperature. In
Sect.~\ref{sec-flex-cons} we apply the conserving-approximation
method, and finally in Sect.~\ref{sec-remarks} we estimate the
fluctuations of the fermion charge $Q_i$\,. The paper closes with a
conclusion and some technical appendices. Calculations are performed
at low temperatures $T< 0.2\,J$\,, where scaling is observed. A
possible crossover to quantum-critical
behavior\cite{chn88,sokol94,kimtro98} at higher temperatures is not
considered. Sections \ref{sec-mf-corr}\,, \ref{sec-flex-cons}\,
\ref{sec-red-anl}\,, and \ref{sec-remarks} contain material not
directly related to the physics and can probably be skipped in a first
reading.

\section{Auxiliary-Fermion Formulation}
\label{sec-slave}
The constraint in Eq.(\ref{eqn-spinop}) requires the fermion charge to
be fixed to 1 at each lattice site $i$ individually. This projection
can be performed exactly in models like the Anderson impurity and
Kondo model\cite{abrmig70,col84,bic87}, which possess just one atomic
orbital with strong electron--electron interaction. In models
containing a whole lattice of such orbitals the exact projection leads
to a loss of the linked-cluster theorem, which prohibits
infinite-order resummation of the perturbation
series\cite{grekei81,kur85} and self-consistent approximations based
on the skeleton-diagram expansion. The limit of infinite spatial
dimension\cite{kimkurkas87,metvol89} has frequently been used to
circumvent this problem. An alternative is to start from a mean-field
theory\cite{newrea87}, where the constraint is observed only in the
thermal average. For the present case this reads,
\begin{equation}  \label{eqn-constr}
  Q_i\to \langle Q_i \rangle= \langle Q_1 \rangle= 
    \sum_\alpha \langle f^\dagger_{1\alpha} f_{1\alpha} \rangle
    = 1
\end{equation}
and is introduced into the Hamiltonian (\ref{eqn-ham}),
(\ref{eqn-spinop}) through a chemical potential $\mu^f$ for the
fermions. Due to particle--hole symmetry\cite{aff88} it is
$\mu^f=0$\,. The partition function can now be written as a standard
coherent-state path integral\cite{negorl},
\begin{equation}  \label{eqn-part}
  Z[h]= \int\mathcal{D}[f, \overline{f}]\,\mathrm{e}^{-A}
\end{equation}
with the action $A= A^0 + A^h + A^J$\,,
\begin{subequations}  \label{eqn-action}
  \begin{eqnarray}
    A^0 & = &  \label{eqn-actnul}
      \int_0^{\beta}\mathrm{d}\tau\,
      \sum_{i,\alpha}
      \overline{f}_{i\alpha}(\tau)[\partial_\tau - \mu^f]f_{i\alpha}(\tau)
      \\
    A^h & = &  \label{eqn-acthh}
      - \int_0^{\beta}\mathrm{d}\tau\,
      \sum_i \mathbf{h}_i(\tau) \mathbf{S}_i(\tau)
      \\
    A^J & = &  \label{eqn-actjj}
      \int_0^{\beta}\mathrm{d}\tau\,
      \frac{1}{2}\sum_{i,j}J_{ij}\,\mathbf{S}_i(\tau) \mathbf{S}_j(\tau)
  \end{eqnarray}
\end{subequations}
with anticommuting Grassmann variables $f, \bar{f}$\,, and $\beta=
1/k_BT\equiv 1/T$\,. An arbitrary magnetic source field $\mathbf{h}$
has been added. The underlying lattice is a $d$-dimensional cubic
lattice with lattice spacing $a\equiv 1$\,; only the two-dimensional
(2D) case $d=2$ will be studied in detail.

In the following the connected spin propagator is considered, 
\begin{eqnarray}
  \chi^{\mu\mu'}_{ij}(\tau,\tau') & = &  \label{eqn-sus}
    \langle \mathcal{T}_\tau S^\mu_i(\tau) S^{\mu'}_j(\tau')
    \rangle_{conn}
    \\
  & = &  \nonumber
    \langle \mathcal{T}_\tau S^\mu_i(\tau) S^{\mu'}_j(\tau') \rangle
    - \langle S^\mu_i(\tau)\rangle \langle S^{\mu'}_j(\tau') \rangle
\end{eqnarray}
where $\mu= x,y,z$ denotes the spin component. It is calculated using
a Feynman-diagram expansion\cite{negorl} with $A^J$ as perturbation.
This will involve the Green's function of fermions,
\begin{eqnarray}
  [\overline{G}_{ij}(\tau,\tau')]^{\alpha\alpha'} & = & \label{eqn-fermgf}
    - \langle \mathcal{T}_\tau f_{i\alpha}(\tau)
              f^\dagger_{j\alpha'}(\tau') \rangle
    \\
  & = &  \nonumber
    - \langle f_{i\alpha}(\tau) \overline{f}_{j\alpha'}(\tau') \rangle
\end{eqnarray}
Time-$\tau$ ordered expectation values are expressed via
\begin{equation}  \label{eqn-tprod}
  \langle M \rangle =
    \frac{1}{Z[h]}\int\mathcal{D}[f,\overline{f}]\,
    \mathrm{e}^{-A}\,M
\end{equation}
Note that the fermion Green's function $\overline{G}$ is not an
observable quantity, since it depends on local $U(1)$ gauge
transformations\cite{iof89} of $f_{i\alpha}, f^\dagger_{i\alpha}$\,.
The susceptibility $\chi$\,, on the contrary, is built from the
gauge-invariant physical observables Eq.(\ref{eqn-spinop})\,.

It is of great help that we can make use of standard Feynman-diagram
techniques for a perturbation expansion in the interaction $A^J$\,.
This requires a mean-field like treatment of the constraint
Eq.(\ref{eqn-constr}), since then perturbation theory starts from
(effectively) free fermions and Wick's theorem can be applied. In
principle the treatment of the constraint can be improved by
generalizing the chemical potential $\mu^f$ to a fluctuating Lagrange
multiplier\cite{bic87}, but we restrict ourselves to the simplest
approach\cite{kroetal92}. Nevertheless, in the approximations to be
discussed below the pseudo fermions do not develop a finite hopping
amplitude connecting different lattice sites $i, j$\,,
\begin{displaymath}
  \langle f^\dagger_{i\alpha} f_{j\alpha'}\rangle\propto\delta_{ij}
\end{displaymath}
and anomalous expectation values of the form
\begin{displaymath}
  \langle f_{i\alpha} f_{j\alpha'}\rangle = 0
\end{displaymath}
are absent. As a consequence the Green's function of fermions is
always local, i.e.,
\begin{equation}
  \label{eqn-gflocal}
  \overline{G}_{ij}= \delta_{ij}\,\overline{G}_{i}
\end{equation}
Since unphysical expectation values are zero the local charge $Q_i$
introduced in Eq.(\ref{eqn-constr}) is most likely conserved.
Therefore spurious unphysical contributions to physical correlators
like Eq.(\ref{eqn-sus}) are due to states with $Q_i\ne 1$ introduced
by the trace in the thermal average Eq.(\ref{eqn-tprod}), but these
are not generated dynamically. We return to this issue later on in
Sect.~\ref{sec-remarks}\,.

%
\section{Mean-Field Theory using Auxiliary Fermions}
\label{sec-mf}
The Heisenberg term $A^J$ is a two-particle interaction.  A mean-field
decomposition of $A^J$ leads to an effective
Hamiltonian\cite{inui88,ubblee92,inaba96,bri02} with
self-consistent order parameters
$\sum_{\alpha,\alpha'}
  \langle f^\dagger_{i\alpha}
  \bm{\sigma}^{\alpha\alpha'} f_{i\alpha'}\rangle
  = 2\langle\mathbf{S}_i\rangle$\,,
$\langle f^\dagger_{i\alpha}f_{j\alpha}\rangle$\,,
$\langle f_{i\uparrow}f_{j\downarrow}\rangle$\,.
These correspond to phases with different spontaneously broken
symmetries, namely magnetic order, the so-called resonating valence
bond\cite{rvbphil} (RVB) or the flux phase\cite{affmar88,mar89}, and a
d-wave paired state of the fermions\cite{ruchir87,naglee92,fuk92},
respectively.  In the undoped case considered here, the latter two are
equivalent by particle--hole symmetry\cite{aff88}. The flux and the
paired state order parameters are unphysical, but may correspond to
physical regimes observed experimentally (the pseudo-gap in the
cuprates). The respective transition temperatures are interpreted as
crossovers. The magnetic order parameter
$\sum_{\alpha,\alpha'}\langle f^\dagger_{i\alpha}
  \bm{\sigma}^{\alpha\alpha'} f_{i\alpha'}\rangle$\,,
  on the other hand, is a physical (gauge invariant) quantity; a
  non-zero value indicates a true phase transition. If the free
  energies of these phases are compared in an unrestricted mean-field
  calculation, it turns out\cite{inui88,inaba96} that only
  $\langle\mathbf{S}\rangle$ becomes finite in the Heisenberg model,
  indicating a N{\'e}el-ordered phase below some transition
  temperature $T_N\sim J$\,. This is consistent with the fact that the
  ground state of the Heisenberg model is close to N{\'e}el
  order\cite{note-manous}. Only for finite hole filling in the
  $t$-$J$-model a phase is observed at low $T$\,, where AF and d-wave
  pairing order coexist\cite{inaba96}. Therefore we do not consider
  flux or pairing instabilities.
  
  Note that the large-$N$ approach\cite{affmar88,mar89} leads to a
  different ground state: The spin-index of the fermions is
  generalized from 2 to $N$ states, and the limit $N\to\infty$ leads
  to a saddle-point theory giving a flux phase with
$\langle f^\dagger_{i\alpha} f_{j\alpha} \rangle\ne 0$\,,
while magnetism is suppressed.

\begin{figure} \large
    \begin{displaymath}
      \overline{\Sigma} = \mygraph{scale=0.5}{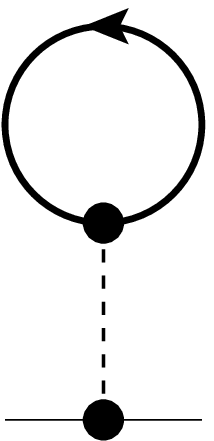}
        \;\;\;\;
      \Pi = \mygraph{scale=0.5}{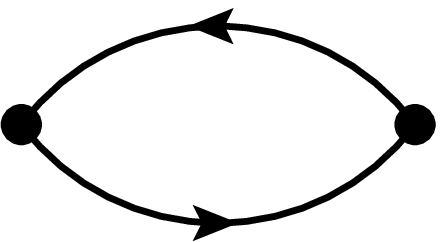}
    \end{displaymath}
    \begin{displaymath}
      \Delta\langle S^z \rangle =
        -\,\,\mygraph{scale=0.5}{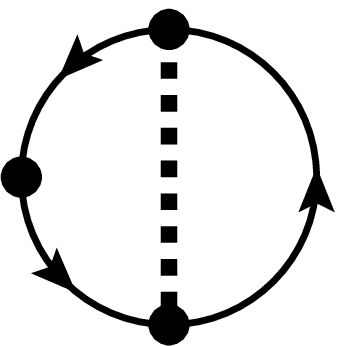}
    \end{displaymath}
  \caption[\ ]{\label{fig-mf}%
    Diagrammatic representation of \protect$\Sigma, \Pi, \Delta\langle
    S^z\rangle\protect$ in mean-field approximation, Sect.\
    \ref{sec-mf-mf}\,. The 
    thin dashed line is the bare interaction $J$ in Eq.(\ref{eqn-actjj}),
    the thick dashed line the renormalized one Eq.(\ref{eqn-effj}),
    and the full lines are the Green's function of pseudo fermions
    Eq.(\ref{eqn-fgfself}) in mean-field approximation,
    Eq.(\ref{eqn-fgfmf})\,. The dots represent Pauli matrices $\times
    1/2$\,. A closed fermion loop requires a trace in spin space.
    }
\end{figure}
\subsection{AF Order and Spin Waves}
\label{sec-mf-mf}
With the source-field $\mathbf{h}$ set to zero, the Green's function
(\ref{eqn-fermgf}), (\ref{eqn-gflocal}) of the fermions reads
\begin{equation}  \label{eqn-fgfself}
  \overline{G}_i(i\omega)= \left[ i\omega + \mu^f - 
    \overline{\Sigma}_i(i\omega) \right]^{-1}
\end{equation}
$\omega= (2n + 1)\pi/\beta$ denotes a fermionic Matsubara frequency,
$\overline{G}$ and $\overline{\Sigma}$ are matrices in spin space, and
the chemical potential is always $\mu^f=0$ by particle--hole symmetry.
In mean-field the self energy $\overline{\Sigma}_i$ is given by the
Hartree approximation depicted in Fig.~\ref{fig-mf} and reads
\begin{eqnarray*}
  \overline{\Sigma}_i(i\omega) & = &  
    \sum_{j} J_{ij}\,\sum_\mu
    \frac{1}{2}\sigma^\mu\frac{1}{2}
    \text{Tr}[\sigma^\mu\overline{G}_j(0, 0_+)]
    \\ 
  & = &
    \sum_{j} J_{ij}\, \sum_\mu\frac{1}{2}\sigma^\mu
    \langle S^\mu_j\rangle
\end{eqnarray*}
Not shown in Fig.~\ref{fig-mf} are the (exchange) Fock and (anomalous)
Bogoliubov diagrams, which correspond to the above-mentioned flux and
pairing order parameters, respectively, and do not contribute. If we
assume N{\'e}el ordering in $z$-direction on sublattices $A$ and
$B$\,, i.e., $\langle S^z_A\rangle= - \langle S^z_B\rangle$\,, the
self energy becomes
\begin{equation}  \label{eqn-mfself}
  \begin{array}{rcl}
    \overline{\Sigma}_A(i\omega) & = & - \sigma^z m_A  \\
    \overline{\Sigma}_B(i\omega) & = & \sigma^z m_A
  \end{array}
    \;\;\;\text{with}\;\;\;
    m_A = \frac{1}{2}\,2dJ\,\langle S^z_A \rangle
\end{equation}
The coordination number $2d$ has been inserted. The Green's function
(\ref{eqn-fgfself}) now reads
\begin{equation}  \label{eqn-fgfmf}
  \overline{G}_A(i\omega)= 
    \frac{i\omega - \sigma^z m_A}{(i\omega)^2 - (m_A)^2}
\end{equation}
and similar for $B$ with $m_A\to m_B= -m_A$\,. Inserting into $\langle
S^z_A \rangle$ delivers an equation for the Weiss-field $m$\,,
\begin{eqnarray}
  m_A & = &  \label{eqn-mfeqn}
    \frac{dJ}{2}\sum_{\alpha=\pm 1}\frac{1}{\beta}\sum_{i\omega}
    \frac{\alpha \,\mathrm{e}^{-i\omega 0_+}}{i\omega + \alpha m_A}
    \\
  & = &  \nonumber
    \frac{dJ}{2}\tanh(\frac{m_A}{2\,T})
\end{eqnarray}
Antiferromagnetic order $m_A\ne 0$ appears below the transition
temperature
$T_N= \frac{d}{4}J= J/2$\,.
The averaged local density of states of the fermions,
$\rho(\omega)= \frac{1}{N_L}\sum_i\frac{1}{\pi}
   \textrm{Im}\frac{1}{2}
   \textrm{Tr}\,[\sigma^z\overline{G}_i(\omega - i0_+)]$\,,
becomes
\begin{equation}  \label{eqn-afdos}
  \rho(\omega)= \frac{1}{2}[ \delta(\omega - m_A) + 
    \delta(\omega + m_A) ]
\end{equation}
The spectrum has a gap $2m_A$ as is known from the spin-density-wave
state. However, our pseudo fermions do not show any dispersion, and
hence $\rho(\omega)$ is $\delta$-like. For $m_A=0$ the bare spectrum
$\rho^0(\omega)= \delta(\omega)$ is recovered.

In the ordered state $T<T_N$ the magnetic excitations should be given
by spin waves. In order to compare to the results from linear
spin-wave theory, we calculate the spin propagator (\ref{eqn-sus})
within the mean-field approximation. It has the RPA form
\begin{equation}  \label{eqn-rpa}
  \chi_{ij}^{\mu\mu'}= \left([1 + J \Pi]^{-1}\,\Pi\right)_{ij}^{\mu\mu'}
\end{equation}
with the irreducible part $\Pi$ given by the fermion bubble shown in
Fig.~\ref{fig-mf},
\begin{equation}  \label{eqn-bubble}
  \Pi_i^{\mu\mu'}(i\nu) = 
    - \frac{1}{\beta}\sum_{i\omega}\frac{1}{4}\text{Tr}
      [\sigma^\mu \overline{G}_i(i\omega + i\nu) 
       \sigma^{\mu'} \overline{G}_i(i\omega)]
\end{equation}
$\nu$ denotes a bosonic Matsubara frequency. Inserting
Eq.(\ref{eqn-fgfmf}) yields the transversal components
\begin{subequations}  \label{eqn-pitrans}
\begin{eqnarray}
  \Pi^{xx}_A(i\nu) & = &  \label{eqn-pixx}
    \Pi^{yy}_A(i\nu) = 
    \frac{2m_A\;\langle S^z_A \rangle}{(2m_A)^2 - (i\nu)^2}
    \\
  \Pi^{xy}_A(i\nu) & = &  \label{eqn-pixy}
    - \Pi^{yx}_A(i\nu) = 
    - i\frac{i\nu\;\langle S^z_A \rangle}{(2m_A)^2 - (i\nu)^2}
\end{eqnarray}
\end{subequations}
and
$\Pi^{xx}_B= \Pi^{xx}_A$\,, $\Pi^{xy}_B= - \Pi^{xy}_A$\,.
The longitudinal response is exponentially small and can be ignored,
$\Pi^{zz}_{A}, \Pi^{zz}_{B}\sim\exp({-\beta |m_{A}|})$\,.
The susceptibility (\ref{eqn-rpa}) is transformed into wave-vector
space using
\begin{eqnarray}
  \chi^{\mu\mu'}(\mathbf{q}, \mathbf{q}') & = &  \label{eqn-qspace}
    \frac{1}{N_L}\sum_{i,j}\chi_{ij}^{\mu\mu'}
    \,\mathrm{e}^{-i(\mathbf{q}\mathbf{R}_i - \mathbf{q}'\mathbf{R}_j)}
    \\
  \Pi^{\mu\mu'}(\mathbf{q}, \mathbf{q}') & = &
    \frac{1}{N_L}\sum_{i}\Pi_{i}^{\mu\mu'}
    \,\mathrm{e}^{-i(\mathbf{q} - \mathbf{q}')\mathbf{R}_i}
\end{eqnarray}
The second line leads to
\begin{displaymath}
  \Pi^{xx}(\mathbf{q}, \mathbf{q}') = 
    \Pi^{xx}_A \delta_{\mathbf{q},\mathbf{q}'}
    \;\;\;,\;\;\;
  \Pi^{xy}(\mathbf{q}, \mathbf{q}') = 
    \Pi^{xy}_A \delta_{\mathbf{q},\mathbf{q}'\pm\mathbf{Q}}
\end{displaymath}
with the AF ordering vector $\mathbf{Q}= (\pi,\pi,\ldots)$ and $\mathrm{e}^{\pm
i\mathbf{Q}\mathbf{R}_i}= +1, -1$ for $\mathbf{R}_i\in A, B$\,. The
interaction becomes
\begin{equation}  \label{eqn-gamq}
  J(\mathbf{q}) = 
    2dJ\,\gamma(\mathbf{q})
    \;\;\;\text{with}\;\;\;
  \gamma(\mathbf{q})= 
    \frac{1}{d}\sum_\mu\cos(q_\mu)
\end{equation}
By transforming Eq.(\ref{eqn-rpa}) in $\mathbf{q}$-space we find
\begin{subequations}  \label{eqn-bothchi}
\begin{eqnarray}
  \chi^{xx}(\mathbf{q}, \mathbf{q}') & = &
    \chi^{yy}(\mathbf{q}, \mathbf{q}') =
    \chi^{xx}(\mathbf{q})\delta_{\mathbf{q},\mathbf{q}'}
    \\
  \chi^{xy}(\mathbf{q}, \mathbf{q}') & = &
    -\chi^{yx}(\mathbf{q}, \mathbf{q}') =
    \chi^{xy}(\mathbf{q})\delta_{\mathbf{q},\mathbf{q}'\pm\mathbf{Q}}
\end{eqnarray}
\end{subequations}
and Eq.(\ref{eqn-rpa}) takes the form (omitting the $\mathbf{q}$
arguments)
\begin{displaymath}
    \left(
    \begin{array}{cc}
      (1 + J\Pi^{xx}_A) & J\Pi^{xy}_A  \\[1ex]
      J\Pi^{xy}_A & (1 - J\Pi^{xx}_A)
    \end{array}\right)
  \left(
  \begin{array}{c}
    \chi^{xx} \\[1ex]
    \chi^{xy}
  \end{array}\right) =
    \left(
    \begin{array}{c}
      \Pi^{xx}_A  \\[1ex]
      \Pi^{xy}_A
    \end{array}\right)
\end{displaymath}
Using Eqs.(\ref{eqn-pitrans}), (\ref{eqn-gamq}),
(\ref{eqn-mfself}) with
$J(\mathbf{q} \pm \mathbf{Q})= - J(\mathbf{q})$
gives
\begin{subequations}  \label{eqn-chiq}
  \begin{eqnarray}
    \chi^{xx}(\mathbf{q}, i\nu) & = &  \label{eqn-chiqxx}
      2dJ(\langle S^z_A \rangle)^2\frac{1 - \gamma(\mathbf{q})}
                                       {\Omega^2(\mathbf{q}) - (i\nu)^2}
      \\ 
    \chi^{xy}(\mathbf{q}, i\nu) & = &  \label{eqn-chiqxy}
      -i\langle S^z_A \rangle\frac{i\nu}{\Omega^2(\mathbf{q}) - (i\nu)^2}
  \end{eqnarray}
\end{subequations}
These response functions have poles at $\pm\Omega(\mathbf{q})$\,,
\begin{equation}  \label{eqn-swdisp}
  \Omega(\mathbf{q})= 
    2dJ\,|\langle S^z_A \rangle|\,\sqrt{1 - \gamma^2(\mathbf{q})}
    \simeq c_0|\mathbf{q} - \mathbf{Q}|
\end{equation}
$\Omega(\mathbf{q})$ resembles the magnon dispersion from linear
spin-wave theory\cite{man91} (SWT). At low energy it has two branches
with linear dispersion for $\mathbf{q}\simeq 0$ and
$\mathbf{q}\simeq\mathbf{Q}$\,. For $T\ll T_N$ the spin-wave velocity
$c_0$ is temperature independent (up to exponentially small terms) and
given by
\begin{displaymath}
  c_0= 2\sqrt{d}J\,|\langle S^z_A \rangle|
     \simeq 2\sqrt{d}J\,S= \sqrt{2}J
\end{displaymath}
This also reproduces exactly the result from SWT. The response
function measured in neutron-scattering experiments is
$\chi''(\mathbf{q}, \omega)= \text{Im}\chi^{xx}(\mathbf{q}, \omega + i0_+)$ 
and comes out from Eq.(\ref{eqn-chiqxx}) as
\begin{eqnarray}
  \chi''(\mathbf{q}, \omega) & = &  \label{eqn-swtchi}
    \frac{\pi}{2}|\langle S^z_A \rangle|
    \frac{1 - \gamma({\bf q})}{\sqrt{1 - \gamma^2(\mathbf{q})}}\times
    \\
  & &  \nonumber
    \mbox{}\times
    [ \delta(\omega - \Omega(\mathbf{q})) - \delta(\omega + \Omega(\mathbf{q}))]
\end{eqnarray}
The spin-wave excitations remain undamped ($\delta$-like) throughout
the whole Brillouin zone, since
$\max(\Omega(\mathbf{q}))= 
   2dJ |\langle S^z_A\rangle|= 2|m_A|$
   at the magnetic zone boundary just reaches the gap $2|m_A|$ for
   charge fluctuations of the auxiliary fermions. The spectral weight
   of the magnons is measured by the transversal static structure
   factor
$S^\perp(\mathbf{q}, \mathbf{q}')=
   \langle S^x_{\mathbf{q}} S^x_{-\mathbf{q}'}\rangle
   = S^\perp(\mathbf{q})\delta_{\mathbf{q},\mathbf{q}'}$
with
\begin{eqnarray}
  S^\perp(\mathbf{q}) & = &  \label{eqn-swtstruct}
    \frac{1}{\pi}\int_{-\infty}^{\infty}\mathrm{d}\omega\,
    [1 + g(\omega)] \chi''(\mathbf{q},\omega)
    \\
  & = &  \nonumber
    \frac{|\langle
    S^z_A\rangle|}{2}\coth(\frac{\Omega(\mathbf{q})}{2T})
    \frac{1 - \gamma(\mathbf{q})}
         {\sqrt{1 - \gamma^2(\mathbf{q})}}
\end{eqnarray}
$g(\omega)$ denotes the Bose function. Eqs.(\ref{eqn-swtchi}),
(\ref{eqn-swtstruct}) result from mean-field theory and apply to
any dimension $d\ge 1$\,. At zero temperature
$|\langle
    S^z_A\rangle|\coth(.)/2\to 1/4$\,,
and $S^\perp(\mathbf{q})$ becomes identical to SWT.

\subsection{\protect$1/S\protect$ Corrections}
\label{sec-mf-corr}
The mean-field theory presented above can be obtained as the first
approximation in an expansion in $1/S$\,. Following
Refs.~\onlinecite{affhal87,sin91,fradkin} the fermion is decorated
with an additional `orbital' quantum number $\kappa$\,,
$f_{i\alpha}\to f_{i\alpha\kappa}$\,, 
with $\kappa= 1, 2, \ldots, N_S$\,, $N_S= S + 1/2$ for half-integer
spin $S$\,. On each lattice site $i$ there are now $2N_S= (2S+1)$
states available. An expansion in $1/N_S\sim 1/S$ is performed by
re-scaling the interaction $J\to J= \widetilde{J}/N_S$ with
$\widetilde{J}=const.$\ in each diagram and in Eq.(\ref{eqn-rpa}), and
counting a prefactor $N_S$ for each closed fermion loop. The limit
$N_S\to\infty$ then produces the mean-field approximation.

To complete the comparison of the auxiliary-fermion approach to
spin-wave theory we also sketch the calculation of the first
correction $\sim 1/N_S$ to the staggered magnetization,
$\langle S^z_i \rangle= \langle S^z_i \rangle^{MF} + 
  \Delta\langle S^z_i \rangle$\,.
Note that $\langle S^z_i\rangle^{MF}\sim (1/N_S)^0$\,.  The
$1/N_S$-correction is shown in Fig.\ \ref{fig-mf} and reads
\begin{eqnarray}  
  \Delta\langle S^z_i \rangle & = &  \label{eqn-corr} 
    \frac{1}{2^3\,\beta^2}\sum_{i\nu, i\omega}
    \sum_{\mu,\mu'} D^{\mu\mu'}_{i}(i\nu)\times
    \\
  & &  \nonumber
    \times\text{Tr}[\sigma^z \overline{G}_i(i\omega)\sigma^\mu 
                    \overline{G}_i(i\omega+i\nu)
                    \sigma^{\mu'} G_i(i\omega)]
\end{eqnarray}
where we already returned to the physical case $S=1/2\leftrightarrow
N_S=1, \widetilde{J}=J$\,. A renormalized spin interaction has been
introduced,
\begin{equation}  \label{eqn-effj}
  D^{\mu\mu'}_{ij}(i\nu)= 
    -J_{ij} + \sum_{l,k}J_{il}\,\chi^{\mu\mu'}_{lk}(i\nu)\,J_{kj}
\end{equation}
The local $D_{i}\equiv D_{ii}$ appearing in Eq.(\ref{eqn-corr}) has no
contribution from the bare interaction, since $J_{ii}=0$\,. However,
propagating spin excitations described by the susceptibility
$\chi_{lk}$ contribute to $D_{i}$\,. With Eq.(\ref{eqn-qspace}) it
follows
\begin{eqnarray}
  D^{\mu\mu'}_{i}(i\nu) & = &    \label{eqn-effjqq}
    \frac{1}{N_L}\sum_{\mathbf{q}, \mathbf{q}'}
    J(\mathbf{q})\chi^{\mu\mu'}(\mathbf{q}, \mathbf{q}')\times
    \\
  & & \nonumber
    \mbox{}\times J(\mathbf{q}')\,\mathrm{e}^{i(\mathbf{q} 
    - \mathbf{q}')\mathbf{R}_i}
\end{eqnarray}
Eq.(\ref{eqn-corr}) becomes for the $A$-sublattice, with the fermion
propagator (\ref{eqn-fgfmf}) inserted,
\begin{eqnarray}
  \Delta\langle S^z_A \rangle & = &    \label{eqn-corr1} 
    \frac{1}{4\,\beta^2}\sum_{i\omega, i\nu}
    \sum_{\alpha=\pm 1}\times
    \\
  & & \nonumber
    \mbox{}\times
    \frac{\alpha D^{xx}_A(i\nu) + iD^{xy}_A(i\nu)}
         {(i\omega + \alpha m_A)^2(i\omega + i\nu - \alpha m_A)}
\end{eqnarray}
$\chi^{yy}=\chi^{xx}$ and $\chi^{yz}= -\chi^{xy}$ have been utilized.
The longitudinal $D^{zz}\sim\Pi^{zz}$ has vanishing weight and does
not contribute. The transversal components read with
Eq.(\ref{eqn-effjqq}) and the relations given in the text below
Eq.(\ref{eqn-qspace}):
\begin{eqnarray*}
  D^{xx}_{A}(i\nu) & = &
    \frac{1}{N_L}\sum_{\mathbf{q}} J^2(\mathbf{q})\,\chi^{xx}(\mathbf{q}, i\nu)
    \\
  D^{xy}_{A}(i\nu) & = & -
    \frac{1}{N_L}\sum_{\mathbf{q}} J^2(\mathbf{q})\,
    \chi^{xy}(\mathbf{q}, i\nu)
\end{eqnarray*}
$D^{xx}_B= D^{xx}_A$\,, $D^{xy}_B= - D^{xy}_A$ on the $B$-sublattice.
These are inserted into Eq.(\ref{eqn-corr1}), using the mean-field
expressions (\ref{eqn-chiq}) for the susceptibilities. Furthermore
Eq.(\ref{eqn-gamq}) and $\sum_{\mathbf{q}}\gamma(\mathbf{q})= 0$ are
taken into account, turning Eq.(\ref{eqn-corr1}) into an expression
where the sums over fermionic $(\omega)$ and bosonic $(\nu)$ Matsubara
frequencies still have to be performed. This is done with
\begin{eqnarray*}
  \frac{1}{\beta}\sum_{i\omega}\mathcal{F}(i\omega) & = &
    \oint\frac{\mathrm{d}z}{2\pi i}\,f(z)\,\mathcal{F}(z)
    \\
  \frac{1}{\beta}\sum_{i\nu}\mathcal{B}(i\nu) & = &
    -\frac{1}{2}\oint\frac{\mathrm{d}z}{2\pi i}\,
     \coth(\beta z/2)\,\mathcal{B}(z)
\end{eqnarray*}
$f(z)$ denotes the Fermi function, and the contour encircles the real
axis, where $\mathcal{F}(z), \mathcal{B}(z)$ may be non-analytic. At
low temperature $\beta\,|m_A|\gg 1$ the result is
\begin{equation}    \label{eqn-corr2}
  \Delta|\langle S^z_A \rangle| = - \frac{\bm{\epsilon}}{2S}
\end{equation}
\begin{displaymath}
    \bm{\epsilon}=
    \frac{1}{2 N_L}\sum_{\mathbf{q}}
    \left[ \frac{\coth(\frac{\Omega(\mathbf{q})}{2T})}
                {\sqrt{1 - \gamma^2(\mathbf{q})}} - 1 \right]
\end{displaymath}
In the ground state at $T=0$\,, where $\coth(\Omega/2T)\to 1$\,,
Eq.(\ref{eqn-corr2}) represents the reduction of the staggered
magnetization due to quantum fluctuations. It resembles exactly the
$1/S$-correction known from SWT. In 2 dimensions it is finite, with
the numerical value $\Delta|\langle S^z_A \rangle| = -\bm{\epsilon}=
-0.197$\,, i.e., for $\langle S^z\rangle>0$ we have $[\langle
S^z\rangle + \Delta\langle S^z\rangle]/\langle S^z\rangle\simeq
0.61$\,. At $T>0$\,, however, the diverging number of thermally
excited magnons with low energies destroy magnetic order in 2
dimensions (theorem of Mermin and Wagner): From Eq.(\ref{eqn-corr2})
we get for wave vectors $\mathbf{q}$ close to $0$ or $\mathbf{Q}$\,,
where $\Omega(\mathbf{q})< T$ and the thermal prefactor
$\coth(\Omega/2T)\simeq 2T/\Omega$\,,
\begin{displaymath}
  T>0\,: \;\;\;
   \Delta\langle S^z_A \rangle 
   \sim \frac{T}{S J}\int\frac{\mathrm{d}^d\,k}{k^2} + reg.
\end{displaymath}
The integral diverges in the infrared $k\to 0$ for $d<3$\,, indicating
the well-known breakdown of perturbation theory at any finite $T$ in
2D.

\section{Self-Consistent Approximation for Finite Temperature in 2D}
\label{sec-flex}
\begin{figure} \large
    \begin{eqnarray*}
      \overline{\Sigma}  & = & \mygraph{scale=0.5}{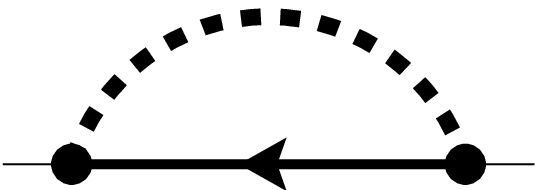}
        \;\;\;\;
      \Pi = \mygraph{scale=0.5}{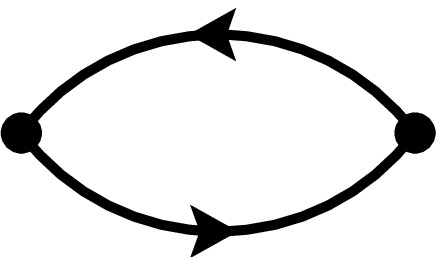}
        \\
      D & = & \mygraph{scale=0.39}{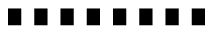} = 
              \mygraph{scale=0.39}{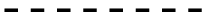} \;+\;
              \mygraph{scale=0.39}{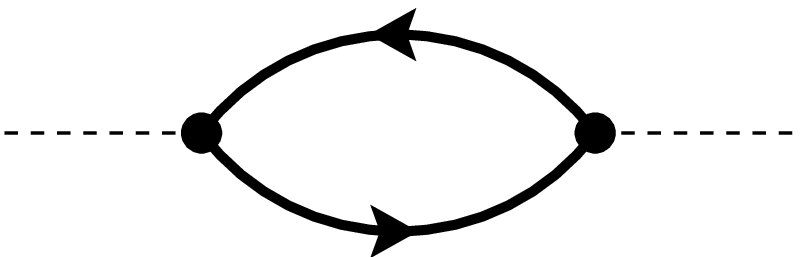} \;+\;
        \\
      & & + \;
              \mygraph{scale=0.39}{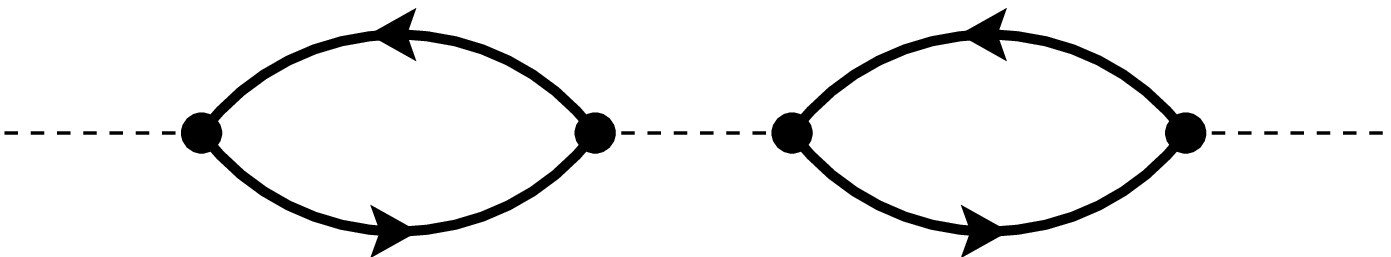} \;+\;
    \end{eqnarray*}
  \caption[\ ]{\label{fig-flex}%
    The self-consistent approximation for the paramagnetic phase at
      $\protect T>0\protect$ discussed in Sect.\
      \ref{sec-flex}\,. Full lines
      denote the fermion propagator \protect$\overline{G}\protect$
      first introduced in Eq.(\protect\ref{eqn-fgfself}), the thick
      dashed line is the renormalized interaction \protect$D\protect$
      defined in 
      Eq.(\protect\ref{eqn-effj}), the thin dashed line represents
      the bare interaction \protect$J\protect$\,. Shown are 
      the fermion self-energy \protect$\overline{\Sigma}\protect$\,,
      Eq.(\protect\ref{eqn-flexself}), the 
      irreducible bubble \protect$\Pi\protect$\,,
      Eq.(\protect\ref{eqn-bubble}), and the renormalized interaction
      \protect$D\protect$\,. Dots are Pauli matrices $\times 1/2$\,.
      In the paramagnetic phase the traces in spin space are easily
      performed, leading to the set of Eqs.(\protect\ref{eqn-scflex})\,.
    }
\end{figure}
In order to fulfill Mermin-Wagner's theorem, i.e., to suppress the
mean-field transition temperature down to zero, we seek an
approximation where the susceptibility $\chi$ is coupled back onto
itself in a self-consistent fashion. Such an approximation has been
proposed originally for the Hubbard model\cite{bicscawhi89}, commonly
referred to as FLEX.  For the present model it takes the form shown in
Fig.\ \ref{fig-flex}\,. The irreducible part $\Pi$ in
Eq.(\ref{eqn-rpa}) is approximated by a bubble of renormalized fermion
Green's functions. The latter are in turn determined by $\chi$ through
the effective interaction $D$\,, which enters the fermion's self
energy on one-loop level. When temperature is lowered and the
mean-field transition at $T\sim J$ is approached, the growing
susceptibility $\chi$ should modify the spectrum of the auxiliary
fermion and therefore $\Pi$\,, such that the increase of $\chi$ and
accordingly the transition is suppressed.

The irreducible part $\Pi$ and the fermion propagator $\overline{G}$
are given by Eqs.(\ref{eqn-bubble}) and (\ref{eqn-fgfself}),
respectively. The self energy depicted in Fig.\ \ref{fig-flex} reads
\begin{equation}  \label{eqn-flexself}
  \overline{\Sigma}(i\omega)= 
    \left(\frac{1}{2}\right)^2 \frac{1}{\beta}\sum_{i\nu}\sum_{\mu,\mu'}
    D^{\mu\mu'}(i\nu)\sigma^\mu\overline{G}(i\omega + i\nu)
    \sigma^{\mu'}
\end{equation}
Without external field and in the paramagnetic phase, $\mathbf{h}=
\langle\mathbf{S}\rangle= 0$\,, the fermions are spin degenerate,
$\overline{G}= \sigma^0 G$\,, $\overline{\Sigma}= \sigma^0 \Sigma$\,,
and the response becomes isotropic, $\Pi^{\mu\mu'}=
\delta_{\mu\mu'}\Pi$\,. Equations (\ref{eqn-fgfself}),
(\ref{eqn-rpa}), (\ref{eqn-bubble}), (\ref{eqn-effj}),
(\ref{eqn-flexself}) now simplify to
\begin{subequations}  \label{eqn-scflex}
\begin{eqnarray}
  \Pi(i\nu) & = &  \label{eqn-sc-pi}
    -\frac{1}{2\,\beta}\sum_{i\omega}G(i\omega + i\nu) G(i\omega)
    \\
  \chi(\mathbf{q}, i\nu) & = &  \label{eqn-sc-chi}
    \frac{\Pi(i\nu)}{1 + J(\mathbf{q})\Pi(i\nu)} 
    \\
  D(i\nu) & = &  \label{eqn-sc-d}
    \frac{1}{N_L}\sum_{\mathbf{q}}J^2(\mathbf{q})
    \chi(\mathbf{q}, i\nu)
    \\
  \Sigma(i\omega) & = &  \label{eqn-sc-self}
    \frac{3}{4\,\beta}\sum_{i\nu}D(i\nu)G(i\nu + i\omega)
    \\
  G(i\omega) & = &  \label{eqn-sc-gf}
    \left[ i\omega - \Sigma(i\omega) \right]^{-1}
\end{eqnarray}
\end{subequations}
Note that the first bare $J$ to $D$ in Fig.\ \ref{fig-flex} does not
contribute, since $J_{ii}= 0$ or equivalently
$\sum_{\mathbf{q}}J(\mathbf{q})= 0$\,.

\begin{figure} \large
    \begin{eqnarray*}
      \overline{\Sigma}^{ex} & = &
        \mygraph{scale=0.35}{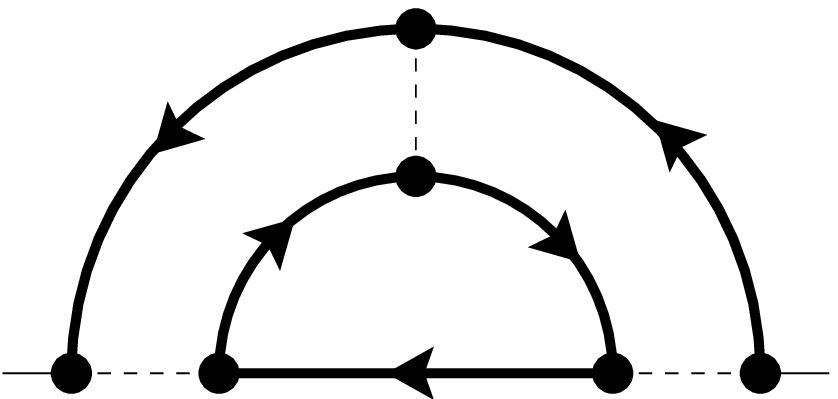} +
        \mygraph{scale=0.35}{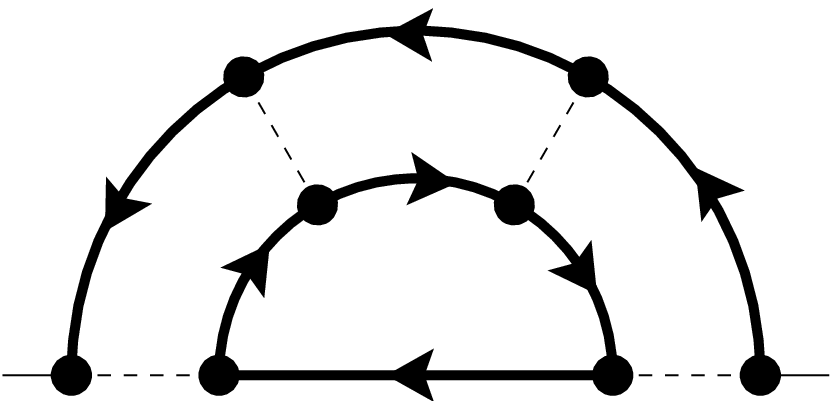} +
        \\
      \overline{\Sigma}^{pp} & = &
        \mygraph{scale=0.35}{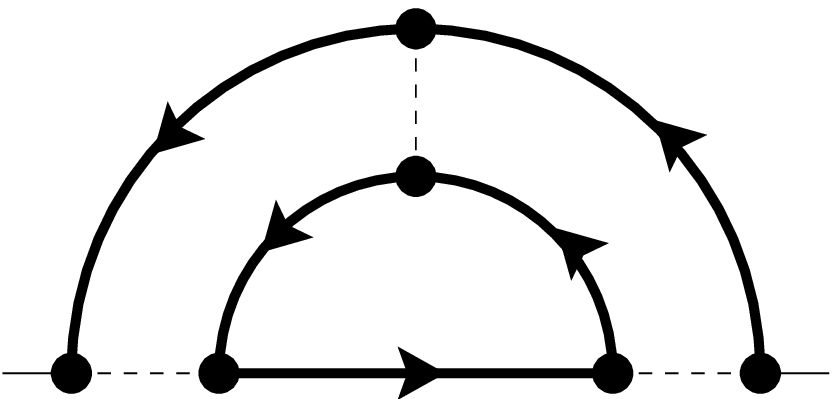} +
        \mygraph{scale=0.35}{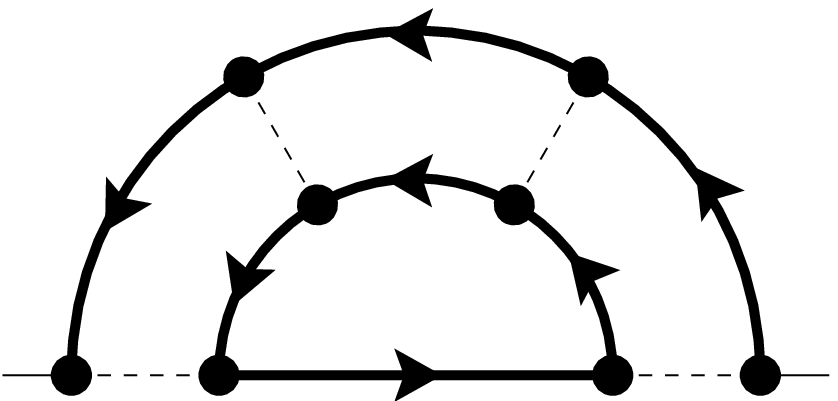} +
    \end{eqnarray*}
  \caption[\ ]{\label{fig-flux}%
    Self-energy diagrams of ladder type for the exchange
      particle--hole (`RVB/flux') channel (top row) and the
      particle--particle (pairing) channel (bottom row).
      The contribution of
      \protect$\overline{\Sigma}^{ex}\protect$ and
      \protect$\overline{\Sigma}^{pp}\protect$ has almost no effect on
      the results, see text, Sect.\ \protect\ref{sec-flex}\,.
    }
\end{figure}
It has been emphasized above that the non-local expectation values
$\langle f^\dagger_{i\alpha} f_{j\alpha}\rangle$ and $\langle
f_{i\uparrow} f_{j\downarrow}\rangle$ are zero at 1/2 filling. Hence
the fermion propagator $G_{ij}= \delta_{ij}\,G_i$\,, $G_i= G$ and the
bubble $\Pi_{ij}= \delta_{ij}\,\Pi_i$\, $\Pi_i= \Pi$ are local, i.e.,
wave-vector independent. The susceptibility $\chi(\mathbf{q})$\,,
however, is $\mathbf{q}$-dependent through the Heisenberg interaction
$J(\mathbf{q})= 2dJ\gamma(\mathbf{q})$ occurring in
Eq.(\ref{eqn-sc-chi})\,. The latter has been introduced in
Eq.(\ref{eqn-gamq})\,. However, the argument for the fermions being
local is based on mean-field theory. If fluctuations are included in
the fermion self-energy, the ladder-diagram series shown in Fig.\ 
\ref{fig-flux} should also be taken into account. While
$\overline{\Sigma}$ in Fig.\ \ref{fig-flex} captures the effect of
magnetic fluctuations on the fermions, $\overline{\Sigma}^{ex}$ and
$\overline{\Sigma}^{pp}$ contain fluctuations in the `RVB/flux'
channel and the pairing channel, respectively: A large numerical
contribution from $\overline{\Sigma}^{ex}$ would indicate a close-by
instability to a flux phase ($\langle f^\dagger_i f_j\rangle\ne 0$),
whereas $\overline{\Sigma}^{pp}$ would dominate if the pseudo fermions
were instable to pairing ($\langle f_i f_j\rangle\ne 0$)\,. In the
1/2-filled case considered in this paper, both
$\overline{\Sigma}^{ex/pp}$ are equivalent by particle--hole symmetry.
The numerical calculation presented in Sect.\ \ref{sec-flex-num} has
been repeated with the two extra channels from Fig.\ \ref{fig-flux}
included. Even at the lowest temperature reached the results were only
altered quantitatively; numbers reported in Sect.\ \ref{sec-flex-num}
came out different by $\simeq 20$\%\,. The ladder sums
(susceptibilities) appearing in $\overline{\Sigma}^{ex/pp}$ remain
small, i.e., $\sim 1/J$\,.  Apparently, the self energies from Fig.\ 
\ref{fig-flux} can be omitted and will no longer be considered.

In the following we calculate the antiferromagnetic correlation length
$\xi(T)$ and the spectrum $S(\mathbf{q},\omega)$ of magnetic
fluctuations from Eqs.(\ref{eqn-scflex})\,.

\subsection{Static Approximation:
            Correlation Length, Energy Scale, Dynamical Scaling}
\label{sec-flex-static}
In order to proceed by analytical calculation we approximate the
effective interaction $D$ in the self energy Eq.(\ref{eqn-sc-self}) by
its static value, $D(i\nu)= D(0)\delta_{\nu,0}$\,. This static
(classical) approximation should be reasonable if the characteristic
energy $\omega_0$ of spin fluctuations is small compared to
temperature. This is the result of earlier studies mentioned in the
introductory Sect.\ \ref{sec-intro} and will also come out of the
following calculation as well as the numerical solution presented in
Sect.\ \ref{sec-flex-num} below.

With $D(i\nu)= D(0)\delta_{\nu,0}$ Eq.(\ref{eqn-sc-self}) turns into
\begin{equation}  \label{eqn-static}
  \Sigma(i\omega)= \frac{3}{4\,\beta} D(0) G(i\omega)
    = \left(\frac{\omega_f}{2}\right)^2 G(i\omega)
\end{equation}
where a coupling constant $\omega_f= \sqrt{3 D(0)/\beta}$ has been
introduced. Eq.(\ref{eqn-sc-d}) requires the static susceptibility
$\chi(\mathbf{q}, 0)$\,. At low temperature wave vectors
$\mathbf{q}\simeq\mathbf{Q}\equiv(\pi,\pi)$ close to the N{\'e}el
ordering-vector $\mathbf{Q}$ give the strongest contribution, allowing
to expand Eq.(\ref{eqn-gamq}) as
\begin{equation}  \label{eqn-gamqexpand}
  J(\mathbf{q})\simeq
    -2dJ + (\mathbf{q} - \mathbf{Q})^2 J
\end{equation}
leading to
\begin{equation}  \label{eqn-statchi}
  \chi(\mathbf{q}, 0)= 
    \frac{\Pi(0)}{1 + J(\mathbf{q})\Pi(0)}
    \simeq \frac{1}{J}\frac{1}{\xi^{-2} + (\mathbf{q} - \mathbf{Q})^2}
\end{equation}
where the antiferromagnetic correlation length has been identified as
\begin{equation}  \label{eqn-xidef}
  \xi^{-2}= 2d\left(\frac{1}{2dJ\Pi(0)} - 1\right)
\end{equation}
The spatial dimension $d$ has been written explicitly for clarity.
In Eq.(\ref{eqn-sc-d}) we may set $J(\mathbf{q})\simeq J(\mathbf{Q})=
-2dJ$\,, and it follows for the coupling constant
\begin{equation}  \label{eqn-omegac}
  (\omega_f)^2= 3(2d)^2J T 
    \int\frac{\mathrm{d}^d k}{(2\pi)^d} \frac{1}{\xi^{-2} + k^2}
\end{equation}
In 2 dimensions this becomes
\begin{equation}  \label{eqn-coupl}
  (\omega_f)^2= \frac{24}{\pi}T J 
    \left\{ \ln(k_c\xi) + \mathcal{O}(\xi^{-2}) \right\}
\end{equation}
where $k_c\sim 1$ denotes a wave-vector cut off coming from the
Brillouin-zone boundary. In higher dimensions $d\ge 3$ we get
\begin{equation}  \label{eqn-couplhigher}
  (\omega_f)^2\sim T J (k_c)^{d-2}
\end{equation}
The Dyson's equation (\ref{eqn-sc-gf}) with Eq.(\ref{eqn-static})
inserted reads after analytical continuation $i\omega\to z$
\begin{displaymath}
  \frac{1}{G(z)}= z - \left(\frac{\omega_f}{2}\right)^2G(z)
\end{displaymath}
At the real axis, $z= \omega + i0_+$\,, this leads to
\begin{eqnarray}  \label{eqn-fgfstatic} 
  \lefteqn{G(\omega + i0_+) =
    \frac{2\,\Theta(\omega_f - |\omega|)}
         {\omega + i\sqrt{(\omega_f)^2 - \omega^2}} +}
    \\[1ex]
  & & \mbox{} +  \nonumber
    \frac{2\,\Theta(|\omega| - \omega_f)}
         {\omega + \textrm{sign}(\omega)\sqrt{\omega^2 - (\omega_f)^2}
            + i0_+}
\end{eqnarray}
In the atomic limit $J=0$ the spectrum of the fermions
$\rho(\omega)= -\frac{1}{\pi}\textrm{Im}G(\omega + i0_+)$
is a delta peak at zero energy, $\rho^0(\omega)= \delta(\omega)$ from
$G^0(z)= 1/z$\,. Through the interaction with spin fluctuations $\rho$
is apparently broadened into a semi-elliptic spectrum with the
coupling constant $\omega_f$ appearing as a high-energy cut off:
Eq.(\ref{eqn-fgfstatic}) gives
\begin{equation}  \label{eqn-fspec}
  \rho(\omega)= \Theta(\omega_f - |\omega|)\frac{2}{\pi\omega_f}
    \sqrt{1 - (\omega/\omega_f)^2}
\end{equation}
Eq.(\ref{eqn-fspec}) is plotted in Fig.\ \ref{fig-flexapr_rhof},
together with the numerical solution described in Sect.\ 
\ref{sec-flex-num}\,. In order to close the set of self-consistency
equations, $\Pi(0)$ has to be calculated from Eq.(\ref{eqn-sc-pi}),
\begin{equation}  \label{eqn-pinull}
  \Pi(0)= -\int_{-\infty}^\infty\mathrm{d}\varepsilon\,
          f(\varepsilon) \rho(\varepsilon)\,\textrm{Re}G(\varepsilon)
        = \frac{1}{\omega_f}\Phi(T/\omega_f)
\end{equation}
with 
\begin{eqnarray*}
  \Phi(t) & = &
    -\frac{4}{\pi}\int_{-1}^1\mathrm{d}x\,
    \frac{x \sqrt{1 - x^2}}{\,\mathrm{e}^{x/t} + 1}
    \\
  & = &
    \frac{4}{3\pi} - \frac{2\pi}{3}t^2 + \mathcal{O}(t^4)
\end{eqnarray*}
The r.h.s.\ results from a Sommerfeld expansion of the integral. 

When temperature is lowered in the 2D system and the correlation
length $\xi$ grows, the fermion's cut-off energy $\omega_f$ increases
(see Eq.(\ref{eqn-coupl})), which in turn reduces the fermion spectrum
$\rho(0)\sim 1/\omega_f$ and the static irreducible part $\Pi(0)\sim
1/\omega_f$\,. Therefore the antiferromagnetic susceptibility
$\chi(\mathbf{Q}, 0)= \Pi(0)/[1 - 2dJ\Pi(0)]$ is shifted away from the
critical point, and the phase transition to AF order is suppressed
self-consistently. In $d\ge 3$ the correlation length $\xi$ does not
enter $\omega_f$\,, Eq.(\ref{eqn-couplhigher}), and this mechanism
does not apply, allowing an AF transition to occur.

An explicit expression for the correlation length is obtained from the
solution of Eqs.(\ref{eqn-xidef}), (\ref{eqn-coupl}),
(\ref{eqn-pinull}), omitting terms of $\mathcal{O}(\xi^{-2})$ and
$\mathcal{O}(T^4)$ at low temperature. For $d=2$ it follows from
Eq.(\ref{eqn-xidef})
\begin{equation}  \label{eqn-sol-pi}
  \Pi(0)= 1/4J + \mathcal{O}(\xi^{-2})
\end{equation}
This is inserted into Eq.(\ref{eqn-pinull}), which is then solved for
$\omega_f$\,,
\begin{equation}  \label{eqn-sol-coupl}
  \frac{\omega_f}{\omega_f^0}=  1 -
    \frac{\pi^2}{2}\left(\frac{T}{\omega_f^0}\right)^2 +
    \mathcal{O}(T^4)
  \;\;\textrm{with}\;\;
  \omega_f^0= \frac{16J}{3\pi}
\end{equation}
Finally, Eq.(\ref{eqn-coupl}) gives the result
$k_c\xi= \exp\{(\pi(\omega_f^0)^2)/(24 TJ) 
   + \mathcal{O}(T/J)\}$\,,
and in leading order,
\begin{equation}  \label{eqn-sol-xi}
  \xi(T)\propto \exp( a_s\,J/T)
    \;\;\;\textrm{with}\;\;\;
    a_s= \frac{32}{27\pi}\simeq 0.38
\end{equation}

Accordingly the correlation length diverges exactly at $T=0$\,. The
exponential behavior of $\xi(T)$ is simply due to the density of
states in 2D,
$\mathcal{N}(\omega)= \int\frac{\mathrm{d}^2q}{4\pi^2}\,
   \delta(\omega - J(\mathbf{q}))\simeq \Theta(4J - |\omega|)/J$\,,
   which is constant near the band edges and therefore leads to the
   logarithm in Eq.(\ref{eqn-coupl})\,. The correlation length
   (\ref{eqn-sol-xi}) is consistent with the theoretical work
   mentioned in Sect.\ \ref{sec-intro}\,. The spin stiffness $\rho_s=
   a_s J/2\pi\simeq 0.06J$ extracted from Eq.(\ref{eqn-sol-xi}) comes
   out too small compared to the literature, where $\rho_s\simeq
   0.18$\,.  The latter value is also the result of exact
   diagonalization studies in the AF ordered ground
   state\cite{note-manous} at $T=0$\,.

In addition to the correlation length, the dynamics of spin
fluctuations is an important issue. The damping of spin fluctuations
is induced by the imaginary part of $\Pi(\omega)$\,, i.e., the
spectrum of particle--hole excitations. From Eq.(\ref{eqn-sc-pi}) we
get at $T\to 0$\,, with the smooth fermion spectrum (\ref{eqn-fspec})
\begin{eqnarray*}
  \textrm{Im}\,\Pi(\omega + i0_+) & = &  
    \frac{\pi}{2}\int_0^\omega\mathrm{d}\varepsilon\,
    \rho(\omega - \varepsilon) \rho(-\varepsilon)
    \\
  & = & \frac{\pi}{2}\rho(0)^2\omega + \mathcal{O}(\omega^3)
\end{eqnarray*}
The real part is approximately calculated using
\begin{eqnarray*}
  \textrm{Re}\,\Pi(\omega) & = &
    \frac{1}{\pi}\int\mathrm{d}\varepsilon\,
    \frac{\textrm{Im}\,\Pi(\varepsilon + i0_+)}{\varepsilon - \omega}
    \\
  & \simeq & 
    \frac{1}{2}\rho(0)^2\int_{-\Omega}^\Omega\mathrm{d}\varepsilon\,
    \frac{\varepsilon}{\varepsilon - \omega}
\end{eqnarray*}
where a high-energy cut off $\Omega\sim J$ for spin-fluctuations has
been introduced. $\Omega$ can be fixed by the condition
$\textrm{Re}\,\Pi(0)= \Pi(0)$\,, and it follows
\begin{displaymath}
  \textrm{Re}\,\Pi(\omega)= \Pi(0)
    \left[ 1 - \left(\omega/\Omega\right)^2 \right] + 
    \mathcal{O}(\omega^4)
\end{displaymath}
with
$\Omega= \Pi(0)/\rho(0)^2= \frac{16}{9}J$\,.
Here $\Pi(0)= 1/4J$ and $\rho(0)= 1/\pi\omega_f^0= 3/16J$ from
Eqs.(\ref{eqn-fspec}), (\ref{eqn-sol-pi}) have been used.  The
susceptibility Eq.(\ref{eqn-sc-chi}) at the real axis $i\nu\to
\omega + i0_+$ becomes
\begin{eqnarray*}
  \chi(\mathbf{q}, \omega)^{-1} & = &
    \Pi(\omega + i0_+)^{-1} + J(\mathbf{q})
    \\ \\
  & = & 
    \frac{1}{\chi(\mathbf{q})} - 
    \frac{1}{\Pi(0)}
    \left[ i\frac{\pi}{2}
      \left(\frac{\omega}{\Omega}\right) +
    \right.
    \\[1ex]
  & & \mbox{}
    + \left.
           \left(\frac{\pi^2}{4} - 1\right)
           \left(\frac{\omega}{\Omega}\right)^2 \right]
    + \mathcal{O}(\omega^3)
\end{eqnarray*}
with the static susceptibility $\chi(\mathbf{q})$ stated in
Eq.(\ref{eqn-statchi}). Introducing a spin-wave velocity $c$ and
-damping constant $\gamma$\,,
\begin{displaymath}
  c^2= \frac{4\Pi(0)\,J \Omega^2}{\pi^2 - 4}\simeq (1.47\,J)^2
    \;\;,\;\;\;
  \gamma= \frac{2\pi\,\Omega}{\pi^4 - 4}\simeq 1.90\,J
\end{displaymath}
the dynamical susceptibility takes the familiar form
\begin{equation}  \label{eqn-susflex}
  \chi(\mathbf{q},\omega)= \frac{c^2}{J}
    \left[ c^2(k^2 + \xi^{-2}) - \omega^2 - i\gamma\omega \right]^{-1}
\end{equation}
where $\mathbf{k}=(\mathbf{q} - \mathbf{Q})$\,.

The calculation seems to yield spin-wave like excitations for wave
vectors $k\gg\xi^{-1}$ far enough off the N{\'e}el vector
$\mathbf{Q}$\,, with the familiar dispersion $\Omega(\mathbf{q})=
ck$\,. However, the damping $\gamma$ appears much larger than the
spin-wave energy, i.e., the `spin waves' are totally overdamped: For
wave vectors well inside the 1.\ B.Z., $k\ll 1$\,, it is
$|\omega|\simeq \Omega(\mathbf{q})\sim Jk\ll J$ and thus
$\gamma\gg|\omega|$\,. Therefore the $\omega^2$ in
Eq.(\ref{eqn-susflex}) does not contribute, and
\begin{equation}  \label{eqn-diffsus}
  \chi(\mathbf{q}, \omega)= 
    \chi(\mathbf{q})
    \frac{i\Gamma(\mathbf{q})}{\omega + i\Gamma(\mathbf{q})}
    \;\;\;,\;\;
  \Gamma(\mathbf{q})= 
    \frac{1}{\chi(\mathbf{q})}\frac{c^2}{\gamma\,J}
\end{equation}
The dynamical structure factor
\begin{equation}  \label{eqn-strucdef}
  S(\mathbf{q}, \omega)= 
    \frac{\textrm{Im}\,\chi(\mathbf{q},\omega)}
         {1 - \mathrm{e}^{-\beta\omega}}
    \simeq \frac{T}{\omega}\textrm{Im}\,\chi(\mathbf{q},\omega)
\end{equation}
shows a single relaxation-like peak at $\omega=0$\,. At long
wavelengths $k\ll\xi^{-1}$ the half-linewidth of this peak is
$\Gamma(\mathbf{q})\propto\Gamma(\mathbf{Q})\sim J\xi^{-2}$\,, which
turns to zero as the AF transition is reached (critical slowing down),
with the dynamical exponent $z=2$\,. This is compatible with
hydrodynamic theory\cite{forster} and the studies of
$S(\mathbf{q},\omega)$ referenced in Sect.\ \ref{sec-intro}\,. Our
result $z=2$\,, however, differs from the $z=1$ found in these
studies. At short wavelengths, $k\gg\xi^{-1}$\,, the structure factor
still consists of a single peak at $\omega=0$\,, with a $k$-dependent
linewidth $\Gamma(\mathbf{q})\sim Jk^2$\,. This is in strong
contradiction to the above-mentioned theories, where for
$k\gg\xi^{-1}$ the structure factor features two peaks with some
dispersion $\omega= \pm\Omega(\mathbf{q})$ and finite width. These
peaks are interpreted as spin-wave like excitations, visible only at
short enough wavelengths $\ll\xi$\,.

Since the $\omega^2$-term in Eq.(\ref{eqn-susflex}) does not
contribute, the dynamical scaling property (\ref{eqn-hypdyn}),
(\ref{eqn-hypstat}) is almost trivially fulfilled: From
Eqs.(\ref{eqn-susflex}), (\ref{eqn-strucdef}) it follows
$S(\mathbf{q}, \omega)= 
   \frac{T\xi^2}{\omega J}\textrm{Im}\,[ 1 + (k\xi)^2 -
   i(\omega/\omega_0) ]^{-1}$\,,
with the energy scale
\begin{equation}  \label{eqn-omnulflex}
  \omega_0= 
    \Gamma(\mathbf{Q})= \frac{c^2}{\gamma\xi^2}\propto J\xi^{-2}
\end{equation}
The static structure factor results from
$S^{st}(\mathbf{q})= 
   \frac{1}{\pi}\int\mathrm{d}\omega\,
   S(\mathbf{q}, \omega)
   \simeq T \chi(\mathbf{q}, 0)
   = \frac{T\xi^2}{J}[1 + (k\xi)^2]^{-1}$\,,
reproducing Eqs.(\ref{eqn-hypdyn}), (\ref{eqn-hypstat}) with the
scaling functions
\begin{equation}  \label{eqn-scaleflex}
  \varphi(x)= \frac{1}{1 + x^2}
    \;\;,\;\;\;
  \Phi(x, y)= \frac{1 + x^2}{(1 + x^2)^2 + y^2}
\end{equation}
For any fixed wave vector $x=k\xi$ the `shape function' $\Phi(x,y)$\,,
$y= \omega/\omega_0$ is a simple Lorentzian located at $\omega=0$\,,
normalized to $\Phi(0,0)= 1$\,. The inset of Fig.\ 
\ref{fig-flexapr_dynscale} shows $\Phi$ from Eq.(\ref{eqn-scaleflex}),
the main figure the corresponding result from the numerical solution.

\subsection{Numerical Results}
\label{sec-flex-num}
\begin{figure}
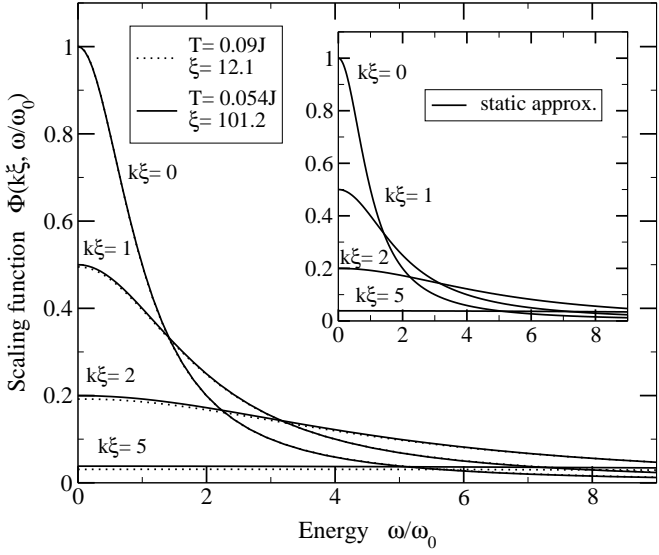

  \mygraph{width=\hsize,clip=true}{fig_flexapr_dynscale}
  \caption[\ ]{\label{fig-flexapr_dynscale}%
\textbf{Main figure:}
Dynamical structure factor from the numerical solution of the
    self-consistent 
    approximation Eqs.(\protect\ref{eqn-scflex})\,. Plotted is the
    scaling function \protect$\Phi(k\xi, \omega/\omega_0)\protect$
    at low energies \protect$\sim\omega_0\protect$\,. \protect$\Phi\protect$ 
    is shown for fixed wave vectors \protect$k= 0, \xi^{-1},
    2\xi^{-1}, 5\xi^{-1}\protect$\,, each for 2 different
    temperatures corresponding to the correlation lengths \protect$\xi= 12.1,
    101.2\protect$\,. The curves lie almost on top of each other,
    indicating that dynamical scaling holds for \protect$\xi> 10\protect$\,.
\textbf{Inset:}
Analytically calculated scaling function \protect$\Phi(k\xi,
    \omega/\omega_0)\protect$\,, Eq.(\protect\ref{eqn-scaleflex}),
    from the static approximation. 
    }
\end{figure}
\begin{figure}
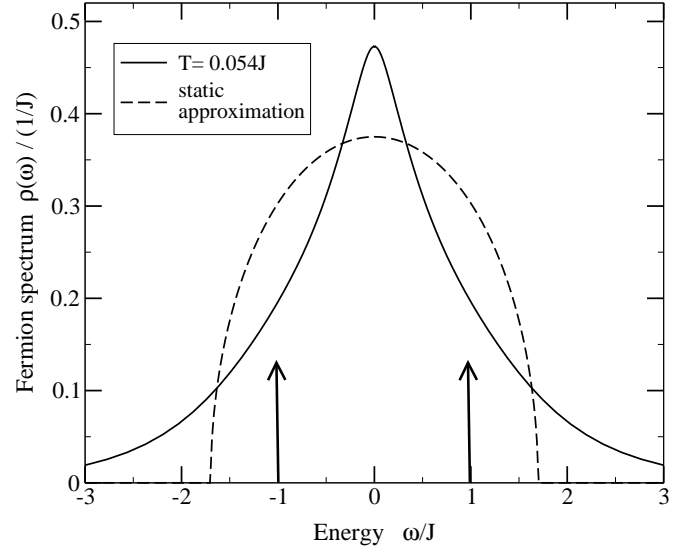

  \mygraph{width=\hsize,clip=true}{fig_flexapr_rhof}
  \caption[\ ]{\label{fig-flexapr_rhof}%
Spectral function \protect$\rho(\omega)\protect$ of the fermion
  propagator Eq.(\protect\ref{eqn-sc-gf})\,. Full line: numerical
  solution of Eqs.(\protect\ref{eqn-scflex})\,. Dashed line:
  static approximation, Eq.(\protect\ref{eqn-fspec})\,. The arrows
  indicate delta peaks, representing
  \protect$\rho(\omega)\protect$ in the antiferromagnetic phase at
  \protect$T= 0\protect$ in mean-field theory,
  Eq.(\protect\ref{eqn-afdos})\,. 
  }
\end{figure}
So far we have considered a static approximation, where the
susceptibility Eq.(\ref{eqn-sc-chi}) has been taken in the classical
limit, $\chi(\mathbf{q}, i\nu)= \chi(\mathbf{q}, 0)\delta_{\nu,0}$
inside the effective local interaction $D$\,, Eq.(\ref{eqn-sc-d})\,.
The set of equations (\ref{eqn-scflex}) has also been solved
numerically, taking all frequency dependencies into account.
Eqs.(\ref{eqn-scflex}) are re-written at the real axis, see
Eq.(\ref{eqn-num-scflex}) in Appendix \ref{sec-app-flex}\,, and solved
by numerical iteration. A stable solution is obtained for temperatures
$0.5J \ge T\ge 0.054J$\,, corresponding to correlation lengths
$0.7\le\xi\le 101$\,, in units of the lattice spacing. $\xi$ is
extracted from the static susceptibility at $\mathbf{q}=\mathbf{Q}$
using Eq.(\ref{eqn-xidef}), it is well reproduced by
$\xi(T)=const.\times\exp(a\,J/T)$ with $a= 0.30$\,. The fit of the
numerical data $\xi(T)$ is performed by a linear regression of
$[t\ln(\xi)]$ to the function $[a + t\ln(const.)]$\,, $t= T/J$\,. The
exponent $a_{s}= 0.38$ from the static approximation,
Eq.(\ref{eqn-sol-xi}), almost agrees with the numerical result.

The numerical solution is also consistent with the dynamical scaling
hypothesis: If the static scaling function $\varphi(x)=
S^{st}(\mathbf{q})/S^{st}(\mathbf{Q})$ is plotted against $x= k\xi$\,,
curves lie on top of each other for temperatures $T<0.1J$ where
$\xi>10$\,. $\varphi(x)$ is well described by the simple Lorentzian
shape (\ref{eqn-scaleflex}) of the static approximation. The energy
scale $\omega_0$ is extracted from the numerical output via
\begin{equation}  \label{eqn-extractomnul}
  \omega_0= \frac{S^{st}(\mathbf{q}= \mathbf{Q})}
                 {S(\mathbf{q}= \mathbf{Q}, \omega= 0)}
\end{equation}
which gives $\omega_0$ up to a constant prefactor, which is chosen
such that $\Phi(0, 0)= 1$\,. The scaling function $\Phi(x, y)$ is
then gained from
\begin{equation}  \label{eqn-extractscalefun}
  \Phi(x, y)= \omega_0\,S(\mathbf{q}, \omega)\,/\,S^{st}(\mathbf{q})
\end{equation}
and plotted against $x= k\xi$\,, $y= \omega/\omega_0$\,. Fig.\ 
\ref{fig-flexapr_dynscale} shows $\Phi(x, y)$ for two temperatures,
where $\xi\ge 10$\,. Apparently scaling is well obeyed, and the shape
of $\Phi$ agrees with the static approximation. In particular, it
consists of a single peak even for $k\xi>1$\,. In Sect.\ \ref{sec-red}
a modified self-consistent approximation will be presented, where
$\Phi(x,y)$ features spin-wave like propagating excitations for
$k\xi>1$\,; also the dynamical exponent will come out correctly as $z=
1$\,.

For completeness the spectrum $\rho(\omega)$ of auxiliary fermions is
shown in Fig.\ \ref{fig-flexapr_rhof}\,. The delta-peak of the bare
($J=0$) fermion is broadened into a continuum of width $\sim 2J$\,,
qualitatively similar to the static approximation,
Eq.(\ref{eqn-fspec})\,. Further comments on $\rho(\omega)$ will be
made in Sect.\ \ref{sec-flex-sum}\,.

\begin{figure*}[t] \large
  $\displaystyle 
      \widehat{\Pi}= \mygraph{scale=0.5}{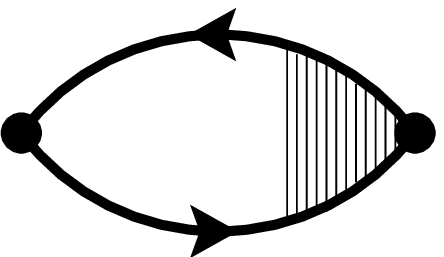}$
      \\[1ex]
  $\displaystyle 
      \mygraph{scale=0.55}{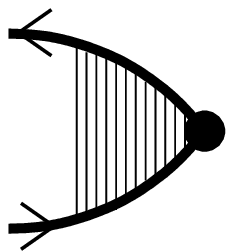}= 
        \overline{\Lambda}= 
        \mygraph{scale=0.55}{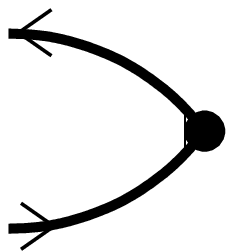} + 
        \mygraph{scale=0.55}{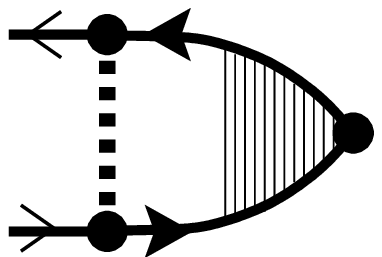} +  
        \mygraph{scale=0.55}{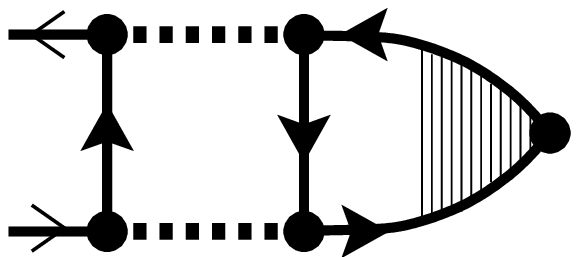} +  
        \mygraph{scale=0.55}{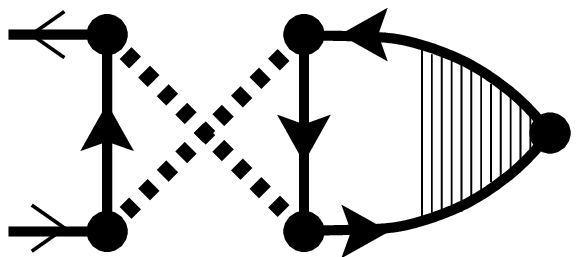}$
  \caption[\ ]{\label{fig-flexvert}%
    {\bf Top:} Irreducible bubble $\widehat{\Pi}$ including the vertex
      corrections from the conserving approximation-scheme.
      {\bf Bottom:} Bethe--Salpeter equation
      for the vertex function $\overline{\Lambda}$\,. Lines
      are those introduced in Fig.\ \ref{fig-flex}, calculated
      self-consistently from the corresponding equations 
      (\protect\ref{eqn-scflex})\,. 
    }
\end{figure*}
\subsection{Conserving-Approximation Approach: \\
            Vertex Corrections}
\label{sec-flex-cons}
We now turn to the so-called conserving-approximation method, which
has been left aside so far: The spin susceptibility (\ref{eqn-sus}) is
given by
\begin{displaymath}
  \widehat{\chi}^{\mu\mu'}_{ij}(\tau,\tau')= 
    \left.\frac{\delta\langle S^\mu_i(\tau) \rangle}
               {\delta h^{\mu'}_j(\tau')}\right|_{\mathbf{h}=0}
    = \left.\frac{- \delta^2 W[\mathbf{h}]}
                 {\delta h^\mu_i(\tau)\,\delta h^{\mu'}_j(\tau')}
      \right|_{\mathbf{h}=0}
\end{displaymath}
$\mu, \mu'= x, y, z$\,. $h^\mu$ are the components of the magnetic
source field $\mathbf{h}$\,. $W$ is the generating functional (free
energy)
$W[T,\mathbf{h}]=
    - \ln\int\mathcal{D}[f,\overline{f}]\,\mathrm{e}^{-A}$
to be calculated from the action (\ref{eqn-action}) of the model. $W$
can be expressed as a functional\cite{lutwar60,bay62} of $G$ and
$\Sigma$\,,
$W[T, \mathbf{h}]= 
   \Phi[G] - 2\textrm{Tr}[\Sigma\,G] 
     - 2\textrm{Tr}\ln[ - G_0^{-1} + \Sigma ]$\,.
The functional $\Phi[{G}]$ appearing here is related to the fermion
self-energy via
$\Sigma[{G}]= \frac{1}{2}\delta\Phi/\delta G$\,.
The existence of a $\Phi$-functional for a particular approximation of
$\Sigma$ is a sufficient condition for $\widehat{\chi}$ to respect
conservation laws\cite{baykad61,bay62}. The self-energy in Fig.\
\ref{fig-flex}, Eq.(\ref{eqn-scflex}) can be derived from a
$\Phi$-functional similar to the FLEX\cite{bicscawhi89}.

$\widehat{\chi}$ is given by the same RPA-formula (\ref{eqn-rpa}) as
$\chi$\,, but the irreducible part $\Pi$ from Fig.\ \ref{fig-flex} has
to be replaced by $\widehat{\Pi}$ shown in Fig.\ \ref{fig-flexvert}\,.
If the theory was exact, $\Pi$ and $\widehat{\Pi}$ would be identical.
In any approximation, however, $\widehat{\Pi}$ contains additional
vertex corrections, and $\widehat{\chi}$ differs from the correlation
function $\chi$ of the self-consistency equations, e.g.,
Eqs.(\ref{eqn-scflex})\,. If now $\widehat{\chi}$ diverges, indicating
a magnetic instability at some finite temperature $\widehat{T}_N$\,,
this is not coupled back into the self-consistency cycle. As a
consequence the spurious phase transition is not suppressed, and
$\widehat{\chi}$ becomes invalid below $\widehat{T}_N$\,.

To elucidate further on this important point, we calculate the
correlation length $\widehat{\xi}(T)$ from the susceptibility
$\widehat{\chi}(\mathbf{q}, 0)$\,. We employ the static approximation
from Sect.\ \ref{sec-flex-static}, using the fermion propagator
(\ref{eqn-fgfstatic}) and the self-consistent $D$\,,
Eq.(\ref{eqn-sc-d}), as input. The bubble including the vertex
corrections is shown in Fig.\ \ref{fig-flexvert} and reads
\begin{displaymath}
  \widehat{\Pi}^{\mu\mu'}_{ij}(i\nu)= 
    -\frac{1}{2\,\beta}\sum_{i\omega}
    \textrm{Tr}[ \sigma^\mu\overline{G}_i(i\omega + i\nu)
                   \overline{\Lambda}^{\mu'}_{ij}(i\omega, i\nu)
                   \overline{G}_i(i\omega) ]
\end{displaymath}
$\overline{G}$ and $\overline{\Lambda}$ denote the renormalized
fermion propagator and the vertex function, respectively; both are
matrices in spin space. Whereas the former is always local, the latter
in general depends on lattice sites $i, j$\,. In the disordered phase
$\langle \mathbf{S}\rangle= 0$ we have $\overline{G}_i= G \sigma^0$
and
$\widehat{\Pi}_{ij}^{\mu\mu'}=\widehat{\Pi}_{ij}^\mu\delta_{\mu\mu'}$\,.
For the correlation length we set $i\nu=0$\,, and $\widehat{\Pi}$
reads in wave-vector space
\begin{eqnarray}
  \widehat{\Pi}^\mu(\mathbf{q}, 0) & = &  \label{eqn-picc} 
    - \frac{1}{\beta}\sum_{i\omega} G(i\omega)^2\lambda^\mu(\mathbf{q}; i\omega)
    \\
  \lambda^\mu(\mathbf{q}; i\omega) & = &  \nonumber 
    \frac{1}{2}\textrm{Tr}[ \sigma^\mu
    \overline{\Lambda}^\mu(\mathbf{q}; i\omega)]
\end{eqnarray}
The Bethe-Salpeter equation for $\overline{\Lambda}$ is depicted in
Fig.\ \ref{fig-flexvert}, for $i\nu= 0$ it takes the form
\begin{eqnarray*}
  \lefteqn{\overline{\Lambda}^\mu(\mathbf{q}; i\omega) =
    \frac{1}{2}\sigma^\mu +} 
    \\ 
  & & \mbox{} +
    \frac{1}{4}\frac{1}{\beta} D(0) G(i\omega)^2
    \sum_{\tilde{\mu}= x,y,z}\sigma^{\tilde{\mu}}
    \overline{\Lambda}^\mu(\mathbf{q}; i\omega)\sigma^{\tilde{\mu}} - \\
  & & \mbox{} 
    - \frac{1}{16}
    \frac{1}{N_L}\sum_{\mathbf{k}} \frac{1}{\beta}D(\mathbf{k} +
    \mathbf{q}, 0) D(\mathbf{k}, 0)\,G(i\omega)^2 \times 
    \\ 
  & & \mbox{} 
    \times \sum_{\tilde{\mu}, \tilde{\mu}'}
    (\sigma^{\tilde{\mu}}\sigma^{\tilde{\mu}'} +
    \sigma^{\tilde{\mu}'}\sigma^{\tilde{\mu}}) \times
    \\
  & & \mbox{}
    \times \frac{1}{\beta}\sum_{i\tilde{\omega}} G(i\tilde{\omega})^3
    \textrm{Tr}[\sigma^{\tilde{\mu}} \overline{\Lambda}^\mu(\mathbf{q};
    i\tilde{\omega}) \sigma^{\tilde{\mu}'} ]
\end{eqnarray*}
$D(\mathbf{k}, i\nu)$ denotes the renormalized interaction
Eq.(\ref{eqn-effj}) and $D(i\nu)$ its local version
Eq.(\ref{eqn-sc-d})\,. In the static approximation set
$D(i\nu)= D(0)\delta_{\nu,0}$\,. The last term in the equation above
corresponds to the last two diagrams in Fig.\ \ref{fig-flexvert}\,. In
the static case $i\nu= 0$ considered here, these diagrams cancel,
since
$(\sigma^{\tilde{\mu}}\sigma^{\tilde{\mu}'} +
    \sigma^{\tilde{\mu}'}\sigma^{\tilde{\mu}})= 2 \sigma^0
    \delta_{\tilde{\mu}\tilde{\mu}'}$\,,
and therefore their contribution to $\lambda^\mu$ is 
$\propto\textrm{Tr}[\sigma^\mu \sigma^0]= 0$\,.
In addition, in the undoped case of the Heisenberg model it is 
$\sum_{i\tilde{\omega}}G(i\tilde{\omega})^l= 0$\,, $l= 3, 5, 7, \ldots$\,,
due to particle--hole symmetry. Thus only the first two terms in
$\overline{\Lambda}$ contribute, giving a local vertex function
\begin{equation}  \label{eqn-lambda}
  \lambda^\mu(\mathbf{q}; i\omega)= 
    \frac{1}{2} - \frac{1}{4\,\beta}D(0) G(i\omega)^2\,
    \lambda^\mu(\mathbf{q}; i\omega)
\end{equation}
We used 
$\textrm{Tr}[\sigma^{\tilde{\mu}}\sigma^\mu \sigma^{\tilde{\mu}}
   \overline{\Lambda}^\mu]
   = (\pm)\,\textrm{Tr}[\sigma^\mu \Lambda^\mu]$
   with $(+)$ for $\tilde{\mu}= \mu$ and $(-)$ for $\tilde{\mu}\ne
   \mu$\,. From Eqs.(\ref{eqn-lambda}) and (\ref{eqn-picc}) it follows
   $\widehat{\Pi}^\mu(\mathbf{q}, 0)= \widehat{\Pi}(0)$\,,
\begin{displaymath}
  \widehat{\Pi}(0)=
    \frac{1}{2\pi}\int\mathrm{d}\varepsilon\,f(\varepsilon)\,
    \textrm{Im}\frac{1}{G(\varepsilon + i0_+)^{-2} +
      \frac{1}{3}(\omega_f/2)^2}
\end{displaymath}
with the coupling constant $\omega_f= \sqrt{3 D(0)/\beta}$ introduced
below Eq.(\ref{eqn-static})\,. For the fermion propagator $G$ the
self-consistent solution Eq.(\ref{eqn-fgfstatic}) has to be inserted,
resulting in
\begin{eqnarray}
  \widehat{\Pi}(0) & = &  \label{eqn-piccstat}
    \frac{1}{\omega_f}\widehat{\Phi}(T/\omega_f)
    \\
  \widehat{\Phi}(t) & = &  \nonumber
    - \frac{3}{\pi}\int_{-1}^1\mathrm{d}x\,
    \frac{x\sqrt{1 - x^2}}{x^2 + 1/3}
    \frac{1}{\mathrm{e}^{x/t} + 1}
    \\
  & = &  \nonumber
    \widehat{\Phi}(0) - \frac{3\pi}{2}t^2 + \mathcal{O}(t^4)
\end{eqnarray}
and
$\widehat{\Phi}(0)\simeq 0.497$\,.
Taking the $T$-dependent cut off $\omega_f$ from
Eq.(\ref{eqn-sol-coupl}), the vertex-renormalized bubble becomes
\begin{eqnarray}
  \widehat{\Pi}(0) & = &  \nonumber
    \frac{1}{\omega_f^0}\left[
      \widehat{\Phi}(0) - \left(T/\omega_f^0\right)^2
      \frac{\pi}{2}(3 - \pi\widehat{\Phi}(0)) \right]
      + \mathcal{O}(T^4)
    \\
  & = &  \nonumber
    \frac{1}{J}\left[
      (0.293) - (0.462)\,(T/J)^2 \right]
\end{eqnarray}
Now the static susceptibility $\widehat{\chi}$ is formed similar to
Eq.(\ref{eqn-statchi}), and the corresponding correlation length is
given by
$\widehat{\xi}^2= J\widehat{\chi}(\mathbf{Q}, 0)= 
   J\widehat{\Pi}(0)\,/\,(1 - 2dJ\,\widehat{\Pi}(0))$\,,
leading to
\begin{equation}  \label{eqn-xicc}
  \widehat{\xi}(T)\sim 
    \left(\frac{\widehat{T}_N}{T - \widehat{T}_N}\right)^{1/2}
    \;\;\textrm{with}\;\;\;\;
  \widehat{T}_N= (0.305)\,J
\end{equation}

The spin susceptibility $\widehat{\chi}$ derived from the
conserving-approximation principle contains a spurious transition to
AF magnetic order at some finite temperature $\widehat{T}_N\simeq
0.3\,J$\,. The mean-field transition $T_N= 0.5\,J$ is only slightly
suppressed. Close to $\widehat{T}_N$ the correlation length
$\widehat{\xi}$ shows a mean-field like behavior, rather than the
exponential one of the 2D system. The vertex corrections included in
$\widehat{\Pi}$ apparently have a dramatic effect: $\Pi(0)$\,, the
irreducible bubble entering Eqs.(\ref{eqn-scflex}), is forced by
self-consistency to be constant,
$J(\mathbf{Q})\Pi(0)= (-1) + \mathcal{O}(\mathrm{e}^{-J/T})$\,,
see Eq.(\ref{eqn-sol-pi})\,. Accordingly the fermion cut-off
$\omega_f$\,, Eq.(\ref{eqn-sol-coupl}), comes out such that this
$\Pi(0)$ is reproduced if $\omega_f$ is inserted into
Eq.(\ref{eqn-pinull}), in particular the $T$-dependence of
$\omega_f$ cancels the one from $\Phi(t)$\,, up to terms
$\mathcal{O}(\mathrm{e}^{-J/T})$\,. However, if the same $\omega_f$ is inserted
into $\widehat{\Pi}(0)$\,, Eq.(\ref{eqn-piccstat}), this
cancellation no longer takes place, leading to the qualitatively
different result for $\widehat{\xi}$ compared to $\xi$\,.

In conclusion, the susceptibility derived from the principle of
conserving approximation leads to very unsatisfactory results. It is
expected that the striking difference between $\xi$ and
$\widehat{\xi}$ is not overcome if a more sophisticated approximation
for the self-consistent $\Pi(0)$ is used, since again $\widehat{\Pi}$
will differ from $\Pi$\,. Therefore we will drop the
conserving-approximation scheme in the following and take the
self-consistent correlation function $\chi$ also as an approximation
for the spin susceptibility, $\widehat{\chi}:= \chi$\,. Since the
sufficient condition for $\widehat{\chi}$ to respect conservation laws
is now lost, this issue has to be discussed separately.

\subsection{Absence of Short-Range Order: Overdamping}
\label{sec-flex-sum}
A self-consistent approximation, designed in close analogy to FLEX,
provides the required suppression of the mean-field transition
temperature $T_N\to 0$ in two dimensions. The correlation length
$\xi(T)$ comes out qualitatively correct. However, the dynamical
structure factor $S(\mathbf{q},\omega)$ does not show any damped
spin-wave peaks expected far off the N{\'e}el vector
$\mathbf{Q}=(\pi,\pi)$\,.

The lack of peak dispersion in $S$ can be traced back to the fermion
density of states $\rho(\omega)$\,, displayed in Fig.\ 
\ref{fig-flexapr_rhof}\,. There is no sign of a suppression around
$\omega= 0$ even as $T\to 0$ and therefore any magnons are completely
damped out. In the AF ordered state at $T=0$ the fermion's spectrum
must show some sort of (pseudo) gap, allowing for well defined spin
waves. In mean-field theory, there is even a true gap, see
Eq.(\ref{eqn-afdos})\,. In the following section a modified (actually
simplified) approximation will be discussed, where a pseudo-gap like
suppression around $\omega=0$ is observed in $\rho(\omega)$\,. In that
approximation the spin dynamics come out quite satisfactorily.

\section{Minimal Self-Consistent Approximation (MSCA) for $T>0$}
\label{sec-red}
\begin{figure} \large
    \begin{eqnarray*}
      \Sigma  & = & \mygraph{scale=0.5}{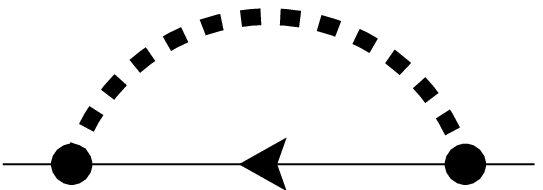}
        \;\;\;\;
      \Pi = \mygraph{scale=0.5}{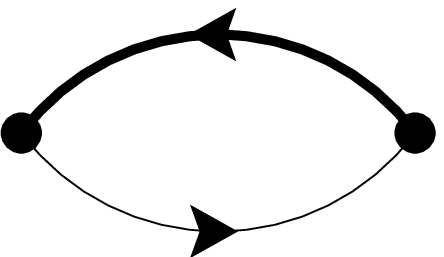}
        \\
      D & = & \mygraph{scale=0.39}{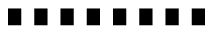} = 
              \mygraph{scale=0.39}{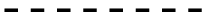} \;+\;
              \mygraph{scale=0.39}{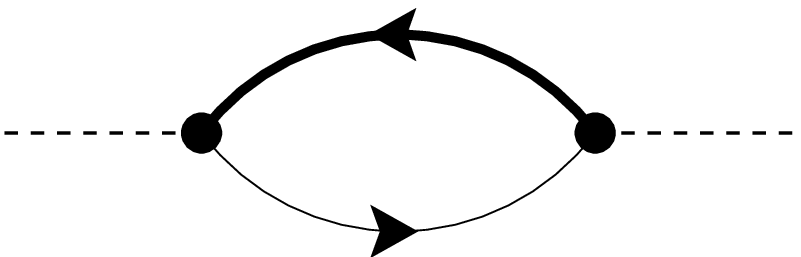} \;+\;
        \\
      & & + \;
              \mygraph{scale=0.39}{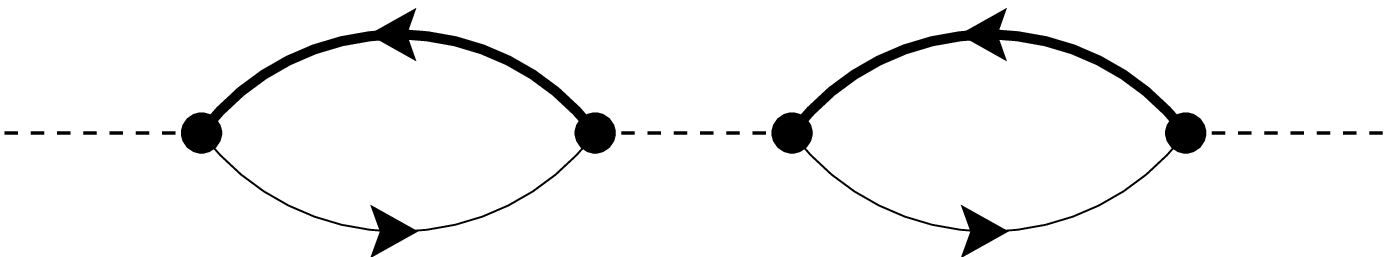} \;+\;
              \cdots
    \end{eqnarray*}
  \caption[\ ]{\label{fig-red}%
    The minimal self-consistent approximation (MSCA) for the
      disordered phase at 
      $\protect T>0\protect$ discussed in Sect.\
      \protect\ref{sec-red}, Eqs.(\protect\ref{eqn-scred})\,. The full
      line denotes the fermion propagator
      Eq.(\protect\ref{eqn-red-gf}), which contains
      the self energy \protect$\Sigma\protect$\,,
      Eq.(\protect\ref{eqn-red-self})\,. The thick dashed line is the
      renormalized interaction D, Eq.(\protect\ref{eqn-red-d})\,. Note that in
      contrast to the fully self-consistent approximation shown in Fig.\
      \ref{fig-flex}, the fermion propagator in the self energy
      $\Sigma$ as well as one of the fermion lines in the irreducible bubble
      $\Pi$\,, Eq.(\protect\ref{eqn-red-pi}), remain unrenormalized
      (bare). These are given by \protect$G^0(i\omega)= 1/
      i\omega\protect$\,. In the MSCA the only
      self-consistent quantity is the susceptibility \protect$\chi\protect$\,,
      Eq.(\protect\ref{eqn-red-chi}), that enters \protect$D\protect$\,.
    }
\end{figure}

The most important property is the self-consistent treatment of the
susceptibility $\chi$\,. This was achieved by renormalizing the
propagator $G$ of the fermion self-consistently, with $\chi$ appearing
in the self energy to one-loop order, see Fig.\ \ref{fig-flex}\,. In
that scheme, the fermion propagator is subject to a self-consistency
loop even if $\chi$ is held fixed, which smears out any pole
structures. Fig.\ \ref{fig-red} shows an approximation, where the
amount of self-consistency is minimized, which we refer to as the
`minimal self-consistent approximation' (MSCA). In Fig.\ \ref{fig-red}
only the susceptibility (the bubble series entering $D$) is a
self-consistent quantity. The auxiliary-fermion propagator, which is
itself not a physical object, is kept unrenormalized. The $\Pi$ shown
in Fig.\ \ref{fig-red} is a bubble of bare fermions, with $D\sim\chi$
inserted to infinite order in the simplest way that preserves the
analytic properties of $\Pi(z)$\,.

The set of equations corresponding to Fig.\ \ref{fig-red} actually
is a simplification of Eqs.(\ref{eqn-scflex})\,. With the bare fermion
propagator $G^0(i\omega)= 1 / i\omega$ they read
\begin{subequations}  \label{eqn-scred}
\begin{eqnarray}
  \Pi(i\nu) & = &  \label{eqn-red-pi}
    -\frac{1}{2\,\beta}\sum_{i\omega}G(i\omega + i\nu) 
     \frac{1}{i\omega}
    \\
  \chi(\mathbf{q}, i\nu) & = &  \label{eqn-red-chi}
    \frac{\Pi(i\nu)}{1 + J(\mathbf{q})\Pi(i\nu)} 
    \\
  D(i\nu) & = &  \label{eqn-red-d}
    \frac{1}{N_L}\sum_{\mathbf{q}}J^2(\mathbf{q})
    \chi(\mathbf{q}, i\nu)
    \\
  \Sigma(i\omega) & = &  \label{eqn-red-self}
    \frac{3}{4\,\beta}\sum_{i\nu}D(i\nu)
    \frac{1}{i\omega + i\nu}
    \\
  G(i\omega) & = &  \label{eqn-red-gf}
    \left[ i\omega - \Sigma(i\omega) \right]^{-1}
\end{eqnarray}
\end{subequations}
Again we first consider the static approximation, i.e., let $D(i\nu)=
D(0)\delta_{\nu,0}$ in the self energy Eq.(\ref{eqn-red-self})\,. This
leads to
$\Sigma(i\omega)= (\bar{\omega}_s)^2\,/\,i\omega$ and thus
\begin{eqnarray}
  G(i\omega) & = &  \label{eqn-fgfredstat}
    \frac{1}{2}\left[
    \frac{1}{i\omega - \bar{\omega}_s} + \frac{1}{i\omega +
    \bar{\omega}_s} \right]
    \\
  \bar{\omega}_s & = &  \nonumber
    \frac{\omega_f}{2}= \sqrt{\frac{3}{4\beta}D(0)}
\end{eqnarray}
That is, the fermion spectrum consists of two $\delta$-peaks at
$\pm\bar{\omega}_s\sim J$\,, which is much more close to the
anticipated (pseudo) gap than the continuous one
Eq.(\ref{eqn-fspec})\,. Within the static approximation the gap comes
out as a hard gap of $2\bar{\omega}_s$\,, it almost resembles the
mean-field result for the ordered state at $T= 0$ given by
Eq.(\ref{eqn-afdos})\,. From Eq.(\ref{eqn-red-pi}) we get
\begin{displaymath}
  \Pi(i\nu)=
    \frac{1}{4}\frac{\bar{\omega}_s}{\bar{\omega}_s^2 - (i\nu)^2}
\end{displaymath}
Here also the similarity to the mean-field expression
Eq.(\ref{eqn-pixx}) is apparent. At the real axis 
$i\nu\to \omega + i0_+$\,, we find using
$J(\mathbf{q})\simeq -2dJ + Jk^2$\,,
$\mathbf{k}= \mathbf{q} - \mathbf{Q}$\,,
\begin{equation}  \label{eqn-redchi}
  \chi(\mathbf{q}, \omega)= 
    \frac{c^2}{J}\left[
    c^2(\xi^{-2} + k^2) - \omega^2 - i\omega\,0_+ \right]^{-1}
\end{equation}
with
$c^2= J \bar{\omega}_s^2 \Pi(0)$
and $\xi^{-2}$ from Eq.(\ref{eqn-xidef})\,. Omitting terms of
$\mathcal{O}(\mathrm{e}^{-\beta\bar{\omega}_s})$ and $\mathcal{O}(\xi^{-2})$\,,
for $d=2$ it is
$\Pi(0)= 1/4J$ from Eq.(\ref{eqn-xidef}), $\Pi(0)=
1/4\bar{\omega}_s$\,, and thus $\bar{\omega}_s= J$\,. The latter has
to be equal to $\bar{\omega}_s= \omega_f/2$\,, with $\omega_f$ already
derived in Eq.(\ref{eqn-coupl})\,. The solution of the resulting
self-consistency equation gives a correlation length and spin-wave
velocity
\begin{equation}  \label{eqn-redcorrl}
  \xi(T)\propto\exp(a_s\,J/T)
  \;\;,\;\;\;
  a_s= \frac{\pi}{6}\simeq 0.52
  \;\;,\;\;\;
  c= \frac{J}{2}
\end{equation}

Accordingly, in the static approximation, the MSCA delivers sharp
`spin-waves'. Their dispersion
$\Omega(\mathbf{q})= \sqrt{\Delta^2 + c^2 k^2}\ge\Delta$
features a gap $\Delta= c\xi^{-1}$\,, since at $T>0$ the spin-rotation
symmetry is not broken. In the static approximation the dynamical
scaling property (\ref{eqn-hypdyn}), (\ref{eqn-hypstat}) is almost
trivially fulfilled by Eq.(\ref{eqn-redchi}), since the linear term
$\sim i\omega$ does not contribute: From
$\chi(\mathbf{q},\omega)= 
   \frac{1}{J}[ 1 + (k\xi)^2 - (\omega/\Delta)^2 - i\omega 0_+ ]^{-1}$
   the structure factor (\ref{eqn-strucdef}) immediately takes the
   form (\ref{eqn-hypdyn}), with the scales $\xi$ and $\omega_0=
   \Delta$\,.  The dynamical exponent comes out correctly as $z=1$\,.
   The infinite lifetime of spin waves originates from the hard gap
   $2\bar{\omega}_s= 2J$ in the fermion propagator
   (\ref{eqn-fgfredstat})\,. If the static approximation in $\Sigma$
   is relaxed, this gap will be partly filled, and spin excitations
   acquire a finite damping. In that case both terms $\sim i\omega$ and
   $\sim\omega^2$ contribute in $\chi(\mathbf{q},\omega)^{-1}$\,, and
   dynamical scaling becomes non-trivial.

\subsection{Relation to Earlier Work}
\label{sec-red-relate}
Our minimal self-consistent approximation bears similarity to a scheme
that has been introduced early on in Refs.\ 
\onlinecite{marschw59,kadmar61,patton71} and later used in Refs.\ 
\onlinecite{levin97,levin99} to study pseudo-gap effects caused by
strong (resonant) Cooper-pair-amplitude fluctuations in
low-dimensional systems.  The physics is similar, except for the type
of long-range fluctuations responsible for the formation of a pseudo
gap.  The fermion self-energy used in, e.g., Ref.\ 
\onlinecite{levin97} consists of particle--particle ladders (pair
susceptibility) convoluted with a bare fermion propagator, whereas in
Fig.\ \ref{fig-red} it is the particle--hole channel (magnetic
susceptibility). In both cases the ladder element (or bubble) is
formed with one bare and one renormalized Green's function.
Apparently, in order to obtain a pseudo-gap in the fermion spectrum
due to strong collective fluctuations, an approximation with `reduced
self consistency' is desirable. This has also been observed in Ref.\ 
\onlinecite{vilktrem97} for the Hubbard model in 2D close to the
antiferromagnetic transition.

Supposedly, the MSCA cannot be obtained from the stationary point of
an approximate free-energy functional ($\Phi$-derivable
approximation\cite{lutwar60,bay62}), since both bare and renormalized
Green's functions enter the self energy\cite{note-phiappr}. However,
exact equations of motion in combination with the use of cumulants
(connected Green's functions) provide an alternative systematic
approach. The basic approximation is to ignore the effective
interaction (vertex function) of 3 and more fermions. That technique
has been introduced in Ref.\ \onlinecite{marschw59} and applied to
superconductors in Refs.\ \onlinecite{kadmar61,patton71}\,. In
Appendix \ref{sec-app-emotion} we re-derive the MSCA for the
Heisenberg model from the equations of motion.

\begin{figure}
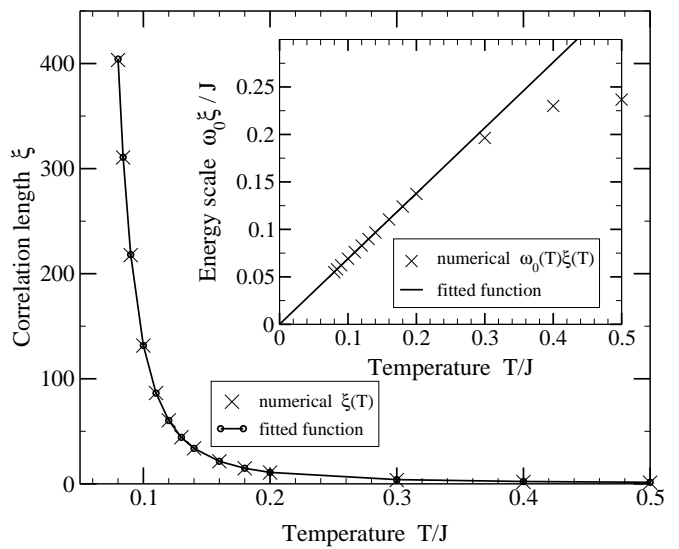

  \mygraph{width=\hsize,clip=true}{fig_minapr_corrl}
  \caption[\ ]{\label{fig-minapr_corrl}%
\textbf{Main figure:}
Correlation length \protect$\xi(T)\protect$ of the minimal
    self-consistent approximation (MSCA). Shown is the data from
    the numerical solution of Eqs.(\protect\ref{eqn-scred})
    and the function \protect$\xi(T)=
    0.362\frac{J}{T}\,\exp(0.359\,J/T)\protect$ resulting from
    a linear regression of the data, plotted as  \protect$[t(\ln(\xi)
    + \ln(t))]\protect$\,, to the function \protect$[a + t\ln(b)], t=
    T/J\protect$ for \protect$T\le 0.18\,J\protect$\,.
\textbf{Inset:}
Energy scale \protect$\omega_0(T)\protect$ of the MSCA,
    extracted from the numerical solution of
    Eqs.(\protect\ref{eqn-scred}) using Eq.(\protect\ref{eqn-extractomnul})\,. 
    Shown is the numerical data for
    \protect$\omega_0(T)\xi(T)\protect$ and the function
    \protect$\omega_0\xi= 0.690\,T\protect$\,, obtained from
    a linear regression for \protect$T\le 0.18\,J\protect$\,. The
    extrapolation to $\protect T=0\protect$ is
    \protect$\omega_0(0)\xi(0)= 0.00003\,J\protect$\,.
    }
\end{figure}
\subsection{Numerical Results}
\label{sec-red-num}
In the following we solve Eqs.(\ref{eqn-scred}) numerically, taking
the frequency dependence of $D(i\nu)$ into account.
Eqs.(\ref{eqn-scred}) are analytically continued to the real axis,
$i\nu\to \omega + i0_+$\,, leading to Eqs.(\ref{eqn-num-scred})\,.
These are solved by numerical iteration. Numerically stable solutions
have been obtained for temperatures $T\agt 0.08J$\,, corresponding to
correlation lengths $\xi\alt 400$ lattice spacings. Correlation length
$\xi(T)$ and energy scale $\omega_0(T)$ are extracted from the
numerical data using Eq.(\ref{eqn-xidef}) and
Eq.(\ref{eqn-extractomnul}), respectively. For $T\alt 0.2J$\,, where
$\xi\agt 10$\,, these are well fitted by
\begin{equation}  \label{eqn-rednumfit}
  \xi(T)\propto\frac{J}{T}\exp(a\,J/T)  \;\;,\;\;\;
  a= 0.36  \;\;,\;\;\;
  \omega_0(T)\propto\frac{T}{\xi}
\end{equation}
Fig.\ \ref{fig-minapr_corrl} shows the numerical data and the fits.

\begin{figure}
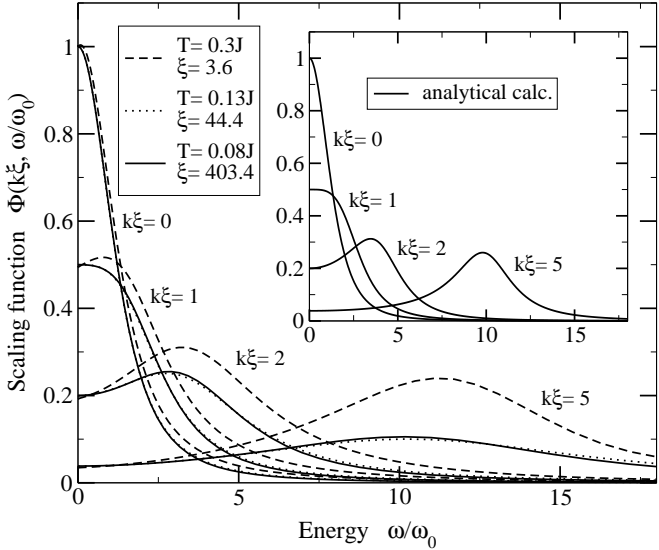

  \mygraph{width=\hsize,clip=true}{fig_minapr_dynscale}
  \caption[\ ]{\label{fig-minapr_dynscale}%
\textbf{Main figure:}
Dynamical structure factor from the numerical solution of the MSCA,
    Eqs.(\protect\ref{eqn-scred}), at low energies 
    \protect$\sim\omega_0\protect$\,. Plotted is the
    scaling function \protect$\Phi(k\xi, \omega/\omega_0)\protect$
    as defined in
    Eq.(\protect\ref{eqn-extractscalefun})\,. \protect$\Phi\protect$
    is shown for fixed wave vectors \protect$k= 0, \xi^{-1},
    2\xi^{-1}, 5\xi^{-1}\protect$\,, each for 3 different
    temperatures corresponding to \protect$\xi= 3.6, 44.4,
    403.4\protect$\,. The curves for \protect$\xi= 44.4,
    403.4\protect$ are almost indistinguishable. For wave
    vectors \protect$k>\xi^{-1}\protect$ a propagating `spin-wave' mode
    becomes visible.
\textbf{Inset:}
The scaling function \protect$\Phi(k\xi,
    \omega/\omega_0)\protect$ of
    the analytical calculation described in Sect.\ 
    \protect\ref{sec-red-anl},
    Eq.(\protect\ref{eqn-scalered}), for \protect$p= 0.25\protect$\,.
    }
\end{figure}

\begin{figure}
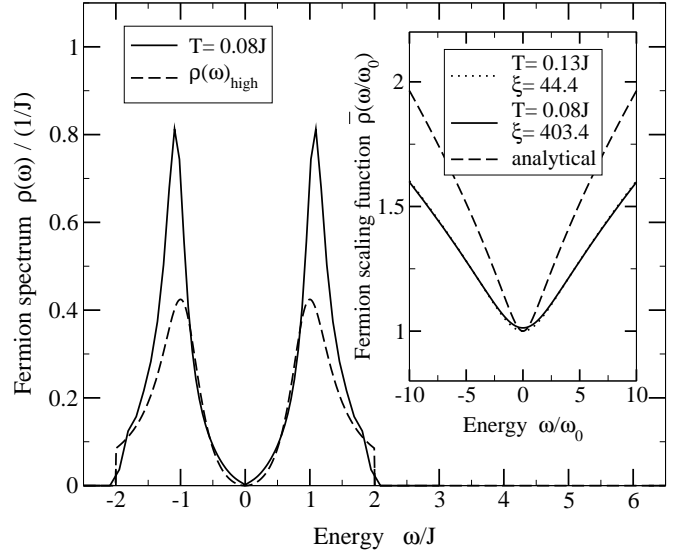

  \mygraph{width=\hsize,clip=true}{fig_minapr_rhof}
  \caption[\ ]{\label{fig-minapr_rhof}%
\textbf{Main figure:}
Numerical spectral function \protect$\rho(\omega)\protect$ of the
    fermion (full line) in the MSCA, Eqs.(\protect\ref{eqn-scred}), in
    the high-energy range \protect$\sim J\protect$\,.
The dashed line is the high-energy approximation
    Eq.(\protect\ref{eqn-rho-high}) from the analytical calculation.
\textbf{Inset:}
Numerical \protect$\rho(\omega)\protect$ at low energies
    \protect$\sim\omega_0\protect$\,, shown is the scaling function
    \protect$\bar{\rho}(\omega/\omega_0)=
    \frac{3JT}{\omega_0}\rho(\omega)\protect$ introduced in
    Eq.(\protect\ref{eqn-rho-low})\,. The curves for 
    \protect$\xi= 44.4\protect$ and \protect$403.4\protect$ (dotted
    and full line, respectively) are almost identical. The dashed line
    is \protect$\bar{\rho}(\omega/\omega_0)\protect$ from the analytical
    calculation, Eq.(\protect\ref{eqn-rho-low})\,.
    }
\end{figure}
The result for $\xi$ contains a factor $1/T$ in front of the usual
exponential. In that respect $\xi$ is different from the static
approximation discussed above. The $1/T$ prefactor has been obtained
for the QNL$\sigma$M\,, as long as the RG calculation is restricted to
1-loop order\cite{chn88}. The energy scale $\omega_0\sim\xi^{-1}$\,,
which we confirm by analytical calculation later on, is now consistent
with a dynamical exponent $z=1$\,. The MSCA also features short-range
order in the dynamical structure factor $S(\mathbf{q}, \omega)$\,: The
scaling function $\Phi(k\xi,\omega/\omega_0)$\,, extracted from the
numerical data using Eq.(\ref{eqn-extractscalefun}), is plotted in
Fig.\ \ref{fig-minapr_dynscale}\,. Apparently the single peak at
$\omega=0$ splits into two peaks located symmetrically to
$\omega=0$\,, if the inverse wave vector is smaller than the
correlation length, $k\xi>1$\,. (Negative energies are not shown in
the figure). These peaks move away from $\omega=0$ with increasing
$k$\,, similar to damped spin waves. According to the analytical
calculation, their dispersion $\bar{\omega}(k\xi)$ becomes linear for
$k\xi\gg 1$\,, $|\bar{\omega}|/\omega_0= 2k\xi$\,. The numerical data
obeys scaling for temperatures below $\sim 0.1 J$\,, where $\xi>
10$\,.  Dynamical scaling is also visible in the spectrum
$\rho(\omega)$ of the fermions. The inset of Fig.\ 
\ref{fig-minapr_rhof} shows the scaling function
$\bar{\rho}(\omega/\omega_0)= 
   \frac{3JT}{\omega_0}\rho(\omega)
   \propto\xi\rho(\omega)$
   at low energies $\sim\omega_0$\,. The curves for $\xi\simeq 40$
   (dotted line) and for $\xi\simeq 400$ (full line) lie on top of
   each other. The definition of $\bar{\rho}$ is a result of the
   analytical calculation, Eq.(\ref{eqn-rho-low})\,. The main Figure
   \ref{fig-minapr_rhof} shows an overall view of $\rho(\omega)$\,,
   the anticipated pseudo-gap structure is apparent, with two broad
   peaks near $\pm J$\,.

\subsection{Analytical Calculation: Origin of Scaling
            and the Propagating Mode}
\label{sec-red-anl}
In order to confirm the numerical result for the energy scale,
$\omega_0\propto T/\xi$\,, and to obtain some insight into the origin
of the dynamical scaling numerically observed, we finally discuss an
approximate analytical solution of Eqs.(\ref{eqn-scred}) for $T\ll J$
and $\xi\gg 1$\,, which does not make use of the static approximation.

We start from the following set of equations, which is derived in
Appendix \ref{sec-app-redan} from Eqs.(\ref{eqn-scred}),
\begin{subequations}  \label{eqn-redan}\label{EQN-REDAN}
\begin{eqnarray}
  \chi(\mathbf{q}, \omega) & = &     \label{eqn-redan-chi}
    \frac{\xi^2}{J}\frac{1}{1 + (\xi k)^2 - R(\omega)}
    \\
  R(\omega) & = &                    \label{eqn-redan-r} 
      16 J \xi^2 \widetilde{\Pi}(\omega)
    \\
  \widetilde{\Pi}(\omega) & = &      \label{eqn-redan-pi}
    i\frac{\pi}{4}\tanh(\omega/2T)\rho(\omega) +
    \\
  & &  \nonumber \mbox{}+
    \frac{\omega^2}{2}\int_0^\infty\!\!\mathrm{d}\varepsilon\,
    \rho(\varepsilon)\frac{\tanh(\varepsilon/2T)}
                          {\varepsilon(\varepsilon^2 - \omega^2)}
    \\
  \rho(\omega) & = &                 \label{eqn-redan-rho}
    -\frac{1}{\pi}\textrm{Im}[ \omega - \Sigma(\omega) ]^{-1}
    \\
  \Sigma(\omega) & = &               \label{eqn-redan-self}
    \frac{3}{8\pi}\int_{-\infty}^\infty\!\!\mathrm{d}\varepsilon\,
    \textrm{Im}D(\varepsilon)\frac{\coth(\varepsilon/2T)}
                                  {\omega - \varepsilon + i0_+}
    \\
  D(\omega) & = &                    \label{eqn-redan-int}
    \frac{4J}{\pi}\int_0^{(k_c\xi)^2}\!\!\mathrm{d}x\,
    \frac{1}{1 + x - R(\omega)}
\end{eqnarray}
The correlation length $\xi$ is given by Eq.(\ref{eqn-xidef}) as
before, $k_c$ denotes a wave-vector cut off of $\mathcal{O}(1)$\,. The
static $\omega=0$ value of $\Pi$ required for Eq.(\ref{eqn-xidef})
reads
\begin{equation}  \label{eqn-redan-pinul} 
  \Pi(0)=
    \frac{1}{2}\int_0^\infty\!\!\mathrm{d}\varepsilon\,
    \tanh(\varepsilon/2T)\frac{\rho(\varepsilon)}{\varepsilon}
\end{equation}
\end{subequations}
In Eq.(\ref{eqn-redan-chi}) the static properties of $\chi$ have been
separated from its dynamics, which are determined by $\Pi(0)$ (through
$\xi$) and $\widetilde{\Pi}(\omega)$ (through $R(\omega)$),
respectively. See Appendix \ref{sec-app-redan} for details.

For the dynamical part $R(\omega)$ we consider the ansatz
\begin{equation}  \label{eqn-ransatz}
  R(\omega)= 
    i\omega/\omega_0 + p\,(\omega/\omega_0)^2
\end{equation}
$\omega_0$ and $p$ are independent of $\omega$\,. This form of
$R(\omega)$\,, inserted in Eq.(\ref{eqn-redan-chi}), is able to
describe the relaxation regime $k\xi\lesssim 1$ as well as propagating
`spin-wave' modes for $k\xi\gtrsim 1$\,.

With Eq.(\ref{eqn-ransatz}) the imaginary part of the renormalized
interaction (\ref{eqn-redan-int}) becomes
\begin{equation}  \label{eqn-dspec}
  \textrm{Im}D(\omega)= 
    \frac{4J}{\pi}\arctan(\omega/\omega_0)\,\Theta(\omega_C - |\omega|)
\end{equation}
$\textrm{Im}D(\omega)$ is essentially the local spectral function for
spin excitations. In Eq.(\ref{eqn-dspec}) all features of
$\textrm{Im}D(\omega)$ for $\omega>\omega_0$ have been replaced by a
constant; $\omega_C\sim 2J$ is a high-energy cut-off.

The energy scale $\omega_0$ is exponentially small, $\omega_0\sim
J/\xi$\,. This already follows from the real part at $\omega=0$\,,
which results from Eq.(\ref{eqn-redan-int}) as
$\textrm{Re}D(0)= 
   \frac{8J}{\pi}\ln(k_c\xi)$
and should roughly compare to the result of a Kramers--Kronig
transformation applied to Eq.(\ref{eqn-dspec}),
$\textrm{Re}D(0)\simeq
   \frac{2J}{\pi}\ln(\omega_C/\omega_0)$\,.

   To calculate $\omega_0$ and $p$ with Eq.(\ref{eqn-dspec}) as input,
   Eq.(\ref{eqn-redan-pi}) is expanded up to $\mathcal{O}(\omega^2)$
   and the parameters are identified as
\begin{eqnarray}
  \frac{1}{\omega_0} & = &          \label{eqn-omnul}
    \frac{2\pi J}{T}\xi^2\rho(0)
    \\
  p & = &                           \label{eqn-pee}
    8J \omega_0^2 \xi^2 
    \int_0^\infty\!\!\mathrm{d}\varepsilon\,
    \left.\frac{\tanh(\varepsilon/2T) \,\rho(\varepsilon)}
          {\varepsilon(\varepsilon^2 - \omega^2)}\right|_{\omega\to 0}
\end{eqnarray}
The energy scale $\omega_0$ is extracted easily with the observation
that the self energy at $\omega=0$ is imaginary,
$\textrm{Re}\Sigma(0)= 0$\,. This comes from the particle--hole
symmetry of the Hamiltonian (\ref{eqn-action}) at 1/2-filling. With
Eqs.(\ref{eqn-redan-self}) and (\ref{eqn-dspec}) we get
$\textrm{Im}\Sigma(0)= 
   - 3JT/\pi\omega_0$\,,
leading to
$\rho(0)= -(\pi\textrm{Im}\Sigma(0))^{-1}
        = \omega_0/3JT$\,.
With Eq.(\ref{eqn-omnul}) a self-consistency condition for $\omega_0$
follows, with the solution
\begin{equation}  \label{eqn-omscale}
  \omega_0= \sqrt{\frac{3}{2\pi}}\frac{T}{\xi}
    \;\;,\;\;\;
  \rho(0)= \frac{1}{J}\sqrt{\frac{1}{6\pi}}\frac{1}{\xi}
\end{equation}
Apparently the dynamical exponent is $z=1$\,.

The parameter $p$ requires an integration over the fermion spectrum
$\rho(\varepsilon)$\,. Therefore we approximate $\Sigma(\omega)$ for
low $\omega\lesssim\omega_0\ll T$ and high energies
$\omega\gg\omega_0$ separately, at low $T\ll J, \omega_C$\,. For
$\omega\lesssim\omega_0$ we have from Eq.(\ref{eqn-redan-self}) and
(\ref{eqn-dspec}),
\begin{equation}  \label{eqn-imself-low}
  \textrm{Im}\Sigma(\omega)_{low}= 
    - \frac{3JT}{\pi}\frac{\arctan(\omega/\omega_0)}{\omega}
\end{equation}
The real part is obtained from the Kramers--Kronig transform
\begin{equation}  \label{eqn-self-kk}
  \textrm{Re}\Sigma(\omega)= 
    \frac{1}{\pi}\int_{-\infty}^\infty\!\!\mathrm{d}\varepsilon\,
    \frac{\textrm{Im}\Sigma(\varepsilon)}{\varepsilon - \omega}
\end{equation}
as
\begin{eqnarray}
  \textrm{Re}\Sigma(\omega)_{low} & = &  \nonumber
    \frac{6JT}{\pi^2}\omega
    \int_0^\infty\!\!\mathrm{d}\varepsilon\,
    \frac{\arctan(\varepsilon/\omega_0)}
         {\varepsilon(\omega^2 - \varepsilon^2)}
    \\
  & = &    \label{eqn-reself-low}
    \frac{3JT}{2\pi}\,\frac{\ln[ 1 + (\omega/\omega_0)^2 ]}{\omega}
\end{eqnarray}
Since $\omega\sim\omega_0$ the self-energy is large,
$|\Sigma(\omega)|_{low}\sim JT/\omega_0\propto J\xi\gg\omega$\,, and the
spectrum (\ref{eqn-redan-rho}) is given by
$\rho(\omega)_{low}= \frac{1}{\pi}
   \textrm{Im}[1/\Sigma(\omega)] + \mathcal{O}(\xi^{-2})$\,.
Accordingly the fermion spectrum at low energy takes the form
\begin{eqnarray}
  \rho(\omega)_{low} & = &  \nonumber
    \frac{\omega_0}{3JT}\,\textrm{Im}
    \frac{y}{\frac{1}{2}\ln(1 + y^2) - i\arctan(y)}
    \\
  & \equiv &   \label{eqn-rho-low}
    \frac{\omega_0}{3JT}\,\bar{\rho}(y)
    \;\;\;,\;\;\;
    y= \omega/\omega_0
\end{eqnarray}
where some scaling function $\bar{\rho}(y)$ has been identified. For
$y\to 0$ the previous result (\ref{eqn-omscale}) for $\rho(0)$ is
recovered. $\rho(\omega)$ shows an incomplete gap, the spectral weight
in the region $\omega\alt\omega_0$ is $\sim 1/J\xi$\,. The scaling
function $\bar{\rho}(\omega/\omega_0)$ is plotted in the inset of
Fig.\ \ref{fig-minapr_rhof}, it qualitatively agrees with the result
from the numerical solution.

In the regime of high energy $\omega\gg\omega_0$ the imaginary part of
$\Sigma$ reads
\begin{equation}  \label{eqn-imself-high}
  \textrm{Im}\Sigma(\omega)_{hi}= 
    -\frac{3J}{4}\coth(|\omega|/2T)\Theta(\omega_C - |\omega|)
\end{equation}
and the real part is calculated using Eq.(\ref{eqn-self-kk}),
\begin{eqnarray*}
  \lefteqn{%
  \textrm{Re}\Sigma(\omega)_{hi} =
    \frac{3J}{2\pi}\omega
    \int_{\omega_0}^{\omega_C}\!\!\mathrm{d}\varepsilon\,
    \frac{\coth(\varepsilon/2T)}{\omega^2 - \varepsilon^2}}
    \\
  & \simeq &
    \frac{3J}{2\pi}\omega
    \left[ \int_{\omega_0}^T\!\!\mathrm{d}\varepsilon\,
           \frac{2T}{\varepsilon(\omega^2 - \varepsilon^2)}
           + \int_T^{\omega_C}\!\!\mathrm{d}\varepsilon\,
             \frac{1}{\omega^2 - \varepsilon^2}
    \right]
\end{eqnarray*}
A lower integral boundary $\omega_0$ has been introduced, since in
Eq.(\ref{eqn-redan-self}) the arctan from Eq.(\ref{eqn-dspec}) cuts
off the $1/\varepsilon$ divergence coming from the
$\coth(\varepsilon/2T)$\,. In the resulting $\textrm{Re}\Sigma_{hi}$
we keep only the terms $\sim 1/\omega$ showing the strongest $\omega$
dependence, that is,
\begin{equation}  \label{eqn-reself-high}
  \textrm{Re}\Sigma(\omega)_{hi}=
    \frac{\bar{\omega}^2}{\omega}
    \;\;,\;\;\;
  \bar{\omega}^2= \frac{3JT}{\pi}\ln(T/\omega_0)
\end{equation}
Thus we arrive at the fermion spectrum in the high-energy regime,
\begin{equation}  \label{eqn-rho-high}
  \rho(\omega)_{hi}= 
    \frac{1}{\pi}\,\textrm{Im}
    \frac{\Theta(\omega_C - |\omega|)}
         {\bar{\omega}^2/\omega - \omega - iJ\frac{3}{4}
          \coth(|\omega|/2T)}
\end{equation}
This spectrum has two peaks at $\omega= \pm\bar{\omega}$\,, similar to
the result from the static approximation appearing below
Eq.(\ref{eqn-scred}), $\bar{\omega}$ and $\bar{\omega}_s$ are of the
same order of magnitude $\sim J$\,. However, here the peaks are no
longer $\delta$-like but show a significant
width. $\rho(\omega)_{hi}$ is plotted and compared to the numerical
solution in Fig.\ \ref{fig-minapr_rhof}\,.

With Eqs.(\ref{eqn-rho-low}) and (\ref{eqn-rho-high}) the parameter
$p$ can now be estimated: The integral in Eq.(\ref{eqn-pee}) is split
into three regions and the $\tanh(\varepsilon/2T)$ approximated,
\begin{eqnarray}
  p & = &  \nonumber
    \frac{12}{\pi}JT^2\left[
    \frac{1}{2T}\int_0^{\alpha\omega_0}\!\!\mathrm{d}\varepsilon\,
    \left.\frac{\rho(\varepsilon)_{low}}
    {\varepsilon^2 - \omega^2}\right|_{\omega\to 0} +
    \right.
    \\
  & &  \nonumber
    \mbox{} + \left.
    \frac{1}{2T}\int_{\alpha\omega_0}^T\!\!\mathrm{d}\varepsilon\,
    \frac{\rho(\varepsilon)_{hi}}{\varepsilon^2} +
    \int_T^{\omega_C}\!\!\mathrm{d}\varepsilon\,
    \frac{\rho(\varepsilon)_{hi}}{\varepsilon^3}
    \right]
    \\
  & \equiv &  \label{eqn-pints}
    \frac{12}{\pi}JT^2\left[
      I_{\omega_0} + I_T + I_J \right]
\end{eqnarray}
The result (\ref{eqn-omscale}) for $\omega_0$ has been
inserted. $\alpha$ is a number, $\alpha\omega_0$ indicates where
$\rho_{low}$ and $\rho_{hi}$ roughly match. We assume that $\alpha$
can be chosen constant at small $T$\,, where $\omega_0\ll
\alpha\omega_0\ll T\ll J, \omega_C$\,. We seek the leading
$T$-dependence of $p$\,.

With Eq.(\ref{eqn-rho-low}) the low-energy contribution becomes
$I_{\omega_0}= \lim\limits_{y\to 0}
   \frac{1}{6JT^2}\int_0^\alpha\mathrm{d}x\,
   \bar{\rho}(x)/(x^2 - y^2)$\,,
that is,
\begin{equation}  \label{eqn-pint-omnul}
  I_{\omega_0}\propto 1/JT^2
\end{equation}
For estimating $I_T$ the spectrum (\ref{eqn-rho-high}) is expanded for
$|\omega|\le T\ll J, \bar{\omega}$\,, i.e.,
\begin{eqnarray}
  \rho(\omega)_{hi} & = &  \nonumber 
  \frac{1}{\pi}\textrm{Im}\left[
  \frac{\omega}{\bar{\omega}^2}
  \left\{1 + i\frac{3J}{2\bar{\omega}}\frac{T}{\bar{\omega}}
  \textrm{sign}(\omega)
  + \mathcal{O}(\frac{\omega^2, T^2}{\bar{\omega}^2})
  \right\}\right]
  \\
  & = &  \label{eqn-rho-hiexp}
  \frac{3JT}{2\pi\bar{\omega}^4}|\omega|
\end{eqnarray}
and we arrive at
\begin{eqnarray*}
  I_T & = &
    \frac{3J}{4\pi\bar{\omega}^4}\left[
      \ln(\frac{T}{\omega_0}) - \ln(\alpha) \right]
    =
    \frac{1}{4T\bar{\omega}^2}
    \left[ 1 + \mathcal{O}(\frac{T}{J})\right]
\end{eqnarray*}
The leading $T$-dependence of $\bar{\omega}$ is a constant,
$\bar{\omega}\sim J$ (see below), and we find
\begin{equation}  \label{eqn-pint-t}
  I_T\propto 1/J^2T
\end{equation}
In the high-energy contribution $I_J$ we use Eq.(\ref{eqn-rho-high})
with $\coth(|\omega|/2T)\simeq 1$\,. The dominant contribution to the
integral in $I_J$ comes from the lower bound $T$\,. That is,
$\rho_{hi}$ may be expanded as above, omitting terms
$\mathcal{O}(\omega/J)$\,, which results in
$\rho(\omega)_{hi}= 
   (3J/4\bar{\omega}^4)\,\omega^2$\,.
With $\bar{\omega}=const.\sim J$ its contribution to $p$ is
\begin{equation}  \label{eqn-pint-j}
  I_J\propto \ln(J/T)/J^3
\end{equation}
Collecting Eqs.(\ref{eqn-pint-omnul}), (\ref{eqn-pint-t}),
(\ref{eqn-pint-j}) together, the leading contribution to $p$ for $T\to
0$ is the lowest-energy part $I_{\omega_0}$\,. Its $T$-dependence
cancels the $T^2$ prefactor in Eq.(\ref{eqn-pints}), and
\begin{equation}  \label{eqn-pee2}
  p\propto JT^2\,I_{\omega_0}= const.
\end{equation}

Hence, the parameter $p$ comes out as a (dimensionless) finite
constant. With $p$ in Eq.(\ref{eqn-ransatz}) being constant, the
dynamical susceptibility (\ref{eqn-redan-chi}) immediately assumes a
scaling form,
$\chi(\mathbf{q}, \omega)= 
   \frac{\xi^2}{J}\left[1 + x^2 - py^2 - iy \right]^{-1}$
with length and energy scales $\xi$ and $\omega_0$ in $x= k\xi$\,, $y=
\omega/\omega_0$ given by
\begin{equation}  \label{eqn-redscales}
  \omega_0\propto T/\xi \;\;,\;\;\; \xi\propto \exp(0.63\,J/T)
\end{equation}
The former has been given in Eq.(\ref{eqn-omscale}), the latter will be
derived below.

The structure factor defined in Eq.(\ref{eqn-strucdef}) then takes the
scaling form Eqs.(\ref{eqn-hypdyn}), (\ref{eqn-hypstat}), with the
scaling functions
\begin{equation}  \label{eqn-scalered}
  \varphi(x)= \frac{1}{1 + x^2}
    \;\;,\;\;\;
  \Phi(x,y)= \frac{1 + x^2}{(1 + x^2 - py^2)^2 + y^2}
\end{equation}
The precise value of $p$ is not known from the above calculation.  The
`shape function' $\Phi(x, y)$ is plotted in the inset of Fig.\ 
\ref{fig-minapr_dynscale} for $p=1/4$\,. It consists of a single peak
at $\omega= 0$ for wave vectors in the relaxation regime $x= k\xi\ll
1$\,, which splits into two peaks located symmetrically to $\omega= 0$
for $x= k\xi> \bar{x}$ in the regime of propagating modes. The
boundary between these regimes is given by
$\left.\partial^2_y\Phi(\bar{x}, y)\right|_{y=0}= 0$ and reads
$\bar{x}= \sqrt{(1 - 2p)/2p}$\,. The scaling function $\Phi$ compares
well to the numerical result shown in the main Figure
\ref{fig-minapr_dynscale}\,. Apparently the choice $p= 1/4$
corresponding to $\bar{x}= 1$ is consistent with the numerics. For
energies $\omega>\omega_0$ the damping (i.e., the width of the peaks
for $k\xi> 1$) comes out larger in the numerical calculation,
supposedly due to higher-order terms omitted in the ansatz
(\ref{eqn-ransatz}) for $R(\omega)$ in Eq.(\ref{eqn-redan-chi}). The
`spin-wave' dispersion $\bar{y}(x)= \bar{\omega}(x)/\omega_0$ of the
peak-maximum is determined from $\partial_y\Phi(x, \bar{y})= 0$\,. At
large wave vectors $k\xi\gg\bar{x}$ it becomes linear,
$\bar{\omega}/\omega_0\simeq k\xi/\sqrt{p}$\,.

Finally the correlation length already stated in (\ref{eqn-redscales})
has to be derived self-consistently from Eq.(\ref{eqn-redan-pinul})
and (\ref{eqn-xidef})\,.  The integral in Eq.(\ref{eqn-redan-pinul})
is split into three parts, with the $\tanh(\varepsilon/2T)$
approximated appropriately,
\begin{eqnarray*}
  \Pi(0) & = &
    \frac{1}{4T}\int_0^{\alpha\omega_0}\!\!\mathrm{d}\varepsilon\,
    \rho_{low}(\varepsilon)
    + \frac{1}{4T}\int_{\alpha\omega_0}^T\!\!\mathrm{d}\varepsilon\,
    \rho_{hi}(\varepsilon) +
    \\
  & & \mbox{}
    + \frac{1}{2}\int_T^{\omega_C}\!\!\mathrm{d}\varepsilon\,
    \frac{\rho_{hi}(\varepsilon)}{\varepsilon}
    \\
  & \equiv &
    \Pi(0)_{\omega_0} + \Pi(0)_{T} + \Pi(0)_{J}
\end{eqnarray*}
With $\rho_{low}$ from Eq.(\ref{eqn-rho-low}) inserted, the low-energy
part becomes
$\Pi(0)_{\omega_0}= 
   \frac{\omega_0^2}{12JT^2}\int_0^\alpha\mathrm{d}y\,
   \widehat{\rho}(y)
   \propto (\omega_0/T)^2/J\propto 1/(\xi^2J)$\,.
Its contribution to the self-consistency condition (\ref{eqn-xidef})
can be ignored, since
$\Pi(0)= 1/4J + \mathcal{O}(\xi^{-2})$\,.
In $\Pi(0)_T$ we can again use Eq.(\ref{eqn-rho-hiexp}), i.e.,
$\rho_{hi}(\varepsilon)= 
   \frac{3JT}{2\pi\bar{\omega}^4}|\varepsilon|$\,,
leading to
$\Pi(0)_T\propto JT^2/\bar{\omega}^4\sim T^2/J^3$\,,
which also can be ignored at small $T\ll J$\,. The dominant
contribution to $\Pi(0)$ comes from high energies: Using
$\rho_{hi}$\,, Eq.(\ref{eqn-rho-high}), for $\omega> T$ results in
\begin{eqnarray*}
  \Pi(0)_J & = &
    \frac{1}{2\pi}\int_T^{\omega_C}\!\!
    \frac{\mathrm{d}\varepsilon}{\varepsilon}\,
    \textrm{Im}\frac{1}{\bar{\omega}^2/\varepsilon - \varepsilon - 
                        i\frac{3}{4}J}
    \\
  & = &
    \frac{2}{3\pi J}\int_{4T/3J}^{4\omega_C/3J}\!\!
    \mathrm{d}u\,
    \frac{u}{(g^2 - u^2)^2 + u^2}
\end{eqnarray*}
where
$g^2= (4\bar{\omega}/3J)^2= 
   \frac{16}{3\pi}\frac{T}{J}\ln(T/\omega_0)$\,.
   The above expression has to fulfill $\Pi(0)_J= \Pi(0)= 1/4J$\,,
   giving an implicit equation for $g^2$ and therefore $\xi\propto
   T/\omega_0$\,.  The exponential part of $\xi$ can be extracted
   using $T\to 0$ and $\omega_C\to\infty$ in the integral boundaries.
   We find $g^2= 1.07$\,, and from Eq.(\ref{eqn-omscale}) the correlation
   length is
$\xi(T)= \sqrt{\frac{3}{2\pi}}\exp(a\,J/T)$\,,
$a= 3\pi g^2/16= 0.63$\,.

\section{Quality of the Constraint}
\label{sec-remarks}
The representation of spin operators in auxiliary particles,
  Eq.(\ref{eqn-spinop}), is valid only if the Fock space of canonical
  fermions is restricted to the subspace with
$Q_i= \sum_\alpha f^\dagger_{i\alpha}f_{i\alpha}= 1$
on each lattice site $i$\,.
We have argued in Sects.~\ref{sec-slave}, \ref{sec-mf}, \ref{sec-flex}
that the local charge density $Q_i$ is a conserved quantity in the
approximations considered in this paper, $\partial_\tau Q_i(\tau)=
0$\,. In particular there is no spontaneous breaking of the
corresponding gauge symmetry in the 1/2-filled Heisenberg model.
However, we have replaced the local constraint $Q_i=1$ by a global
one,
$\langle Q_i\rangle= 1
  \Leftrightarrow
  Q_{tot}= \sum_1^{N_L}Q_i= N_L$\,,
  since in that case we were able to use standard many-body
  techniques.
  
  The true ground state of the model Hamiltonian belongs to the
  physical sector of the Hilbert space and thus features a total
  charge $Q_{tot}= N_L$ with a homogeneous density $Q_i= 1$\,. With
  just the global constraint $Q_{tot}= N_L$ in effect, this ground
  state might be (almost) degenerate with multiplets of unphysical
  states where $Q_{tot}$ is distributed inhomogeneously on the
  lattice. In that case the unphysical states mix spectral weight into
  correlation functions like the spin susceptibility. However, one can
  hope that the separation in energy of these unphysical states from
  the ground state is finite, and at $T\to 0$ the correlation
  functions actually measure physical processes. Since the $Q_i$ are
  conserved, the excited states generated in the correlation function
  (e.g., through a spin flip) remain in the physical sector.

The gap of the lowest unphysical state to the ground state will depend
on the physics of the latter. For the AF ordered state of the
mean-field theory (see Sect.~\ref{sec-mf}) chances are good: The
N{\'e}el state $|AF\rangle= |AF; \{Q_i=1\}\rangle$ is compared to some
state $|AF'\rangle= |AF; Q_l=0, Q_{l'}=2, \{Q_i= 1, i\ne
l,l'\}\rangle$\,, where a fermion has been transfered from some site
$l$ to another site $l'$\,. Thus the spins on sites $l$ and $l'$ have
disappeared, and the unphysical state $|AF'\rangle$ is higher in
energy by $2J$\,, equivalent to 8 lost $J$-bonds. The gap $2J$ is just
the charge-transfer gap $\lim_{T\to 0}2|m_A|= 2J$ in the spectrum
Eq.(\ref{eqn-afdos}) of the auxiliary fermions.

The above reasoning can easily be verified by calculation, at least
for the mean-field theory. The local charge (constraint) fluctuations on
an arbitrary site $i$ are
\begin{eqnarray*}
  \langle (\Delta Q_i)^2\rangle & = &
    \langle Q_i \, Q_i \rangle
    - \langle Q_i \rangle \langle Q_i \rangle
    \\
  & = &
    -\frac{1}{\beta}\sum_{i\nu}\frac{1}{\beta}\sum_{i\omega}
    \textrm{Tr}[\overline{G}_i(i\omega + i\nu)\,
      \overline{G}_i(i\omega)]
\end{eqnarray*}
involving a charge-response bubble built similar to the spin response
Eq.(\ref{eqn-bubble})\cite{note-cfluctmf}. With the fermion Green's
function in mean field, Eq.(\ref{eqn-fgfmf}), this becomes
\begin{equation}  \label{eqn-qfluct}
  \langle (\Delta Q_i)^2\rangle =
    2 f(m_A)[1 - f(m_A)]
\end{equation}
for $i\in A$ or $B$ sublattice. Above the mean-field transition
temperature, $T> T_N= J/2$\,, we have $m_A= 0$ and therefore
$\langle (\Delta Q_i)^2\rangle= 1/2$\,.
That is, the constraint is significantly violated. However, in the AF
ordered phase $T< T_N$ we have $|m_A|>0$ and charge fluctuations are
suppressed; at $T\ll T_N$ Eq.(\ref{eqn-qfluct}) leads to activated
behavior,
$\langle (\Delta Q_i)^2\rangle=
   2\exp(-2J/T)$\,,
   revealing a gap $2J$ between the N{\'e}el ground
   state $|AF\rangle$ and the lowest unphysical one $|AF'\rangle$\,.
   At $T= 0$ the constraint is exactly fulfilled.

\begin{figure}
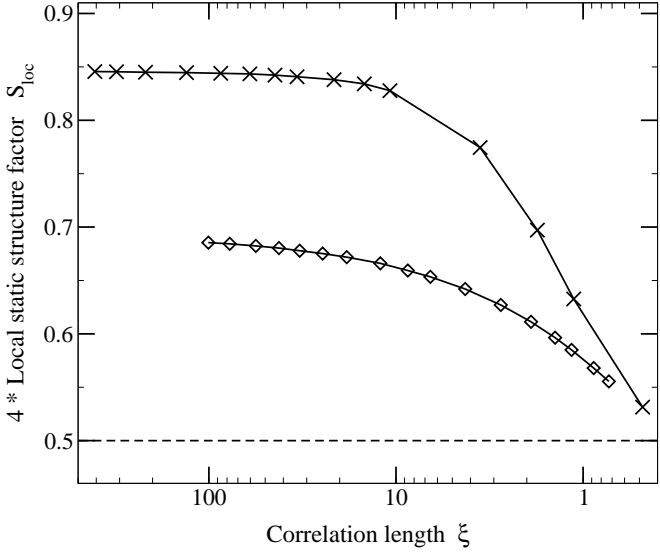

  \mygraph{width=\hsize,clip=true}{fig_bothapr_locmom}
  \caption[\ ]{\label{fig-minapr_locmom}%
    Influence of constraint fluctuations on the local spin moment
    \protect$S^{st}_{loc}= \langle (S^x_i)^2\rangle\protect$\,,
    calculated numerically using Eq.(\ref{eqn-num-locmom})\,. Shown is
    \protect$4S^{st}_{loc}\protect$ as function of the correlation
    length \protect$\xi(T)\protect$ for the MSCA (crosses) and the
    fully
    self-consistent approximation (diamonds).
    For \protect$\xi>10\protect$ the MSCA reaches $85$\% of the exact value
    \protect$4\,S^{st}_{loc}= \frac{4}{3}S(S + 1)= 1\protect$\,.
    The dashed line indicates the high-$T$ limit $1/2$\,.
    }
\end{figure}
For the minimal self-consistent approximation we do not have an
expression for the free energy\cite{note-phiappr} and thus do not know
how to calculate $\langle (\Delta Q_i)^2\rangle$ consistently. An
alternative test for the constraint is the local spin moment, which
reads in the paramagnetic phase $\langle S^\mu_i\rangle= 0$\,,
\begin{displaymath}
  \langle(\mathbf{S}_i)^2\rangle=
    \sum_\mu\langle S^\mu_i S^\mu_i\rangle
    = 3 S^{st}_{loc}
    \;\;,\;\;\;
  S^{st}_{loc}= \langle S^x_i S^x_i\rangle
\end{displaymath}
Since our system of interacting auxiliary fermions represents a pure
spin-1/2 model, we have
$\langle(\mathbf{S}_i)^2\rangle= S(S + 1)= 3/4$\,,
and the local static structure factor should be
$S^{st}_{loc}= 1/4$\,.
Fluctuations of the fermion occupancy $Q_i$ will reduce
$S^{st}_{loc}$ from this exact value. At high temperature $T\gg J$ the
susceptibility Eq.(\ref{eqn-rpa}) is given by the simple fermion
bubble Eq.(\ref{eqn-bubble}), i.e.,
\begin{equation}  \label{eqn-locm-high}
  S^{st}_{loc} =
    \frac{1}{\beta}\sum_{i\nu}
    \Pi_i^{x x}(i\nu)
    =
    \frac{1}{2}\langle f^\dagger_{i\uparrow} f_{i\uparrow}\rangle\,
    \langle f_{i\downarrow} f^\dagger_{i\downarrow}\rangle
    = \frac{1}{8}
\end{equation}
where Eq.(\ref{eqn-fermgf}) and
$\langle f^\dagger_{i\uparrow} f_{i\uparrow}\rangle=
   \langle f^\dagger_{i\downarrow} f_{i\downarrow}\rangle
   = \langle Q_i\rangle/2= 1/2$
   has been used. That is, constraint fluctuations reduce the spin
   moment to $50$\% of its physical value. Note that this holds for
   any density of states of the fermions, including the case of free
   spins ($J= 0$). When temperature is lowered below $\simeq J$ in the
   interacting system, we expect the spin moment to be restored as a
   consequence of a strong reduction of $\langle (\Delta
   Q_i)^2\rangle$ in the short-range orderd regions. $S^{st}_{loc}$
   has been calculated numerically using Eq.(\ref{eqn-num-locmom}),
   the result for the MSCA is shown in Fig.~\ref{fig-minapr_locmom} as
   crosses.  In fact the spin moment saturates to $\simeq 0.85$\,. The
   fully self-consistent approximation, which does not show
   short-range order, reaches $\simeq 0.68$ (shown as diamonds in
   Fig.~\ref{fig-minapr_locmom}).
   
   For comparison we also discuss the spin moment in mean-field
   theory: At high $T\gg T_N$ Eq.(\ref{eqn-locm-high}) holds, and we
   have a reduced moment as above,
$\langle(\mathbf{S}_i)^2\rangle= 3\frac{1}{8}$\,.
  At $T\ll T_N$\,, in the presence of long-range AF order, the spin
  moment is
\begin{displaymath}
  \langle(\mathbf{S}_i)^2\rangle= 
    S^{st||}_{loc} + 2 S^{st\perp}_{loc}
    \;\;,\;\;\;
  S^{st\perp}_{loc}= 
    \frac{1}{N_L}\sum_{\mathbf{q}}S^\perp(\mathbf{q})
\end{displaymath}
In the longitudinal direction just the condensate contributes,
$S^{st||}_{loc}= \langle S^z_A\rangle^2$\,,
since fluctuations have vanishing weight,
$\langle S^z_i S^z_i\rangle_{conn}\sim\exp(-J/T)$\,.
The transversal contribution $S^{st\perp}_{loc}$ measures the spectral
weight of spin waves and reads, from Eq.(\ref{eqn-swtstruct}),
\begin{displaymath}
  S^{st\perp}_{loc}=
    \frac{1}{2}|\langle S^z_A\rangle|
    (1 + 2\bm{\epsilon})
    \;\;\Rightarrow\;\; 
  \langle(\mathbf{S}_i)^2\rangle= 
    \frac{1}{4}[1 + 2(1 + 2\bm{\epsilon})]
\end{displaymath}
with
$|\langle S^z_A\rangle|= 1/2 + \mathcal{O}(\mathrm{e}^{-J/T})$
and the $\bm{\epsilon}$ introduced in Eq.(\ref{eqn-corr2})\,. At $T=
0$ the loop integral $\bm{\epsilon}= 0.197$ is finite in 2D\,.
The spin moment is fully restored, the contribution of
(transversal) magnons is even overestimated in spin-wave theory.

\section{Summary and Conclusion}
\label{sec-concl}
This paper presented a study of critical fluctuations in the quantum
Heisenberg antiferromagnet in two dimensions. Our starting point is to
re-write the spin-$1/2$ operators in (unphysical) fermionic auxiliary
particles, which enables the use of standard many-body techniques
based on Feynman diagrams. Although the Heisenberg and the
$t$-$J$-model have frequently been studied on mean-field level, the
problem of strong long-range fluctuations in the two-dimensional plane
has not been considered before by this type of approach.

The ground state at zero temperature $T=0$ shows long-range
antiferromagnetic order and can be treated in mean-field theory and
perturbative (in $1/S$) corrections. The auxiliary fermions form a
simple spin-density-wave state, their density of states consists of
two delta peaks separated by a large gap $2J$\,. Results for
magnetization, spin-wave velocity and -weight have been derived in
Sect.~\ref{sec-mf}\,. We found even quantitative agreement with
conventional spin-wave theory.

At finite temperature the perturbation theory in $1/S$ breaks down,
and the susceptibility has to be treated self-consistently in order to
suppress the mean-field transition temperature $T_N= J/2$ down to
zero, as required by the theorem of Mermin and Wagner. In
Sect.~\ref{sec-flex} we considered the simplest skeleton-diagram
expansion in the renormalized fermion-propagator that provides this
feature. Within a static (classical) approximation (see
Sect.~\ref{sec-flex-static}) the self-consistency problem can be
solved analytically, leading to a correlation length
$\xi(T)\propto\exp(a\,J/T)$ and an exponentially small energy scale
$\omega_0\propto J\xi^{-z}$\,, in agreement with the existing
theoretical and experimental work. The so-called dynamical scaling
relation\cite{chn89} is also fulfilled. However, the dynamical
exponent $z=2$ is incorrect, and the spin-excitation spectrum lacks a
spin-wave like propagating mode. The analytical results have been
confirmed by numerical calculation in Sect.~\ref{sec-flex-num}\,.

We argued that the lack of `spin waves' is due to a large density of
states $\rho(\omega\simeq 0)\sim 1/J$ of the auxiliary particles at
the Fermi energy $\omega= 0$\,, which causes an extreme overdamping.
It turns out that this problem can be cured by reducing the amount of
self-consistency: In the skeleton-diagram expansion both the
susceptibility and the Green's function of fermions are subject to
their respective self-consistency loops, while being coupled to each
other through the fermion self-energy. In the MSCA presented in
Sect.~\ref{sec-red}\,, which is not based on skeleton diagrams, the
additional loop for the fermion propagator is avoided. The results
obtained from this approximation are very satisfactory.  In particular
we find $\xi(T)\propto \frac{1}{T}\exp(0.359\,J/T)$ and the energy
scale $\omega_0\sim 1/\xi$\,, i.e., the dynamical exponent is $z=1$\,.
The spin-excitation spectrum now features dynamical scaling as well as
a propagating mode for $k\xi>1$\,. $k$ denotes the offset from the
N{\'e}el wave vector. The MSCA equations have also been solved by
analytical calculation in Sect.~\ref{sec-red-anl}\,, confirming in
particular the energy scale $\omega_0= \sqrt{3/2\pi}\,T/\xi$\,. In the
MSCA the spectral function $\rho(\omega)$ of auxiliary fermions
features a strong suppression (pseudo gap) at the Fermi energy. The
spectral weight at $\omega= 0$ is exponentially small, $\rho(0)\propto
1/J\xi$\,, and at low energies $\rho(\omega)$ shows scaling,
$\rho(\omega)= \frac{\omega_0}{3JT} \bar{\rho}(\omega/\omega_0)$\,.

Our MSCA is technically and physically related to earlier work on low
dimensional superconductors and the 2D Hubbard model, where long-range
fluctuations of the superconducting or magnetic order-parameter
amplitude cause a pseudo gap in the single-particle density of states.
References have been given in Sect.~\ref{sec-red-relate}\,. Supposedly
there is no $\Phi$-functional available\cite{note-phiappr} which could
have served as a systematic basis for the approximation scheme.
However, these schemes follow quite naturally from a sequence of
equations of motion in combination with a cumulant decomposition of
higher-order vertex functions. This type of method has been used in
the above-mentioned studies on superconductors and is applied to the
present problem in App.~\ref{sec-app-emotion}\,.

During all calculations performed in this paper the auxiliary-particle
constraint $Q_i= \sum_\alpha f^\dagger_{i\alpha}f_{i\alpha}= 1$ has
been treated in a mean-field fashion, $\langle Q_i\rangle= 1$\,.
Although it turns out that the fermions remain strictly local at
$1/2$-filling (no intersite hopping), on-site fluctuations $\langle
(\Delta Q_i)^2\rangle$ into unphysical sectors of Hilbert space are
not a priori suppressed.  We did not attempt to improve on the
constraint, e.g., by including a fluctuating Lagrange multiplier
$\lambda_i(\tau)$ in the action Eq.(\ref{eqn-action}) via the term
$\int_0^\beta\mathrm{d}\tau\,
   \sum_i\lambda_i(\tau)[1 - Q_i(\tau)]$\,.
   However, as has been explored in Sect.~\ref{sec-remarks}\,, the
   actual amount of constraint fluctuations at low temperature
   depends on the physics of the ground state.  Measuring the local
   spin moment $\langle(\mathbf{S}_i)^2\rangle$ in the MSCA revealed
   reasonably small fluctuations $\langle(\Delta Q_i)^2\rangle$ at
   $T\ll J$\,.  This is due to the magnetic order appearing on short
   length scales. At $T=0$\,, with long-range order present, we even
   find $\langle(\Delta Q_i)^2\rangle= 0$ in mean-field theory.
   
   A major question remaining open is whether the theory violates
   spin-rotation symmetry. In Sect.~\ref{sec-flex-cons} it has been
   shown that a straight-forward application of the
   conserving-approximation scheme in two dimensions leads to a
   spurious phase transition at some $\widehat{T}_N\sim J$\,, because
   the susceptibility is modified by additional vertex corrections
   that are not part of the self-consistency loop. However, since
   approximations constructed from this scheme do respect conservation
   laws only in the sense of a sufficient condition, it is worth
   testing our MSCA for spin-rotation symmetry `by hand', e.g.,
   through confirming the Goldstone peak at $\mathbf{q}= (\pi,\pi)$
   and $\omega= 0$ in the long-range ordered state at $T= 0$\,. This
   requires a self-consistent calculation at
   $\langle\mathbf{S}\rangle\ne 0$\,, which we postpone to future
   work.
   
   We think that the auxiliary-fermion approach proved quite useful
   for dealing with strong critical fluctuations in the
   two-dimensional quantum model. Using standard many-particle
   techniques a suitable self-consistent approximation can be found
   and treated with moderate effort, even by analytical means. The
   approach may serve as a starting point for studies of strong
   fluctuations in the $t$-$J$-model at small doping levels.  This
   would require to add bosons representing the doped holes (standard
   `slave-boson' formulation).  Also the connection of the present
   study to the more general problem of pseudo-gap formation from
   long-range fluctuations might be helpful for the development of
   theories for cuprate superconductors. Other possible applications
   include spin systems like the 2D Heisenberg model with frustration
   (e.g., the $J_1$-$J_2$-model) or different lattice geometry.

\begin{acknowledgments}
  We acknowledge fruitful discussions on several stages of this work
  with A.~Millis, A.~Rosch, A.~Ruckenstein, O.~F.~Sylju\r{a}sen, and
  A.-M.~S.~Tremblay\,. This work has been partially supported by the
  Deutsche Forschungsgemeinschaft.
\end{acknowledgments}

\appendix
\section{Self-Consistency Equations}
\subsection{Equations for Numerical Solution}
\label{sec-app-flex}
For a solution of the Eqs.(\ref{eqn-scflex}) by numerical iteration,
retarded propagators are introduced by analytic continuation to the
real axis,
\begin{displaymath}
  A(i\nu)\to A(\omega + i0_+)\equiv A(\omega)= 
    A'(\omega) + iA''(\omega)
\end{displaymath}
where $A= \chi, \Pi, D, \Sigma$\,. In order to avoid the numerical
treatment of Bose functions $g(\omega)$ we introduce the structure
factors
\begin{eqnarray*}
  S(\mathbf{q}, \omega) & = & 
    [1 + g(\omega)]\,\chi''(\mathbf{q}, \omega)
    \\
  S^0(\omega) & = & 
    [1 + g(\omega)]\,\Pi''(\omega)
    \\
  U(\omega) & = & 
    [1 + g(\omega)]\,D''(\omega)
\end{eqnarray*}
As will be shown further below, the Eqs.(\ref{eqn-scflex}) can now be
written as
\begin{subequations}  \label{eqn-num-scflex}
\begin{eqnarray}
  S(\mathbf{q}, \omega) & = &        \label{eqn-num-sc-struct}
    \frac{S^0(\omega)}
         {|1 + J(\mathbf{q})\Pi(\omega)|^2}
    \\
  \Pi''(\omega) & = &                \label{eqn-num-sc-piim}
    S^0(\omega) - S^0(-\omega)
    \\
  \Pi'(\omega) & = &                  \label{eqn-num-sc-pire}
    \frac{1}{\pi}P\!\!\int_{-\infty}^\infty\!\!\mathrm{d}\varepsilon\,
    \frac{\Pi''(\varepsilon)}{\varepsilon - \omega}
    \\
  S^0(\omega) & = &                  \label{eqn-num-sc-snull}
    \frac{\pi}{2}\int_{-\infty}^{\infty}\!\!\mathrm{d}\varepsilon\,
    \rho^{+}(\varepsilon)\rho^{-}(\varepsilon - \omega)
    \\
  \rho^{\{{+\atop -}\}}(\omega) & = &  \label{eqn-num-sc-rho}
    \left\{{1 - f(\omega) \atop f(\omega)}\right\}
    \frac{1}{\pi}\frac{-\Sigma''(\omega)}
                      {|\omega - \Sigma(\omega)|^2}
    \\
  -\Sigma''(\omega) & = &            \label{eqn-num-sc-selfim}
    \frac{3}{4}\int_{-\infty}^{\infty}\!\!\mathrm{d}\varepsilon\,
    \left[ U(\varepsilon)\rho^{-}(\varepsilon + \omega) + \right.
    \\[-0.5em]
  & & \mbox{}\hspace*{15mm}  \nonumber
    \left. + U(-\varepsilon)\rho^{+}(\varepsilon + \omega) \right]
    \\
  \Sigma'(\omega) & = &              \label{eqn-num-sc-selfre}
    \frac{1}{\pi}P\!\!\int_{-\infty}^\infty\!\!\mathrm{d}\varepsilon\,
    \frac{-\Sigma''(\varepsilon)}{\omega - \varepsilon}
    \\
  U(\omega) & = &                    \label{eqn-num-sc-int} 
    S^0(\omega)\!\!\int_{-4J}^{4J}\!\!\mathrm{d}\varepsilon\,
    \frac{\mathcal{N}(\varepsilon)\,\varepsilon^2}
    {\left|1 + \varepsilon\,\Pi(\omega)\right|^2}
\end{eqnarray}
with the density of states for the nearest-neighbor interaction
$J(\mathbf{q})$ in two dimensions,
\begin{displaymath}
  \mathcal{N}(\varepsilon)= 
    \int\!\!\!\!\int_{-\pi}^\pi\frac{\mathrm{d}^2 q}{(2\pi)^2}
    \delta(\varepsilon - 2J[\cos(q_x) + \cos(q_y)])
\end{displaymath}
The temperature enters through the Fermi function $f(\omega)$ in
$\rho^{+}\,, \rho^{-}$\,. The output from the numerical iteration of
these equations is the dynamical spin-structure factor $S(\mathbf{q},
\omega)$\,, Eq.(\ref{eqn-num-sc-struct}), and the spectrum of the
auxiliary fermions $\rho(\omega)= [\rho(\omega)^+ + \rho(\omega)^-]$\,,
Eq.(\ref{eqn-num-sc-rho})\,. Furthermore, the static structure factor
(equal-time correlation function) is obtained using
\begin{equation}  \label{eqn-num-sc-statstrct}
  S^{st}(\mathbf{q})= 
    \langle S^x_{\mathbf{q}}\,S^x_{-\mathbf{q}} \rangle
    = \frac{1}{\pi}\int_{-\infty}^\infty\!\!\mathrm{d}\omega\,
      S(\mathbf{q}, \omega)
\end{equation}

In order to save computation time, the wave-vector integral in
Eq.(\ref{eqn-num-sc-int}) is done analytically: $\mathcal{N}$ is
approximated by its value at the band edges,
$\mathcal{N}(\varepsilon)\simeq\frac{1}{4\pi J}
   \Theta(4J - |\varepsilon|)$\,.
That is, the log-singularity at $\varepsilon=0$ is ignored, which is
anyway suppressed by the factor $\varepsilon^2$ in $U$\,. The integral
is now easily done, with the result (not writing the
$\omega$-arguments)
\begin{eqnarray}
  U & = &  \nonumber
    \frac{S^0}{\pi4J|\Pi|^2}
    \left[8J - \frac{\Pi'}{|\Pi|^2}\ln
      \left|\frac{(M^+)^2 + (\Pi'')^2}{(M^-)^2 + (\Pi'')^2}\right|
      + \right.
    \\
  & &                  \label{eqn-num-sc-int2}
    \hspace*{ 2cm}\mbox{}+ \left.
    \left(1 - \frac{2(\Pi'')^2}{|\Pi|^2}\right) K\right]
    \\
  K & = &  \nonumber
    [\arctan(M^+/\Pi'') - \arctan(M^-/\Pi'')]/\Pi''
    \\
  M^\pm & = &  \nonumber
    \Pi' \pm 4J\,|\Pi|^2
\end{eqnarray}
\end{subequations}
At $|\omega|\to 0$\,, where $\Pi''(\omega)\propto\omega\to 0$\,, we use
$K(\omega)= [1/M^-(\omega) - 1/M^+(\omega)]$\,.

The derivation of Eqs.(\ref{eqn-num-scflex}) goes as follows. Using
the spectral representation of the fermion propagator $G(i\omega)$ in
Matsubara space,
\begin{displaymath}
  G(i\omega)= \int_{-\infty}^{\infty}\!\!\mathrm{d}\varepsilon\,
    \frac{\rho(\varepsilon)}{i\omega - \varepsilon}
\end{displaymath}
the fermion bubble (\ref{eqn-sc-pi}) becomes
\begin{displaymath}
  \Pi(i\nu)=
    \frac{1}{2}\int_{-\infty}^{\infty}\!\!\mathrm{d}\varepsilon\,
    \mathrm{d}\varepsilon'\,
    \rho(\varepsilon)\rho(\varepsilon')
    \frac{f(\varepsilon) - f(\varepsilon')}{i\nu - \varepsilon +
    \varepsilon'}
\end{displaymath}
At the real axis, $i\nu\to \omega + i0_+$ its imaginary part is
multiplied by $[1 + g(\omega)]$ to form the `bare' structure factor,
\begin{displaymath}
  S^0(\omega)= 
    -\frac{\pi}{2}\int_{-\infty}^{\infty}\!\!\mathrm{d}\varepsilon\,
    \rho(\varepsilon)\rho(\varepsilon - \omega)
    [f(\varepsilon) - f(\varepsilon - \omega)]
    [1 + g(\omega)]
\end{displaymath}
Using 
$[f(\varepsilon) - f(\varepsilon - \omega)][1 + g(\omega)]= 
   -[1 - f(\varepsilon)]f(\varepsilon - \omega)$\,,
Eq.(\ref{eqn-num-sc-snull}) immediately
follows. Eq.(\ref{eqn-num-sc-piim}) is just a result of the definition
of $S^0(\omega)$ and time-reversal symmetry,
$\chi''(-\mathbf{q}, -\omega)= -\chi''(\mathbf{q}, \omega)$\,,
$\Pi''(-\omega)= -\Pi''(\omega)$\,.
Similarly the imaginary part of the self energy (\ref{eqn-sc-self})
reads
\begin{displaymath}
  -\Sigma''(\omega)= 
    \frac{3}{4}\int_{-\infty}^{\infty}\!\!\mathrm{d}\varepsilon\,
    D''(\varepsilon)\rho(\varepsilon + \omega)
    [g(\varepsilon) + f(\varepsilon + \omega)]
\end{displaymath}
With $g + f= g(1 - f) + f(1 + g)$ and $D''(\varepsilon)g(\varepsilon)=
D''(-\varepsilon)[1 + g(-\varepsilon)]$\,, which follows from
$D''(-\varepsilon)= -D''(\varepsilon)$\,, Eq.(\ref{eqn-num-sc-selfim})
is obtained. The real parts $\Pi'$ and $\Sigma'$ are just
Kramers--Kronig transformations.

\subsection{Equations for Numerical Solution:
  Minimal Approximation (MSCA)}
\label{sec-app-red}
The set of equations representing the minimal self-consistent
approximation (\ref{eqn-scred}) is identical to
Eqs.(\ref{eqn-num-scflex}), except that the convolutions in $S^0$ and
$\Sigma''$ are missing, i.e., Eqs.(\ref{eqn-num-sc-snull}) and
(\ref{eqn-num-sc-selfim}) are to be replaced by
\begin{equation}  \label{eqn-num-scred}
\begin{array}{rcl}
  S^0(\omega) & = &  \displaystyle 
    \frac{\pi}{4}\rho^+(\omega)
    \\ \\
  -\Sigma''(\omega) & = &  \displaystyle 
    \frac{3}{8}[U(\omega) + U(-\omega)]
\end{array}
\end{equation}

\section{Local Spin Moment}
\label{sec-app-locmom}
The spin moment
$\langle(\mathbf{S}_i)^2\rangle=
   3\langle S^x_i S^x_i\rangle \equiv 3 S^{st}_{loc}$
is measured through the local static structure factor
\begin{eqnarray}
  S^{st}_{loc} & = &  \nonumber
    \frac{1}{N_L}\sum_{\mathbf{q}}S^{st}(\mathbf{q})
    \\
  & = &  \nonumber
    \int_{-\infty}^{\infty}\frac{\mathrm{d}\omega}{\pi}
    \frac{1}{N_L}\sum_{\mathbf{q}} S(\mathbf{q}, \omega)
    \\
  & = &  \label{eqn-num-locmom}
    \int_{-\infty}^{\infty}\frac{\mathrm{d}\omega}{\pi}
    S^0(\omega)\,
    \int_{-4J}^{4J}\mathrm{d}\varepsilon\,
    \frac{\mathcal{N}(\varepsilon)}
         {|1 + \varepsilon\,\Pi(\omega)|^2}
\end{eqnarray}
using Eq.(\ref{eqn-num-sc-struct}) and the density of states for 2D,
introduced in App.~\ref{sec-app-flex}\,. The latter can be written
\begin{displaymath}
  \mathcal{N}(\varepsilon)= 
    \frac{1}{2\pi^2 J}K(m)\Theta(4J - |\varepsilon|)
    \;\;,\;\;\;
  m= 1 - (\varepsilon/4J)^2
\end{displaymath}
with the complete elliptic integral
$K(m)= 
   \int_0^1\mathrm{d}t\,
   [(1 - t^2)(1 - m t^2)]^{-1/2}$\,.
In contrast to the calculation of $U(\omega)$ in
App.~\ref{sec-app-flex} above, it is not suitable to approximate
$\mathcal{N}(\varepsilon)$ by a constant, and $S^{st}_{loc}$ is
computed numerically from Eq.(\ref{eqn-num-locmom})\,.

\section{Derivation of Eqs.(\protect\ref{eqn-redan}) For The MSCA}
\label{sec-app-redan}
The equations (\ref{eqn-redan}) for the analytical treatment of the
minimal approximation in Sect.\ \ref{sec-red-anl} is derived from
Eqs.(\ref{eqn-scred}) as follows: For wave vectors $k\ll 1$\,,
$\mathbf{k}= \mathbf{q} - \mathbf{Q}$\,, we have
$J(\mathbf{q})\simeq -2dJ + Jk^2$\,,
and the susceptibility (\ref{eqn-red-chi}) can be written
\begin{displaymath}
  \chi= 
    \frac{1}{J}\left[
    \left(\frac{1}{J\Pi(0)} - 2d\right) + k^2 + 
    \left(\frac{1}{J\Pi(\omega)} - \frac{1}{J\Pi(0)}\right)
      \right]^{-1}
\end{displaymath}
Introducing the correlation length $\xi$ as in Eq.(\ref{eqn-xidef}),
representing the static and $\widetilde{\Pi}(\omega)$ containing the
dynamical behavior of $\chi$\,,
\begin{displaymath}
  \xi^{-2}= \left(\frac{1}{J\Pi(0)} - 2d\right)
    \;\;,\;\;\;
  \widetilde{\Pi}(\omega)= \Pi(\omega) - \Pi(0)
\end{displaymath}
it follows Eq.(\ref{eqn-redan-chi}), with
\begin{displaymath}
  R(\omega)= -\frac{\xi^2}{J}
    \left(\frac{1}{\Pi(\omega)} - \frac{1}{\Pi(0)}\right)
    = 2d\xi^2\frac{2dJ\widetilde{\Pi}(\omega)}
                  {1 + 2dJ\widetilde{\Pi}(\omega)}
\end{displaymath}
Here $\Pi(0)= 1/2dJ$ has been inserted, omitting terms of
$\mathcal{O}(\xi^{-2})$\,. For dispersing peaks (i.e., propagating
modes) to be able to appear in $\textrm{Im}\chi$ for $\xi k> 1$ we may
assume that $|R(\omega)|\sim\xi^2k^2$ and therefore
$|\widetilde{\Pi}(\omega)|\sim k^2\ll 1$\,. In $d=2$ this leads to
Eq.(\ref{eqn-redan-r})\,.

The fermion propagator (\ref{eqn-red-gf}) reads
\begin{equation}  \label{eqn-gfrho}
  G(i\omega)= \int_{-\infty}^\infty\!\!\mathrm{d}\varepsilon\,
    \frac{\rho(\varepsilon)}{i\omega - \varepsilon}
\end{equation}
its spectral function $\rho(\varepsilon)$ has been defined in
Eq.(\ref{eqn-redan-rho})\,. Now the irreducible bubble
(\ref{eqn-red-pi}) turns into
\begin{displaymath}
  \Pi(i\nu)=
    \frac{1}{4}\int_{-\infty}^\infty\!\!\mathrm{d}\varepsilon\,
    \rho(\varepsilon)\frac{\tanh(\varepsilon/2T)}
                          {\varepsilon - i\nu}
\end{displaymath}
At $i\nu\to \omega + i0_+$\,, the imaginary part becomes
$\text{Im}\Pi(\omega)= \text{Im}\widetilde{\Pi}(\omega)= 
   \frac{\pi}{4}\tanh(\omega/2T)\rho(\omega)$\,,
which forms the imaginary part of Eq.(\ref{eqn-redan-pi})\,. The
expression for $\Pi(0)$ given in Eq.(\ref{eqn-redan-pinul}) follows
immediately with $\nu=0$\,, if the particle--hole symmetry
$\rho(\varepsilon)= \rho(-\varepsilon)$ is exploited. The real part of
Eq.(\ref{eqn-redan-pi}) comes from
$\textrm{Re}\widetilde{\Pi}(\omega)= 
   \textrm{Re}\Pi(\omega)- \Pi(0)$\,,
\begin{displaymath}
  \textrm{Re}\widetilde{\Pi}(\omega)= 
    \frac{1}{4}\int_{-\infty}^\infty\!\!\mathrm{d}\varepsilon\,
    \tanh(\varepsilon/2T)\rho(\varepsilon)
    \left[\frac{1}{\varepsilon - \omega} - \frac{1}{\varepsilon}
    \right]
\end{displaymath}
using again $\rho(\varepsilon)= \rho(-\varepsilon)$\,.

The renormalized interaction (i.e., the local susceptibility) is
also written in spectral representation,
\begin{equation}  \label{eqn-drho}
  D(i\nu)= \frac{1}{\pi}\int_{-\infty}^\infty\!\!\mathrm{d}\varepsilon\,
    \frac{\textrm{Im}D(\varepsilon)}{\varepsilon - i\nu}
\end{equation}
Inserting this into the self energy (\ref{eqn-red-self}) and
performing the Matsubara sum results in
Eq.(\ref{eqn-redan-self})\,.

Finally, $D$ as defined in Eq.(\ref{eqn-red-d}) can be written
\begin{displaymath}
  D(\omega)= 
    \int\!\!\!\int\frac{\mathrm{d}^2q}{4\pi^2}
    J(\mathbf{q})^2\frac{\xi^2}{J}
    \frac{1}{1 + \xi^2|\mathbf{q} - \mathbf{Q}|^2 - R(\omega)}
\end{displaymath}
Here only the contribution from wave vectors $\mathbf{q}$ close to the
N\'{e}el vector $\mathbf{Q}= (\pi,\pi)$ is reproduced accurately. It
is then consistent to let
$J(\mathbf{q})^2\simeq J(\mathbf{Q})^2= (2dJ)^2$\,,
which leads to Eq.(\ref{eqn-redan-int}) if $d=2$\,.

\section{Derivation of the MSCA from the Equations Of Motion}
\label{sec-app-emotion}
We follow the line of Ref.\ \onlinecite{marschw59} and consider the
general Hamiltonian $H= H_0 + H_1$ for interacting fermions,
\begin{equation}  \label{eqn-em-ham}
  H_0= 
    \sum_r \varepsilon_r c^\dagger_r c_r  \;\;,\;\;\;
  H_1=
    \frac{1}{2}\sum_{r,r',s,s'}
    V^{r'r}_{s's} c^\dagger_{r'} c^\dagger_{s'} c_s c_r
\end{equation}
$r$ collects all single-particle quantum numbers; for the Heisenberg
model Eq.(\ref{eqn-ham}) it is $r\equiv (i,\alpha)$\,, and the
interaction matrix element reads
$V^{i\alpha'\alpha}_{j\beta'\beta}=
    J_{ij}\frac{1}{4}\bm{\sigma}^{\alpha'\alpha}
    \bm{\sigma}^{\beta'\beta}$\,.
The commutator of a fermion operator with the Hamiltonian is given by
\begin{equation}  \label{eqn-em-comm}
  [ c_r\,, H_0 ]=
    \varepsilon_r\,c_r  \;\;,\;\;\;
  [ c_r\,, H_1 ]=
    \sum_{r_1,r_2,r'_2}V^{r r_1}_{r'_2 r_2}
    c^\dagger_{r'_2} c_{r_2} c_{r_1}
\end{equation}
The propagator for $n$ fermions is defined as
\begin{displaymath}
  G_n(1 2 \ldots n; 1' 2' \ldots n')=
    (-1)^n\langle\mathcal{T}_\tau
    c_1 c_2 \ldots c_n c^\dagger_{n'} \ldots c^\dagger_{1'}
    \rangle
\end{displaymath}
with the short-hand notation
$c_1\equiv c_{r_1}(\tau_1)\,,
 c^\dagger_{1'}\equiv c^\dagger_{r'_1}(\tau'_1)$ etc.
The equations of motion for the 1- and 2-particle Green's function are
given by (see, e.g., Ref.\ \onlinecite{rick}),
\begin{displaymath}
  (\partial_{\tau_1} + \varepsilon_{r_1})\,
  G(1;1') =
    \langle\mathcal{T}_\tau
    \,[ c_{r_1}, H_1 ](\tau_1)\, c^\dagger_{1'} \rangle
    - \delta(1,1')
\end{displaymath}
\begin{eqnarray*}
  (\partial_{\tau_1} + \varepsilon_{r_1})\,
  G_2(12;1'2') & = &
    - \langle\mathcal{T}_\tau
    \,[ c_{r_1}, H_1 ](\tau_1)\, 
    c_2 c^\dagger_{2'} c^\dagger_{1'} \rangle
    \\
  & & \mbox{}\hspace*{-2cm}
    - \delta(1,1')\,G(2;2') + \delta(1,2')\,G(2;1')
\end{eqnarray*}
with
$\delta(1,1')\equiv\delta(\tau_1 - \tau'_1)
   \delta_{r_1, r'_1}$\,.
Inserting the commutator (\ref{eqn-em-comm}) and applying
$\int\mathrm{d}1\,\widetilde{G}(3;1)$
to both sides of the resulting equations, we obtain after re-labeling
$1\to\bar{1}$\,, $3\to 1$\,,
\begin{eqnarray}
  \lefteqn{%
  G(1;1') =  \label{eqn-em-1gf}
    \widetilde{G}(1;1') +}
    \\
  & & \mbox{}  \nonumber
    + \int\mathrm{d}\bar{1}'\mathrm{d}\bar{1}
          \mathrm{d}\bar{2}'\mathrm{d}\bar{2}\,
    \widetilde{G}(1;\bar{1}')\,V(\bar{1}'\bar{1};\bar{2}'\bar{2})\,
    G_2(\bar{1}_{-}\bar{2};1'\bar{2}'_+)
\end{eqnarray}
Here $l_\pm\equiv (r_l, \tau_l\pm 0_+)$\,,
$\int\mathrm{d}l\equiv\int_0^\beta\mathrm{d}\tau_l\sum_{r_l}$\,,
$V(1'1;2'2)\equiv
  V^{r'_1 r_1}_{r'_2 r_2}\delta(\tau'_1 - \tau_1)
  \delta(\tau'_2 - \tau_2)\delta(\tau_1 - \tau_2)$\,,
and $\widetilde{G}$ denotes
the bare ($V=0$) Green's function, which fulfills
\begin{displaymath}
  (\partial_{\tau_1} + \varepsilon_{r_1})\,
  \widetilde{G}(1;1')= - \delta(1,1')
\end{displaymath}
Similarly,
\begin{eqnarray}
  \lefteqn{%
  G_2(12;1'2') =  \label{eqn-em-2gf}
    \widetilde{G}(1;1')\,G(2;2') -
    \widetilde{G}(1;2')\,G(2;1') +}
    \\
  & & \mbox{}  \nonumber
    + \int\mathrm{d}\bar{1}'\mathrm{d}\bar{1}
          \mathrm{d}\bar{2}'\mathrm{d}\bar{2}\,
    \widetilde{G}(1;\bar{1}')\,V(\bar{1}'\bar{1};\bar{2}'\bar{2})\,
    G_3(\bar{1}_{-}2\bar{2};1'2'\bar{2}'_+)
\end{eqnarray}

The Equations (\ref{eqn-em-1gf}) and (\ref{eqn-em-2gf}) are
exact. Self-consistent approximations for the 1-particle Green's
function $G$ result from a decomposition of the 3-particle function
$G_3$ in the r.h.s.\ of Eq.(\ref{eqn-em-2gf}) into lower order
propagators. For that we note the generating functional for the
$n$-particle Green's function\cite{negorl},
\begin{displaymath}
\begin{array}{c}  \displaystyle 
  Z= \int\mathcal{D}[c, \bar{c}]\,\mathrm{e}^{-A}
    \\ \\  \displaystyle 
  A= \int\mathrm{d}1\,\bar{c}_1
    (\partial_{\tau_1})c_1
    + \int_0^\beta\!\!\mathrm{d}\tau\,H(\tau)
    + \int\mathrm{d}1\,(\bar{c}_1\eta_1 + \bar{\eta}_1c_1)
    \\ \\  \displaystyle
  G_n(1\ldots n;1'\ldots n')= 
    \left.\frac{1}{Z}
    \frac{\delta^{2n}\,Z}
         {\delta\bar{\eta}_1\delta{\eta}_{1'}\ldots
          \delta\bar{\eta}_n\delta{\eta}_{n'}}
    \right|_{\eta=0}
    \\ \\
\end{array}
\end{displaymath}
The corresponding connected Green's function is defined
as\cite{negorl}
\begin{equation}  \label{eqn-em-congf}
  G^C_n(1\ldots n;1'\ldots n')= 
    \left.
    \frac{\delta^{2n}\,\ln(Z)}
         {\delta\bar{\eta}_1\delta{\eta}_{1'}\ldots
          \delta\bar{\eta}_n\delta{\eta}_{n'}}
    \right|_{\eta=0}
\end{equation}
Applying Eq.(\ref{eqn-em-congf}) for $n= 1, 2$ leads to the familiar
results
$G^C(1;1')= G(1;1')$ and
\begin{eqnarray*}
  \lefteqn{%
  G^C_2(12;1'2')= G_2(12;1'2') -}
    \\
  & & \mbox{}
    - \left\{G(1;1')\,G(2;2') - G(1;2')\,G(2;1')\right\}
\end{eqnarray*}
whereas the expression for $n=3$ reads
\begin{eqnarray}
  \lefteqn{%
  G^C_3(123;1'2'3')= G_3(123;1'2'3') -}  \label{eqn-em-cumul}
    \\[1ex]  \nonumber
  & & \mbox{}-
  \{
    \underbrace{G(1;1')\,G_2(23;2'3')}_{\textrm{(rpa)}}\,
    \underbrace{-G(1;2')\,G_2(23;1'3')}_{\textrm{(rpaex)}} +
    \\  \nonumber
  & & \mbox{}+
    G(1;3')\,G_2(23;1'2') + 
    \,\textrm{all other contractions}\} +
    \\[1ex]  \nonumber
  & & \mbox{}+
  2\{
    G(1;1')\,G(2;2')\,G(3;3') - 
    \\ \nonumber
  & & \mbox{}- 
    G(1;1')\,G(2;3')\,G(3;2') + 
    \,\textrm{all other contractions}\}
\end{eqnarray}
The l.h.s.\ of Eq.(\ref{eqn-em-cumul}) represents effective 3-particle
interactions and is ignored\cite{marschw59,kadmar61}\,. From the
r.h.s.\ we keep only $G_3$ and the terms (rpa) and (rpaex), and
Eq.(\ref{eqn-em-2gf}) turns into a Bethe--Salpeter equation,
\begin{eqnarray}  \nonumber
  \lefteqn{%
  G_2(12;1'2')=
    \widetilde{G}(1;1')\,G(2;2') - 
    \widetilde{G}(1;2')\,G(2;1') + }
    \\
  & & \mbox{}+  \label{eqn-em-ladder}
    \int\mathrm{d}\bar{1}'\mathrm{d}\bar{1}
        \mathrm{d}\bar{2}'\mathrm{d}\bar{2}\,
    \widetilde{G}(1;\bar{1}')\,V(\bar{1}'\bar{1};\bar{2}'\bar{2})\,
    \times
    \\
  & & \mbox{}\times \nonumber
    \{ G(\bar{1}_-;1')\,G_2(2\bar{2};2'\bar{2}'_+) - 
       G(\bar{1}_-;2')\,G_2(2\bar{2};1'\bar{2}'_+) \}
\end{eqnarray}
This equation can now be iterated, leading to an expansion of $G_2$ in
powers of $V$ through the series of particle--hole bubble diagrams
shown in Fig.\ \ref{fig-em-bubble}\,.

\def\myscale{0.4}
\begin{figure} \large
  \begin{eqnarray*}
    \lefteqn{%
    \mygraph{scale=\myscale}{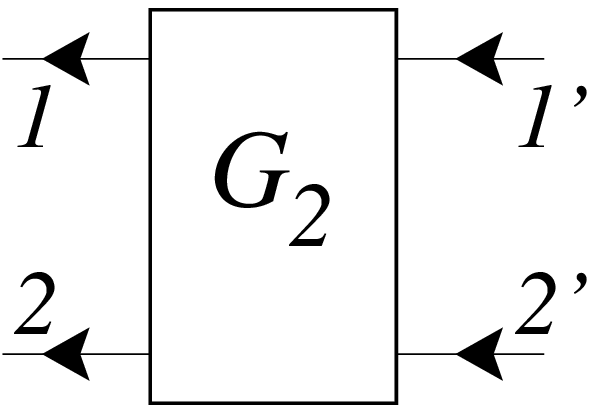}
          = \mygraph{scale=\myscale}{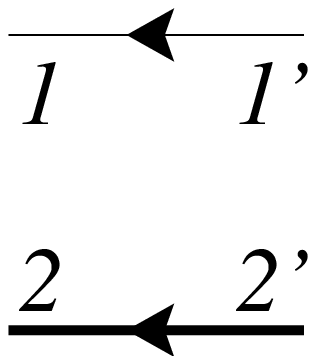}
          + \mygraph{scale=\myscale}{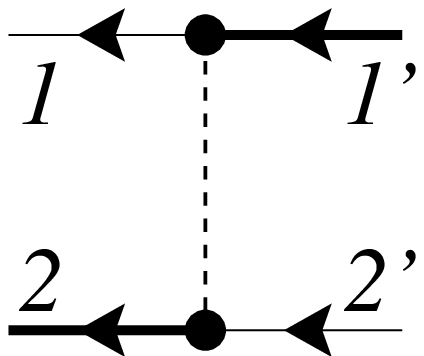} +}
        \\
      & & \mbox{}
          + \mygraph{scale=\myscale}{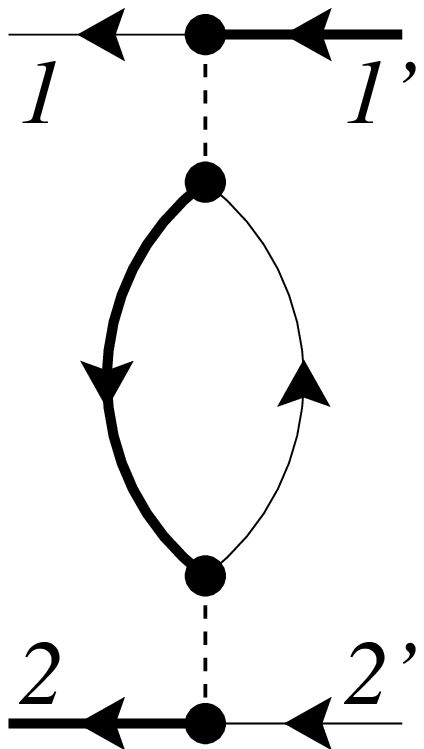}
          + \mygraph{scale=\myscale}{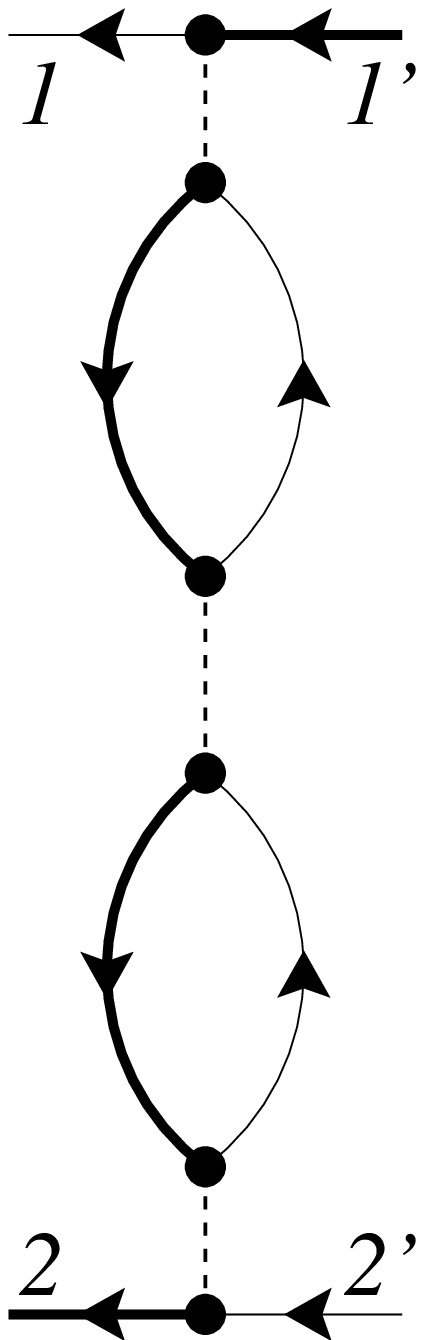} +
        \\
      & & \mbox{}
          + \;\ldots\;
          - \textrm{\normalsize exchange diagrams}
  \end{eqnarray*}
  \caption[\ ]{\label{fig-em-bubble}%
    Diagrammatic representation of Eq.(\ref{eqn-em-ladder}), which has
    been iterated. All diagrams corresponding to vertex corrections
    have been omitted, see text. Thin and thick continuous lines
    denote Green's functions $\widetilde{G}$ and $G$\,,
    respectively. The dashed line is $-V$\,, fermion
    loops contain a $-$ sign. The exchange diagrams, obtained via
    $1'\leftrightarrow 2'$\,, are not shown.
    }
\end{figure}
Note that the bubbles contain one bare $\widetilde{G}$ (thin line) and
one full Green's function $G$ (full line), which is a direct
consequence of the combination $\widetilde{G}\,G$ appearing in the
first line of Eq.(\ref{eqn-em-2gf}) and accordingly in
(\ref{eqn-em-ladder})\,. The iteration of Eq.(\ref{eqn-em-ladder})
additionally produces diagrams which either contribute to vertex
corrections to the simple bubbles, or form self-energy insertions of
Hartree type.  The former have been omitted in Fig.\
\ref{fig-em-bubble}\,. The latter contain loops of the form
$\int\mathrm{d}\bar{1}'\mathrm{d}\bar{1}\,
   V(2'2;\bar{1}'\bar{1})\,g(\bar{1};\bar{1}'_+)
   \propto\bm{\sigma}\textrm{Tr}[\bm{\sigma}]= 0$\,,
$g= \widetilde{G}, G$\,,
which are zero for the Heisenberg model without magnetic order.  For the
same reason the other terms (except (rpa) and (rpaex)) in
Eq.(\ref{eqn-em-cumul}) are not considered.

Finally, the $G_2$ shown in Fig.\ \ref{fig-em-bubble} is inserted in
Eq.(\ref{eqn-em-1gf}), omitting again all vertex-correction diagrams
that emerge from the contraction of $G_2$ in
(\ref{eqn-em-1gf})\,. Eq.(\ref{eqn-em-1gf}) now takes the form of a
Dyson's equation,
\begin{eqnarray}
  \lefteqn{%
    G(1;1') =  \label{eqn-em-fgf}
    \widetilde{G}(1;1') +}
    \\
  & &  \nonumber
    \mbox{}+
    \int\mathrm{d}\bar{1}'\mathrm{d}\bar{1}\,
    \widetilde{G}(1;\bar{1}')\,\Sigma(\bar{1}';\bar{1})\,
    G(\bar{1};1')
\end{eqnarray}
where the self-energy $\Sigma$ is identified as the series of diagrams
obtained from
\begin{subequations}  \label{eqn-em-self}
\begin{equation}
  \Sigma(1';1) =
    \int\mathrm{d}\bar{1}'\mathrm{d}\bar{1}\,
    \widehat{V}(1'\bar{1};\bar{1}'1)\,
    \widetilde{G}(\bar{1};\bar{1}')
\end{equation}
with the renormalized interaction
\begin{eqnarray}
  \lefteqn{%
    \widehat{V}(1'1;2'2) =  \label{eqn-em-effint}
    V(1'1;2'2) +}
    \\
  & &  \nonumber
    \mbox{}+
    \int\mathrm{d}\bar{1}'\mathrm{d}\bar{1}
        \mathrm{d}\bar{2}'\mathrm{d}\bar{2}\,
    V(1'1;\bar{1}'\bar{1})\,
    G(\bar{1};\bar{2}')\,\widetilde{G}(\bar{2};\bar{1}')\,
    \widehat{V}(\bar{2}'\bar{2};2'2)
\end{eqnarray}
\end{subequations}
Eqs.(\ref{eqn-em-fgf}) and (\ref{eqn-em-self}) represent exactly the
minimal self-consistent approximation given in Fig.\
\ref{fig-red}\,.


\begin{thebibliography}{73}
\expandafter\ifx\csname natexlab\endcsname\relax\def\natexlab#1{#1}\fi
\expandafter\ifx\csname bibnamefont\endcsname\relax
  \def\bibnamefont#1{#1}\fi
\expandafter\ifx\csname bibfnamefont\endcsname\relax
  \def\bibfnamefont#1{#1}\fi
\expandafter\ifx\csname citenamefont\endcsname\relax
  \def\citenamefont#1{#1}\fi
\expandafter\ifx\csname url\endcsname\relax
  \def\url#1{\texttt{#1}}\fi
\expandafter\ifx\csname urlprefix\endcsname\relax\def\urlprefix{URL }\fi
\providecommand{\bibinfo}[2]{#2}
\providecommand{\eprint}[2][]{\url{#2}}

\bibitem[{not({\natexlab{a}})}]{note-manous}
\bibinfo{note}{For review and references see Ref.~\onlinecite{man91}.}

\bibitem[{\citenamefont{Beard et~al.}(1998)\citenamefont{Beard, Birgeneau,
  Greven, and Wiese}}]{bea98}
\bibinfo{author}{\bibfnamefont{B.~B.} \bibnamefont{Beard}},
  \bibinfo{author}{\bibfnamefont{R.~J.} \bibnamefont{Birgeneau}},
  \bibinfo{author}{\bibfnamefont{M.}~\bibnamefont{Greven}}, \bibnamefont{and}
  \bibinfo{author}{\bibfnamefont{U.-J.} \bibnamefont{Wiese}},
  \bibinfo{journal}{Phys.~Rev.~Lett.~} \textbf{\bibinfo{volume}{80}},
  \bibinfo{pages}{1742} (\bibinfo{year}{1998}).

\bibitem[{\citenamefont{Kim and Troyer}(1998)}]{kimtro98}
\bibinfo{author}{\bibfnamefont{J.-K.} \bibnamefont{Kim}} \bibnamefont{and}
  \bibinfo{author}{\bibfnamefont{M.}~\bibnamefont{Troyer}},
  \bibinfo{journal}{Phys.~Rev.~Lett.~} \textbf{\bibinfo{volume}{80}},
  \bibinfo{pages}{2705} (\bibinfo{year}{1998}).

\bibitem[{\citenamefont{Makivi\'{c} and Ding}(1991)}]{makdin91}
\bibinfo{author}{\bibfnamefont{M.~S.} \bibnamefont{Makivi\'{c}}}
  \bibnamefont{and} \bibinfo{author}{\bibfnamefont{H.-Q.} \bibnamefont{Ding}},
  \bibinfo{journal}{Phys.~Rev.~B} \textbf{\bibinfo{volume}{43}},
  \bibinfo{pages}{3562} (\bibinfo{year}{1991}).

\bibitem[{\citenamefont{Elstner et~al.}(1995)\citenamefont{Elstner, Sokol,
  Singh, Greven, and Birgeneau}}]{elst95}
\bibinfo{author}{\bibfnamefont{N.}~\bibnamefont{Elstner}},
  \bibinfo{author}{\bibfnamefont{A.}~\bibnamefont{Sokol}},
  \bibinfo{author}{\bibfnamefont{R.~R.~P.} \bibnamefont{Singh}},
  \bibinfo{author}{\bibfnamefont{M.}~\bibnamefont{Greven}}, \bibnamefont{and}
  \bibinfo{author}{\bibfnamefont{R.~J.} \bibnamefont{Birgeneau}},
  \bibinfo{journal}{Phys.~Rev.~Lett.~} \textbf{\bibinfo{volume}{75}},
  \bibinfo{pages}{938} (\bibinfo{year}{1995}).

\bibitem[{\citenamefont{Arovas and Auerbach}(1988)}]{aaa88}
\bibinfo{author}{\bibfnamefont{D.~P.} \bibnamefont{Arovas}} \bibnamefont{and}
  \bibinfo{author}{\bibfnamefont{A.}~\bibnamefont{Auerbach}},
  \bibinfo{journal}{Phys.~Rev.~B} \textbf{\bibinfo{volume}{38}},
  \bibinfo{pages}{316} (\bibinfo{year}{1988}).

\bibitem[{\citenamefont{Takahashi}(1987)}]{tak87}
\bibinfo{author}{\bibfnamefont{M.}~\bibnamefont{Takahashi}},
  \bibinfo{journal}{Phys.~Rev.~B} \textbf{\bibinfo{volume}{36}},
  \bibinfo{pages}{3791} (\bibinfo{year}{1987}).

\bibitem[{\citenamefont{Nelson and Pelkovits}(1977)}]{nelpel77}
\bibinfo{author}{\bibfnamefont{D.~R.} \bibnamefont{Nelson}} \bibnamefont{and}
  \bibinfo{author}{\bibfnamefont{R.~A.} \bibnamefont{Pelkovits}},
  \bibinfo{journal}{Phys.~Rev.~B} \textbf{\bibinfo{volume}{16}},
  \bibinfo{pages}{2191} (\bibinfo{year}{1977}).

\bibitem[{ami()}]{amitremark}
\bibinfo{note}{See, e.g., Ref.\ \onlinecite{amitbook} and references therein.}

\bibitem[{\citenamefont{Chakravarty et~al.}(1988)\citenamefont{Chakravarty,
  Halperin, and Nelson}}]{chn88}
\bibinfo{author}{\bibfnamefont{S.}~\bibnamefont{Chakravarty}},
  \bibinfo{author}{\bibfnamefont{B.~I.} \bibnamefont{Halperin}},
  \bibnamefont{and} \bibinfo{author}{\bibfnamefont{D.~R.}
  \bibnamefont{Nelson}}, \bibinfo{journal}{Phys.~Rev.~Lett.~}
  \textbf{\bibinfo{volume}{60}}, \bibinfo{pages}{1057} (\bibinfo{year}{1988}).

\bibitem[{\citenamefont{Chakravarty et~al.}(1989)\citenamefont{Chakravarty,
  Halperin, and Nelson}}]{chn89}
\bibinfo{author}{\bibfnamefont{S.}~\bibnamefont{Chakravarty}},
  \bibinfo{author}{\bibfnamefont{B.~I.} \bibnamefont{Halperin}},
  \bibnamefont{and} \bibinfo{author}{\bibfnamefont{D.~R.}
  \bibnamefont{Nelson}}, \bibinfo{journal}{Phys.~Rev.~B}
  \textbf{\bibinfo{volume}{39}}, \bibinfo{pages}{2344} (\bibinfo{year}{1989}).

\bibitem[{\citenamefont{Hasenfratz and Niedermayer}(1993)}]{hn93}
\bibinfo{author}{\bibfnamefont{P.}~\bibnamefont{Hasenfratz}} \bibnamefont{and}
  \bibinfo{author}{\bibfnamefont{F.}~\bibnamefont{Niedermayer}},
  \bibinfo{journal}{Z.~Phys.~B} \textbf{\bibinfo{volume}{92}},
  \bibinfo{pages}{91} (\bibinfo{year}{1993}).

\bibitem[{\citenamefont{Kim et~al.}(2001{\natexlab{a}})\citenamefont{Kim,
  Birgeneau, Chou, Greven, Kastner, Lee, Wells, Aharony, Entin-Wohlmann,
  Korenblit et~al.}}]{kim01a}
\bibinfo{author}{\bibfnamefont{Y.~J.} \bibnamefont{Kim}},
  \bibinfo{author}{\bibfnamefont{R.~J.} \bibnamefont{Birgeneau}},
  \bibinfo{author}{\bibfnamefont{F.~C.} \bibnamefont{Chou}},
  \bibinfo{author}{\bibfnamefont{M.}~\bibnamefont{Greven}},
  \bibinfo{author}{\bibfnamefont{M.~A.} \bibnamefont{Kastner}},
  \bibinfo{author}{\bibfnamefont{Y.~S.} \bibnamefont{Lee}},
  \bibinfo{author}{\bibfnamefont{B.~O.} \bibnamefont{Wells}},
  \bibinfo{author}{\bibfnamefont{A.}~\bibnamefont{Aharony}},
  \bibinfo{author}{\bibfnamefont{O.}~\bibnamefont{Entin-Wohlmann}},
  \bibinfo{author}{\bibfnamefont{I.~Y.} \bibnamefont{Korenblit}},
  \bibnamefont{et~al.}, \bibinfo{journal}{Phys.~Rev.~B}
  \textbf{\bibinfo{volume}{64}}, \bibinfo{pages}{024435}
  (\bibinfo{year}{2001}{\natexlab{a}}).

\bibitem[{\citenamefont{Greven et~al.}(1994)\citenamefont{Greven, Birgeneau,
  Endoh, Kastner, Keimer, Matsuda, Shirane, and Thurston}}]{gre94}
\bibinfo{author}{\bibfnamefont{M.}~\bibnamefont{Greven}},
  \bibinfo{author}{\bibfnamefont{R.~J.} \bibnamefont{Birgeneau}},
  \bibinfo{author}{\bibfnamefont{Y.}~\bibnamefont{Endoh}},
  \bibinfo{author}{\bibfnamefont{M.~A.} \bibnamefont{Kastner}},
  \bibinfo{author}{\bibfnamefont{B.}~\bibnamefont{Keimer}},
  \bibinfo{author}{\bibfnamefont{M.}~\bibnamefont{Matsuda}},
  \bibinfo{author}{\bibfnamefont{G.}~\bibnamefont{Shirane}}, \bibnamefont{and}
  \bibinfo{author}{\bibfnamefont{T.~R.} \bibnamefont{Thurston}},
  \bibinfo{journal}{Phys.~Rev.~Lett.~} \textbf{\bibinfo{volume}{72}},
  \bibinfo{pages}{1096} (\bibinfo{year}{1994}).

\bibitem[{\citenamefont{Birgeneau et~al.}(1999)\citenamefont{Birgeneau, Greven,
  Kastner, Lee, Wells, Endoh, Yamada, and Shirane}}]{bir99}
\bibinfo{author}{\bibfnamefont{R.~J.} \bibnamefont{Birgeneau}},
  \bibinfo{author}{\bibfnamefont{M.}~\bibnamefont{Greven}},
  \bibinfo{author}{\bibfnamefont{M.~A.} \bibnamefont{Kastner}},
  \bibinfo{author}{\bibfnamefont{Y.~S.} \bibnamefont{Lee}},
  \bibinfo{author}{\bibfnamefont{B.~O.} \bibnamefont{Wells}},
  \bibinfo{author}{\bibfnamefont{Y.}~\bibnamefont{Endoh}},
  \bibinfo{author}{\bibfnamefont{K.}~\bibnamefont{Yamada}}, \bibnamefont{and}
  \bibinfo{author}{\bibfnamefont{G.}~\bibnamefont{Shirane}},
  \bibinfo{journal}{Phys.~Rev.~B} \textbf{\bibinfo{volume}{59}},
  \bibinfo{pages}{13788} (\bibinfo{year}{1999}).

\bibitem[{\citenamefont{Keimer et~al.}(1992)\citenamefont{Keimer, Belk,
  Birgeneau, Cassaho, Chen, Greven, Kastner, Aharony, Endoh, Erwin
  et~al.}}]{keim92}
\bibinfo{author}{\bibfnamefont{B.}~\bibnamefont{Keimer}},
  \bibinfo{author}{\bibfnamefont{N.}~\bibnamefont{Belk}},
  \bibinfo{author}{\bibfnamefont{R.~J.} \bibnamefont{Birgeneau}},
  \bibinfo{author}{\bibfnamefont{A.}~\bibnamefont{Cassaho}},
  \bibinfo{author}{\bibfnamefont{C.~Y.} \bibnamefont{Chen}},
  \bibinfo{author}{\bibfnamefont{M.}~\bibnamefont{Greven}},
  \bibinfo{author}{\bibfnamefont{M.~A.} \bibnamefont{Kastner}},
  \bibinfo{author}{\bibfnamefont{A.}~\bibnamefont{Aharony}},
  \bibinfo{author}{\bibfnamefont{Y.}~\bibnamefont{Endoh}},
  \bibinfo{author}{\bibfnamefont{R.~W.} \bibnamefont{Erwin}},
  \bibnamefont{et~al.}, \bibinfo{journal}{Phys.~Rev.~B}
  \textbf{\bibinfo{volume}{46}}, \bibinfo{pages}{14034} (\bibinfo{year}{1992}).

\bibitem[{\citenamefont{Endoh et~al.}(1988)\citenamefont{Endoh, Yamada,
  Birgeneau, Gabbe, Jennsen, Kastner, Peters, Picone, Thurston, Tranquada
  et~al.}}]{end88}
\bibinfo{author}{\bibfnamefont{Y.}~\bibnamefont{Endoh}},
  \bibinfo{author}{\bibfnamefont{K.}~\bibnamefont{Yamada}},
  \bibinfo{author}{\bibfnamefont{R.~J.} \bibnamefont{Birgeneau}},
  \bibinfo{author}{\bibfnamefont{D.~R.} \bibnamefont{Gabbe}},
  \bibinfo{author}{\bibfnamefont{H.~P.} \bibnamefont{Jennsen}},
  \bibinfo{author}{\bibfnamefont{M.~A.} \bibnamefont{Kastner}},
  \bibinfo{author}{\bibfnamefont{C.~J.} \bibnamefont{Peters}},
  \bibinfo{author}{\bibfnamefont{P.~J.} \bibnamefont{Picone}},
  \bibinfo{author}{\bibfnamefont{T.~R.} \bibnamefont{Thurston}},
  \bibinfo{author}{\bibfnamefont{J.~M.} \bibnamefont{Tranquada}},
  \bibnamefont{et~al.}, \bibinfo{journal}{Phys.~Rev.~B}
  \textbf{\bibinfo{volume}{37}}, \bibinfo{pages}{7443} (\bibinfo{year}{1988}).

\bibitem[{\citenamefont{Makivi\'{c} and Jarrell}(1992)}]{makjar92}
\bibinfo{author}{\bibfnamefont{M.~S.} \bibnamefont{Makivi\'{c}}}
  \bibnamefont{and} \bibinfo{author}{\bibfnamefont{M.}~\bibnamefont{Jarrell}},
  \bibinfo{journal}{Phys.~Rev.~Lett.~} \textbf{\bibinfo{volume}{68}},
  \bibinfo{pages}{1770} (\bibinfo{year}{1992}).

\bibitem[{\citenamefont{Wysin and Bishop}(1990)}]{wysbis90}
\bibinfo{author}{\bibfnamefont{G.~M.} \bibnamefont{Wysin}} \bibnamefont{and}
  \bibinfo{author}{\bibfnamefont{A.~R.} \bibnamefont{Bishop}},
  \bibinfo{journal}{Phys.~Rev.~B} \textbf{\bibinfo{volume}{42}},
  \bibinfo{pages}{810} (\bibinfo{year}{1990}).

\bibitem[{\citenamefont{Auerbach and Arovas}(1988)}]{aaa88a}
\bibinfo{author}{\bibfnamefont{A.}~\bibnamefont{Auerbach}} \bibnamefont{and}
  \bibinfo{author}{\bibfnamefont{D.~P.} \bibnamefont{Arovas}},
  \bibinfo{journal}{Phys.~Rev.~Lett.~} \textbf{\bibinfo{volume}{61}},
  \bibinfo{pages}{617} (\bibinfo{year}{1988}).

\bibitem[{\citenamefont{Kopietz}(1990)}]{kop90}
\bibinfo{author}{\bibfnamefont{P.}~\bibnamefont{Kopietz}},
  \bibinfo{journal}{Phys.~Rev.~Lett.~} \textbf{\bibinfo{volume}{64}},
  \bibinfo{pages}{2587} (\bibinfo{year}{1990}).

\bibitem[{\citenamefont{Ty\v{c} et~al.}(1989)\citenamefont{Ty\v{c}, Halperin,
  and Chakravarty}}]{thc89}
\bibinfo{author}{\bibfnamefont{S.}~\bibnamefont{Ty\v{c}}},
  \bibinfo{author}{\bibfnamefont{B.~I.} \bibnamefont{Halperin}},
  \bibnamefont{and}
  \bibinfo{author}{\bibfnamefont{S.}~\bibnamefont{Chakravarty}},
  \bibinfo{journal}{Phys.~Rev.~Lett.~} \textbf{\bibinfo{volume}{62}},
  \bibinfo{pages}{835} (\bibinfo{year}{1989}).

\bibitem[{\citenamefont{Grempel}(1988)}]{grempel88}
\bibinfo{author}{\bibfnamefont{D.~R.} \bibnamefont{Grempel}},
  \bibinfo{journal}{Phys.~Rev.~Lett.~} \textbf{\bibinfo{volume}{61}},
  \bibinfo{pages}{1041} (\bibinfo{year}{1988}).

\bibitem[{\citenamefont{Yamada et~al.}(1989)\citenamefont{Yamada, Kakurai,
  Endoh, Thurston, Kastner, Birgeneau, Shirane, Hidaka, and Murakami}}]{yam89}
\bibinfo{author}{\bibfnamefont{K.}~\bibnamefont{Yamada}},
  \bibinfo{author}{\bibfnamefont{K.}~\bibnamefont{Kakurai}},
  \bibinfo{author}{\bibfnamefont{Y.}~\bibnamefont{Endoh}},
  \bibinfo{author}{\bibfnamefont{T.~R.} \bibnamefont{Thurston}},
  \bibinfo{author}{\bibfnamefont{M.~A.} \bibnamefont{Kastner}},
  \bibinfo{author}{\bibfnamefont{R.~J.} \bibnamefont{Birgeneau}},
  \bibinfo{author}{\bibfnamefont{G.}~\bibnamefont{Shirane}},
  \bibinfo{author}{\bibfnamefont{Y.}~\bibnamefont{Hidaka}}, \bibnamefont{and}
  \bibinfo{author}{\bibfnamefont{T.}~\bibnamefont{Murakami}},
  \bibinfo{journal}{Phys.~Rev.~B} \textbf{\bibinfo{volume}{40}},
  \bibinfo{pages}{4557} (\bibinfo{year}{1989}).

\bibitem[{\citenamefont{Hayden et~al.}(1990)\citenamefont{Hayden, Aeppli, Mook,
  Cheong, and Fisk}}]{hay90}
\bibinfo{author}{\bibfnamefont{S.~M.} \bibnamefont{Hayden}},
  \bibinfo{author}{\bibfnamefont{G.}~\bibnamefont{Aeppli}},
  \bibinfo{author}{\bibfnamefont{H.~A.} \bibnamefont{Mook}},
  \bibinfo{author}{\bibfnamefont{S.-W.} \bibnamefont{Cheong}},
  \bibnamefont{and} \bibinfo{author}{\bibfnamefont{Z.}~\bibnamefont{Fisk}},
  \bibinfo{journal}{Phys.~Rev.~B} \textbf{\bibinfo{volume}{42}},
  \bibinfo{pages}{10220} (\bibinfo{year}{1990}).

\bibitem[{\citenamefont{Kim et~al.}(2001{\natexlab{b}})\citenamefont{Kim,
  Birgeneau, Chou, Erwin, and Kastner}}]{kim01}
\bibinfo{author}{\bibfnamefont{Y.~J.} \bibnamefont{Kim}},
  \bibinfo{author}{\bibfnamefont{R.~J.} \bibnamefont{Birgeneau}},
  \bibinfo{author}{\bibfnamefont{F.~C.} \bibnamefont{Chou}},
  \bibinfo{author}{\bibfnamefont{R.~W.} \bibnamefont{Erwin}}, \bibnamefont{and}
  \bibinfo{author}{\bibfnamefont{M.~A.} \bibnamefont{Kastner}},
  \bibinfo{journal}{Phys.~Rev.~Lett.~} \textbf{\bibinfo{volume}{86}},
  \bibinfo{pages}{3144} (\bibinfo{year}{2001}{\natexlab{b}}).

\bibitem[{\citenamefont{Carretta et~al.}(1997)\citenamefont{Carretta,
  Rigamonti, and Sala}}]{carr97}
\bibinfo{author}{\bibfnamefont{P.}~\bibnamefont{Carretta}},
  \bibinfo{author}{\bibfnamefont{A.}~\bibnamefont{Rigamonti}},
  \bibnamefont{and} \bibinfo{author}{\bibfnamefont{R.}~\bibnamefont{Sala}},
  \bibinfo{journal}{Phys.~Rev.~B} \textbf{\bibinfo{volume}{55}},
  \bibinfo{pages}{3734} (\bibinfo{year}{1997}), \bibinfo{note}{and references
  therein}.

\bibitem[{\citenamefont{Abrikosov}(1965)}]{abr65}
\bibinfo{author}{\bibfnamefont{A.~A.} \bibnamefont{Abrikosov}},
  \bibinfo{journal}{Physics} \textbf{\bibinfo{volume}{2}}, \bibinfo{pages}{5}
  (\bibinfo{year}{1965}).

\bibitem[{\citenamefont{Affleck and Marston}(1988)}]{affmar88}
\bibinfo{author}{\bibfnamefont{I.}~\bibnamefont{Affleck}} \bibnamefont{and}
  \bibinfo{author}{\bibfnamefont{J.~B.} \bibnamefont{Marston}},
  \bibinfo{journal}{Phys.~Rev.~B} \textbf{\bibinfo{volume}{37}},
  \bibinfo{pages}{3774} (\bibinfo{year}{1988}).

\bibitem[{not({\natexlab{b}})}]{note-rvb}
\bibinfo{note}{See, e.g., Ref.~\onlinecite{philrev03} and references therein.}

\bibitem[{\citenamefont{Inui et~al.}(1988)\citenamefont{Inui, Doniach,
  Hirschfeld, and Ruckenstein}}]{inui88}
\bibinfo{author}{\bibfnamefont{M.}~\bibnamefont{Inui}},
  \bibinfo{author}{\bibfnamefont{S.}~\bibnamefont{Doniach}},
  \bibinfo{author}{\bibfnamefont{P.~J.} \bibnamefont{Hirschfeld}},
  \bibnamefont{and} \bibinfo{author}{\bibfnamefont{A.~E.}
  \bibnamefont{Ruckenstein}}, \bibinfo{journal}{Phys.~Rev.~B}
  \textbf{\bibinfo{volume}{37}}, \bibinfo{pages}{2320} (\bibinfo{year}{1988}).

\bibitem[{\citenamefont{Inaba et~al.}(1996)\citenamefont{Inaba, Matsukawa,
  Saitoh, and Fukuyama}}]{inaba96}
\bibinfo{author}{\bibfnamefont{M.}~\bibnamefont{Inaba}},
  \bibinfo{author}{\bibfnamefont{H.}~\bibnamefont{Matsukawa}},
  \bibinfo{author}{\bibfnamefont{M.}~\bibnamefont{Saitoh}}, \bibnamefont{and}
  \bibinfo{author}{\bibfnamefont{H.}~\bibnamefont{Fukuyama}},
  \bibinfo{journal}{Physica C} \textbf{\bibinfo{volume}{257}},
  \bibinfo{pages}{299} (\bibinfo{year}{1996}).

\bibitem[{\citenamefont{Bickers et~al.}(1989)\citenamefont{Bickers, Scalapino,
  and White}}]{bicscawhi89}
\bibinfo{author}{\bibfnamefont{N.~E.} \bibnamefont{Bickers}},
  \bibinfo{author}{\bibfnamefont{D.~J.} \bibnamefont{Scalapino}},
  \bibnamefont{and} \bibinfo{author}{\bibfnamefont{S.~R.} \bibnamefont{White}},
  \bibinfo{journal}{Phys.~Rev.~Lett.~} \textbf{\bibinfo{volume}{62}},
  \bibinfo{pages}{961} (\bibinfo{year}{1989}).

\bibitem[{\citenamefont{Kampf and Schrieffer}(1990)}]{kamsch90b}
\bibinfo{author}{\bibfnamefont{A.~P.} \bibnamefont{Kampf}} \bibnamefont{and}
  \bibinfo{author}{\bibfnamefont{J.~R.} \bibnamefont{Schrieffer}},
  \bibinfo{journal}{Phys.~Rev.~B} \textbf{\bibinfo{volume}{42}},
  \bibinfo{pages}{7967} (\bibinfo{year}{1990}).

\bibitem[{\citenamefont{Vilk and Tremblay}(1997)}]{vilktrem97}
\bibinfo{author}{\bibfnamefont{Y.~M.} \bibnamefont{Vilk}} \bibnamefont{and}
  \bibinfo{author}{\bibfnamefont{A.-M.~S.} \bibnamefont{Tremblay}},
  \bibinfo{journal}{J.~Phys.~I~France} \textbf{\bibinfo{volume}{7}},
  \bibinfo{pages}{1309} (\bibinfo{year}{1997}).

\bibitem[{\citenamefont{Kadanoff and Martin}(1961)}]{kadmar61}
\bibinfo{author}{\bibfnamefont{L.~P.} \bibnamefont{Kadanoff}} \bibnamefont{and}
  \bibinfo{author}{\bibfnamefont{P.~C.} \bibnamefont{Martin}},
  \bibinfo{journal}{Phys.~Rev.~} \textbf{\bibinfo{volume}{124}},
  \bibinfo{pages}{670} (\bibinfo{year}{1961}).

\bibitem[{\citenamefont{Patton}(1971)}]{patton71}
\bibinfo{author}{\bibfnamefont{B.~R.} \bibnamefont{Patton}},
  \bibinfo{journal}{Phys.~Rev.~Lett.~} \textbf{\bibinfo{volume}{27}},
  \bibinfo{pages}{1273} (\bibinfo{year}{1971}).

\bibitem[{\citenamefont{Jank{\'o} et~al.}(1997)\citenamefont{Jank{\'o}, Maly,
  and Levin}}]{levin97}
\bibinfo{author}{\bibfnamefont{B.}~\bibnamefont{Jank{\'o}}},
  \bibinfo{author}{\bibfnamefont{J.}~\bibnamefont{Maly}}, \bibnamefont{and}
  \bibinfo{author}{\bibfnamefont{K.}~\bibnamefont{Levin}},
  \bibinfo{journal}{Phys.~Rev.~B} \textbf{\bibinfo{volume}{56}},
  \bibinfo{pages}{R11407} (\bibinfo{year}{1997}).

\bibitem[{\citenamefont{Manousakis}(1991)}]{man91}
\bibinfo{author}{\bibfnamefont{E.}~\bibnamefont{Manousakis}},
  \bibinfo{journal}{Rev.~Mod.~Phys.~} \textbf{\bibinfo{volume}{63}},
  \bibinfo{pages}{1} (\bibinfo{year}{1991}).

\bibitem[{\citenamefont{Sokol et~al.}(1994)\citenamefont{Sokol, Glenister, and
  Singh}}]{sokol94}
\bibinfo{author}{\bibfnamefont{A.}~\bibnamefont{Sokol}},
  \bibinfo{author}{\bibfnamefont{R.~L.} \bibnamefont{Glenister}},
  \bibnamefont{and} \bibinfo{author}{\bibfnamefont{R.~R.~P.}
  \bibnamefont{Singh}}, \bibinfo{journal}{Phys.~Rev.~Lett.~}
  \textbf{\bibinfo{volume}{72}}, \bibinfo{pages}{1549} (\bibinfo{year}{1994}).

\bibitem[{\citenamefont{Abrikosov and Migdal}(1970)}]{abrmig70}
\bibinfo{author}{\bibfnamefont{A.~A.} \bibnamefont{Abrikosov}}
  \bibnamefont{and} \bibinfo{author}{\bibfnamefont{A.~A.}
  \bibnamefont{Migdal}}, \bibinfo{journal}{J.~Low~Temp.~Phys.~}
  \textbf{\bibinfo{volume}{3}}, \bibinfo{pages}{519} (\bibinfo{year}{1970}).

\bibitem[{\citenamefont{Coleman}(1984)}]{col84}
\bibinfo{author}{\bibfnamefont{P.}~\bibnamefont{Coleman}},
  \bibinfo{journal}{Phys.~Rev.~B} \textbf{\bibinfo{volume}{29}},
  \bibinfo{pages}{3035} (\bibinfo{year}{1984}).

\bibitem[{\citenamefont{Bickers}(1987)}]{bic87}
\bibinfo{author}{\bibfnamefont{N.~E.} \bibnamefont{Bickers}},
  \bibinfo{journal}{Rev.~Mod.~Phys.~} \textbf{\bibinfo{volume}{59}},
  \bibinfo{pages}{846} (\bibinfo{year}{1987}).

\bibitem[{\citenamefont{Grewe and Keiter}(1981)}]{grekei81}
\bibinfo{author}{\bibfnamefont{N.}~\bibnamefont{Grewe}} \bibnamefont{and}
  \bibinfo{author}{\bibfnamefont{H.}~\bibnamefont{Keiter}},
  \bibinfo{journal}{Phys.~Rev.~B} \textbf{\bibinfo{volume}{24}},
  \bibinfo{pages}{4420} (\bibinfo{year}{1981}).

\bibitem[{\citenamefont{Kuramoto}(1985)}]{kur85}
\bibinfo{author}{\bibfnamefont{Y.}~\bibnamefont{Kuramoto}}, in
  \emph{\bibinfo{booktitle}{{Theory of Heavy Fermions {and} Valence
  Fluctuations}}}, edited by
  \bibinfo{editor}{\bibfnamefont{T.}~\bibnamefont{Kasuya}} \bibnamefont{and}
  \bibinfo{editor}{\bibfnamefont{T.}~\bibnamefont{Saso}}
  (\bibinfo{publisher}{Springer}, \bibinfo{address}{Berlin},
  \bibinfo{year}{1985}), p. \bibinfo{pages}{152}.

\bibitem[{\citenamefont{{C.~I.~Kim} et~al.}(1987)\citenamefont{{C.~I.~Kim},
  {Y.~Kuramoto}, and {T.~Kasuya}}}]{kimkurkas87}
\bibinfo{author}{\bibnamefont{{C.~I.~Kim}}},
  \bibinfo{author}{\bibnamefont{{Y.~Kuramoto}}}, \bibnamefont{and}
  \bibinfo{author}{\bibnamefont{{T.~Kasuya}}},
  \bibinfo{journal}{Sol.~State~Comm.~} \textbf{\bibinfo{volume}{62}},
  \bibinfo{pages}{627} (\bibinfo{year}{1987}).

\bibitem[{\citenamefont{Metzner and Vollhardt}(1989)}]{metvol89}
\bibinfo{author}{\bibfnamefont{W.}~\bibnamefont{Metzner}} \bibnamefont{and}
  \bibinfo{author}{\bibfnamefont{D.}~\bibnamefont{Vollhardt}},
  \bibinfo{journal}{Phys.~Rev.~Lett.~} \textbf{\bibinfo{volume}{62}},
  \bibinfo{pages}{324} (\bibinfo{year}{1989}).

\bibitem[{\citenamefont{Newns and Read}(1987)}]{newrea87}
\bibinfo{author}{\bibfnamefont{D.~M.} \bibnamefont{Newns}} \bibnamefont{and}
  \bibinfo{author}{\bibfnamefont{N.}~\bibnamefont{Read}},
  \bibinfo{journal}{Adv.~Physics} \textbf{\bibinfo{volume}{36}},
  \bibinfo{pages}{799} (\bibinfo{year}{1987}).

\bibitem[{\citenamefont{Affleck et~al.}(1988)\citenamefont{Affleck, Zou, Hsu,
  and Anderson}}]{aff88}
\bibinfo{author}{\bibfnamefont{I.}~\bibnamefont{Affleck}},
  \bibinfo{author}{\bibfnamefont{Z.}~\bibnamefont{Zou}},
  \bibinfo{author}{\bibfnamefont{T.}~\bibnamefont{Hsu}}, \bibnamefont{and}
  \bibinfo{author}{\bibfnamefont{P.~W.} \bibnamefont{Anderson}},
  \bibinfo{journal}{Phys.~Rev.~B} \textbf{\bibinfo{volume}{38}},
  \bibinfo{pages}{745} (\bibinfo{year}{1988}).

\bibitem[{\citenamefont{Negele and Orland}(1988)}]{negorl}
\bibinfo{author}{\bibfnamefont{J.~W.} \bibnamefont{Negele}} \bibnamefont{and}
  \bibinfo{author}{\bibfnamefont{H.}~\bibnamefont{Orland}},
  \emph{\bibinfo{title}{Quantum Many-Particle Systems}}
  (\bibinfo{publisher}{Addison-Wesley}, \bibinfo{address}{Menlo Part etc.~},
  \bibinfo{year}{1988}).

\bibitem[{\citenamefont{Ioffe and Larkin}(1989)}]{iof89}
\bibinfo{author}{\bibfnamefont{L.~B.} \bibnamefont{Ioffe}} \bibnamefont{and}
  \bibinfo{author}{\bibfnamefont{A.~I.} \bibnamefont{Larkin}},
  \bibinfo{journal}{Phys.~Rev.~B} \textbf{\bibinfo{volume}{39}},
  \bibinfo{pages}{8988} (\bibinfo{year}{1989}).

\bibitem[{\citenamefont{Kroha et~al.}(1992)\citenamefont{Kroha, Hirschfeld,
  Muttalib, and W{\"o}lfle}}]{kroetal92}
\bibinfo{author}{\bibfnamefont{J.}~\bibnamefont{Kroha}},
  \bibinfo{author}{\bibfnamefont{P.~J.} \bibnamefont{Hirschfeld}},
  \bibinfo{author}{\bibfnamefont{K.~A.} \bibnamefont{Muttalib}},
  \bibnamefont{and}
  \bibinfo{author}{\bibfnamefont{P.}~\bibnamefont{W{\"o}lfle}},
  \bibinfo{journal}{Sol.~State~Comm.~} \textbf{\bibinfo{volume}{83}},
  \bibinfo{pages}{1003} (\bibinfo{year}{1992}).

\bibitem[{\citenamefont{Ubbens and Lee}(1992)}]{ubblee92}
\bibinfo{author}{\bibfnamefont{M.~U.} \bibnamefont{Ubbens}} \bibnamefont{and}
  \bibinfo{author}{\bibfnamefont{P.~A.} \bibnamefont{Lee}},
  \bibinfo{journal}{Phys.~Rev.~B} \textbf{\bibinfo{volume}{46}},
  \bibinfo{pages}{8434} (\bibinfo{year}{1992}).

\bibitem[{\citenamefont{Brinckmann and A.Lee}(2002)}]{bri02}
\bibinfo{author}{\bibfnamefont{J.}~\bibnamefont{Brinckmann}} \bibnamefont{and}
  \bibinfo{author}{\bibfnamefont{P.}~\bibnamefont{A.Lee}},
  \bibinfo{journal}{Phys.~Rev.~B} \textbf{\bibinfo{volume}{65}},
  \bibinfo{pages}{014502} (\bibinfo{year}{2002}).

\bibitem[{\citenamefont{Anderson}(1987)}]{rvbphil}
\bibinfo{author}{\bibfnamefont{P.~W.} \bibnamefont{Anderson}},
  \bibinfo{journal}{Science} \textbf{\bibinfo{volume}{235}},
  \bibinfo{pages}{1196} (\bibinfo{year}{1987}).

\bibitem[{\citenamefont{Marston and Affleck}(1989)}]{mar89}
\bibinfo{author}{\bibfnamefont{J.~B.} \bibnamefont{Marston}} \bibnamefont{and}
  \bibinfo{author}{\bibfnamefont{I.}~\bibnamefont{Affleck}},
  \bibinfo{journal}{Phys.~Rev.~B} \textbf{\bibinfo{volume}{39}},
  \bibinfo{pages}{11538} (\bibinfo{year}{1989}).

\bibitem[{\citenamefont{Ruckenstein et~al.}(1987)\citenamefont{Ruckenstein,
  Hirschfeld, and Appel}}]{ruchir87}
\bibinfo{author}{\bibfnamefont{A.~E.} \bibnamefont{Ruckenstein}},
  \bibinfo{author}{\bibfnamefont{P.~J.} \bibnamefont{Hirschfeld}},
  \bibnamefont{and} \bibinfo{author}{\bibfnamefont{J.}~\bibnamefont{Appel}},
  \bibinfo{journal}{Phys.~Rev.~B} \textbf{\bibinfo{volume}{36}},
  \bibinfo{pages}{857} (\bibinfo{year}{1987}).

\bibitem[{\citenamefont{Nagaosa and Lee}(1992)}]{naglee92}
\bibinfo{author}{\bibfnamefont{N.}~\bibnamefont{Nagaosa}} \bibnamefont{and}
  \bibinfo{author}{\bibfnamefont{P.~A.} \bibnamefont{Lee}},
  \bibinfo{journal}{Phys.~Rev.~B} \textbf{\bibinfo{volume}{45}},
  \bibinfo{pages}{966} (\bibinfo{year}{1992}).

\bibitem[{\citenamefont{Fukuyama}(1992)}]{fuk92}
\bibinfo{author}{\bibfnamefont{H.}~\bibnamefont{Fukuyama}},
  \bibinfo{journal}{Prog.~Theoret.~Phys.~Suppl.~}
  \textbf{\bibinfo{volume}{108}}, \bibinfo{pages}{287} (\bibinfo{year}{1992}).

\bibitem[{\citenamefont{Affleck and Haldane}(1987)}]{affhal87}
\bibinfo{author}{\bibfnamefont{I.}~\bibnamefont{Affleck}} \bibnamefont{and}
  \bibinfo{author}{\bibfnamefont{F.~D.~M.} \bibnamefont{Haldane}},
  \bibinfo{journal}{Phys.~Rev.~B} \textbf{\bibinfo{volume}{36}},
  \bibinfo{pages}{5291} (\bibinfo{year}{1987}).

\bibitem[{\citenamefont{Singh}(1991)}]{sin91}
\bibinfo{author}{\bibfnamefont{A.}~\bibnamefont{Singh}},
  \bibinfo{journal}{Phys.~Rev.~B} \textbf{\bibinfo{volume}{43}},
  \bibinfo{pages}{3617} (\bibinfo{year}{1991}).

\bibitem[{\citenamefont{Fradkin}(1991)}]{fradkin}
\bibinfo{author}{\bibfnamefont{E.}~\bibnamefont{Fradkin}},
  \emph{\bibinfo{title}{Field Theories of Condensed Matter Systems}}
  (\bibinfo{publisher}{Addison-Wesley}, \bibinfo{address}{Reading, MA},
  \bibinfo{year}{1991}).

\bibitem[{\citenamefont{Forster}(1975, 1990)}]{forster}
\bibinfo{author}{\bibfnamefont{D.}~\bibnamefont{Forster}},
  \emph{\bibinfo{title}{Hydrodynamic Fluctuations, Broken Symmetry, and
  Correlation Functions}} (\bibinfo{publisher}{Benjamin},
  \bibinfo{address}{Reading, MA}, \bibinfo{year}{1975, 1990}).

\bibitem[{\citenamefont{Luttinger and Ward}(1960)}]{lutwar60}
\bibinfo{author}{\bibfnamefont{J.~M.} \bibnamefont{Luttinger}}
  \bibnamefont{and} \bibinfo{author}{\bibfnamefont{J.~C.} \bibnamefont{Ward}},
  \bibinfo{journal}{Phys.~Rev.~} \textbf{\bibinfo{volume}{118}},
  \bibinfo{pages}{1417} (\bibinfo{year}{1960}).

\bibitem[{\citenamefont{Baym}(1962)}]{bay62}
\bibinfo{author}{\bibfnamefont{G.}~\bibnamefont{Baym}},
  \bibinfo{journal}{Phys.~Rev.~} \textbf{\bibinfo{volume}{127}},
  \bibinfo{pages}{1391} (\bibinfo{year}{1962}).

\bibitem[{\citenamefont{Baym and Kadanoff}(1961)}]{baykad61}
\bibinfo{author}{\bibfnamefont{G.}~\bibnamefont{Baym}} \bibnamefont{and}
  \bibinfo{author}{\bibfnamefont{L.~P.} \bibnamefont{Kadanoff}},
  \bibinfo{journal}{Phys.~Rev.~} \textbf{\bibinfo{volume}{124}},
  \bibinfo{pages}{287} (\bibinfo{year}{1961}).

\bibitem[{\citenamefont{Martin and Schwinger}(1959)}]{marschw59}
\bibinfo{author}{\bibfnamefont{P.~C.} \bibnamefont{Martin}} \bibnamefont{and}
  \bibinfo{author}{\bibfnamefont{J.}~\bibnamefont{Schwinger}},
  \bibinfo{journal}{Phys.~Rev.~} \textbf{\bibinfo{volume}{115}},
  \bibinfo{pages}{1342} (\bibinfo{year}{1959}).

\bibitem[{\citenamefont{Chen et~al.}(1999)\citenamefont{Chen, Kosztin,
  Jank{\'o}, and Levin}}]{levin99}
\bibinfo{author}{\bibfnamefont{Q.}~\bibnamefont{Chen}},
  \bibinfo{author}{\bibfnamefont{I.}~\bibnamefont{Kosztin}},
  \bibinfo{author}{\bibfnamefont{B.}~\bibnamefont{Jank{\'o}}},
  \bibnamefont{and} \bibinfo{author}{\bibfnamefont{K.}~\bibnamefont{Levin}},
  \bibinfo{journal}{Phys.~Rev.~B} \textbf{\bibinfo{volume}{59}},
  \bibinfo{pages}{7083} (\bibinfo{year}{1999}).

\bibitem[{not({\natexlab{c}})}]{note-phiappr}
\bibinfo{note}{Usually the free energy $W$ is constructed as a functional of
  the full (renormalized) Green's function $G$ and the self energy $\Sigma$ of
  the fermions\cite{lutwar60,bay62}, $W= W[G,\Sigma]= \Phi[G] + \ldots$\,. This
  leads to $\Phi$-derivable approximations where $\Sigma=
  \frac{1}{2}\delta\Phi/\delta G$ is a functional of $G$\,, while the bare
  propagator $G^0$ does not enter $\Phi$\,, $\Sigma$\,. Alternatively, $W$ can
  be written as a functional of the renormalized interaction $D$ and its
  irreducible part $\Pi$\,, $W= W[D,\Pi]= \Phi[D] + \ldots$\,, following the
  line of Ref.\ \onlinecite{lutwar60}\,. $G^0$ is considered a constant, like
  the interaction $J$\,, and may enter $\Phi[D]$ and therefore $\Pi=
  \delta\Phi/\delta D$ to arbitrary order. However, for the particular
  approximation shown in Fig.\ \ref{fig-red} we were not able to find the
  corresponding functional $\Phi[D]$\,.}

\bibitem[{not({\natexlab{d}})}]{note-cfluctmf}
\bibinfo{note}{Note that neither a RPA-like bubble series nor vertex
  corrections to the bubble need to be considered, since the fermion
  self-energy merely consists of a Hartree diagram with spin--spin
  interaction.}

\bibitem[{\citenamefont{Rickayzen}(1980)}]{rick}
\bibinfo{author}{\bibfnamefont{G.}~\bibnamefont{Rickayzen}},
  \emph{\bibinfo{title}{Green's Functions {and} Condensed Matter}}
  (\bibinfo{publisher}{Academic Press}, \bibinfo{address}{London},
  \bibinfo{year}{1980}).

\bibitem[{\citenamefont{Amit}(1984)}]{amitbook}
\bibinfo{author}{\bibfnamefont{D.~J.} \bibnamefont{Amit}},
  \emph{\bibinfo{title}{Field Theory, the Renormalization Group, and Critical
  Phenomena}} (\bibinfo{publisher}{World Scientific}, \bibinfo{year}{1984}),
  \bibinfo{edition}{2nd} ed.

\bibitem[{\citenamefont{Anderson et~al.}(2004)\citenamefont{Anderson, Lee,
  Randeria, Rice, Trivedi, and Zhang}}]{philrev03}
\bibinfo{author}{\bibfnamefont{P.~W.} \bibnamefont{Anderson}},
  \bibinfo{author}{\bibfnamefont{P.~A.} \bibnamefont{Lee}},
  \bibinfo{author}{\bibfnamefont{M.}~\bibnamefont{Randeria}},
  \bibinfo{author}{\bibfnamefont{T.~M.} \bibnamefont{Rice}},
  \bibinfo{author}{\bibfnamefont{N.}~\bibnamefont{Trivedi}}, \bibnamefont{and}
  \bibinfo{author}{\bibfnamefont{F.~C.} \bibnamefont{Zhang}},
  \bibinfo{journal}{J.~Phys.:~Cond.~Mat.~} \textbf{\bibinfo{volume}{16}},
  \bibinfo{pages}{R755} (\bibinfo{year}{2004}).

\end{thebibliography}
\end{document}